\documentclass[superscriptaddress,showpacs,twocolumn,notitlepage,nofootinbib,aps,pre, 10pt]{revtex4-2}
\usepackage[normalem]{ulem}
\usepackage{graphicx}
\usepackage{color}
\usepackage{xcolor}
\usepackage{verbatim}
\usepackage{subfigure}
\usepackage{amssymb}
\usepackage{amsmath}
\usepackage{amsfonts}
\usepackage{comment}
\usepackage{chngcntr}
\usepackage{appendix}
\usepackage{lipsum}
\usepackage{dsfont}
\usepackage{xfrac}
\usepackage[makeroom]{cancel}
\usepackage{makecell}
\usepackage{lipsum}
\usepackage{tikz}
\usepackage{circledsteps}
\usepackage{CJKutf8}
\usepackage{soul}
\usepackage{tocvsec2}

\newcommand{\subsectionnotoc}[1]{%
  {\let\addcontentsline\relax\subsection{#1}}%
}

\newcommand{\bGamma}{\boldsymbol{\Gamma}}
\newcommand{\blambda}{\boldsymbol{\lambda}}

\newcommand{\dd}{\text{d}}
\newcommand{\bz}{{\bm{z}}}

\newcommand{\ee}{\text{e}}

\newcommand{\p}{\partial}
\newcommand{\bx}{{\bm{x}}}

\newcommand{\bT}{\text{\bf T}}

\newcommand{\bt}{{\bm{t}}}

\newcommand{\bu}{{\bm{u}}}

\newcommand{\bB}{\text{\bf B}}

\newcommand{\bJ}{\text{\bf J}}
\newcommand{\bJWPE}{\text{\bf J}^\text{\tiny{WPE}}}

\newcommand{\bH}{{\bf H}}
\newcommand{\bv}{{\bm{v}}}
\newcommand{\bs}{{\bm{s}}}

\newcommand{\bSigma}{\boldsymbol{\Sigma}}
\newcommand{\bphi}{\boldsymbol{\phi}}
\newcommand{\bpsi}{\boldsymbol{\psi}}

\newcommand{\bnabla}{\boldsymbol{\nabla}}

\newcommand{\bmm}{{\bm{m}}}
\newcommand{\bQ}{\text{\bf Q}}
\newcommand{\ERSWPE}{E_\text{\tiny{eff}}^\text{\tiny{WPE}}}
\newcommand{\PRSWPE}{P_\text{\tiny{RS}}^\text{\tiny{WPE}}}
\newcommand{\EWPE}{E^\text{\tiny{WPE}}}
\newcommand{\EWPEMF}{E^\text{\tiny{WPE}}_\text{\tiny{MF}}}
\newcommand{\EWPERS}{E^\text{\tiny{WPE}}_\text{\tiny{eff}}}
\newcommand{\bJSK}{\text{\bf J}^\text{\tiny{SK}}}
\newcommand{\bHSK}{\text{\bf H}^\text{\tiny{SK}}}
\newcommand{\OmegaSK}{\Omega^\text{\tiny{SK}}}
\newcommand{\ZOmega}{Z_\Omega^\text{\tiny{SK}}}

\newcommand{\by}{\text{\bf y}}
\newcommand{\bw}{{\bm{w}}}

\newcommand{\bA}{\text{\bf A}}
\newcommand{\bP}{\text{\bf P}}
\newcommand{\bHWPE}{\text{\bf H}^{\text{\tiny{WPE}}}}
\newcommand{\ZRSSK}{Z^{\text{\tiny{SK}}}_\text{\tiny{eff}}}
\newcommand{\ERSSK}{E^{\text{\tiny{SK}}}_\text{\tiny{eff}}}
\newcommand{\mIMF}{\mathcal{I}_\text{\tiny{MF}}}

\newcommand{\bM}{\text{\bf M}}

\newcommand{\bK}{\text{\bf K}}

\newcommand{\bepsilon}{\boldsymbol{\epsilon}}

\newcommand{\mI}{\mathcal{I}}

\newcommand{\EOmegaeff}{E_{\scriptstyle \Omega,\text{\tiny{eff}}}}

\DeclareMathOperator*{\argmax}{arg\,max}
\DeclareMathOperator*{\argmin}{arg\,min}
\DeclareMathOperator{\Tr}{Tr}

\DeclareMathOperator{\sgn}{sgn}
\DeclareMathOperator{\Crt}{Crt}

\newcommand{\nocontentsline}[3]{}
\let\origcontentsline\addcontentsline
\newcommand\stoptoc{\let\addcontentsline\nocontentsline}
\newcommand\resumetoc{\let\addcontentsline\origcontentsline}

\usepackage{booktabs}
\usepackage{pifont}
\usepackage{xcolor}

\newcommand{\subfigref}[2]{\hyperref[#1]{\ref*{#1}#2}} 
\usepackage{amsthm}
\usepackage{bm}
\usepackage[hang, flushmargin]{footmisc}
\usepackage[colorlinks=true, linkcolor=blue, citecolor=blue]{hyperref}
\usepackage{cleveref}
\crefname{equation}{Eq.}{Eqs.} 
\crefrangelabelformat{equation}{(#3#1#4--#5#2#6)}

\newcommand{\tocless}[2]{\bgroup\let\addcontentsline=\nocontentsline#1{#2}\egroup}

\begin{document}

\title{The geometry and dynamics of annealed optimization \\ in the coherent Ising machine with hidden and planted solutions}

\author{Federico Ghimenti}
\affiliation{Department of Applied Physics, Stanford University, Stanford, CA 94305, USA}
\author{Adithya Sriram}
\affiliation{Department of Physics, Stanford University, Stanford, CA 94305, USA}
\author{Atsushi Yamamura (\begin{CJK}{UTF8}{ipxm}山村 篤志\end{CJK})}
\affiliation{Department of Applied Physics, Stanford University, Stanford, CA 94305, USA}
\author{Hideo Mabuchi}
\affiliation{Department of Applied Physics, Stanford University, Stanford, CA 94305, USA}
\author{Surya Ganguli}
\affiliation{Department of Applied Physics, Stanford University, Stanford, CA 94305, USA}

\setcounter{tocdepth}{0}
\begin{abstract}
The coherent Ising machine (CIM) is a nonconventional hardware architecture for finding approximate solutions to large-scale combinatorial optimization problems. It operates by annealing a laser gain parameter to adiabatically deform a high-dimensional energy landscape over a set of soft spins, going from a simple convex landscape to the more complex optimization landscape of interest. We address how the evolving energy landscapes guides the optimization dynamics against problems with hidden planted solutions. We study the Sherrington-Kirkpatrick spin-glass with ferromagnetic couplings that favor a hidden configuration by combining the replica method, random matrix theory, the Kac-Rice method and dynamical mean field theory. We characterize energy, number, location, and Hessian eigenspectra of global minima, local minima, and critical points as the landscape evolves increasing the laser gain. We find that low energy global minima develop soft-modes which the optimization dynamics can exploit to descend the energy landscape. Even when these global minima are aligned to the hidden configuration, there can be exponentially many higher energy local minima that are all unaligned with the hidden solution. Nevertheless, the annealed optimization dynamics can evade this cloud of unaligned high energy local minima and descend near to aligned lower energy global minima. Eventually, as the landscape is further annealed, these global minima become rigid, terminating any further optimization gains from annealing. We further consider a second optimization problem, the Wishart planted ensemble, which contains a hidden planted solution in a landscape with tunable ruggedness. We describe CIM phase transitions between recoverability and non-recoverability of the hidden solution. Overall, we find intriguing relations between high-dimensional geometry and dynamics in analog machines for combinatorial optimization.
\end{abstract}

\maketitle
\section{Statistical physics \\for analog computing}
The solution of combinatorial optimization problems~\cite{papadimitriou1998combinatorial} is central to many domains of mathematics and science, ranging from information theory and computer science to robotics and circuit design. The development of algorithms and technologies capable of reducing the computational cost of finding solutions is thus of fundamental importance. In recent years, unconventional hardware architectures~\cite{wang2019oim,molnar2020accelerating,mohseni2022ising, mallick2020using} that find approximate solutions through the nonlinear dynamical evolution of analog systems, have been developed and implemented to tackle NP-hard problems. The potential advantage of these approaches over conventional digital systems lies in exploiting the exotic analog dynamics they can implement, as well as in the possibility of scaling up these devices to unprecedented sizes. Among these architectures, the coherent Ising machine (CIM)~\cite{wang2013coherent} stands out, as in the last decade the number of degrees of freedom implemented in a single device has grown from a hundred~\cite{mcmahon2016fully} to thousands~\cite{inagaki2016coherent} to tens of thousands~\cite{honjo2021100}. 

The CIM~\cite{yamamoto2020coherent} is a heuristic solver for quadratic, unconstrained binary optimization (QUBO) problems, involving the minimization of an energy function over $N$ pairwise interacting binary spins. Many problems of practical interest can be formulated in this way, including the traveling salesman problem, the partitioning problem and the graph coloring problem~\cite{lucas2014ising}. Rather than minimizing an energy function over a set of discrete variables, the CIM minimizes an energy function over a set of continuous degrees of freedom, or soft spins, within an optical network. Moreover, as a certain laser gain parameter is slowly annealed over time, the single spin energy landscape evolves from a single well quartic potential, to a soft, wide double well potential, to a sharp or rigid double well potential.  This in turn induces the high dimensional CIM energy landscape  geometry over all spins to also adiabatically evolve, as the soft spins are forced to become binary spins. The final endpoint of this evolution is mapped back to a binary spin configuration yielding an approximate solution to the original optimization problem of interest. 

Despite the proliferation of a diverse set of technological advances~\cite{yamamoto2020coherent,reifenstein2021coherent,lu2023speed,syed2023physics, mastiyage2023mean, zhou2025frustration}, we currently lack a theoretical framework to understand the success and failure modes of the CIM, and of analog computing systems in general.  On the other hand, from the perspective of statistical physics, the problem of solving combinatorial optimization problems is equivalent to the problem of finding ground-state energy configurations in high-dimensional disordered systems. Indeed an arsenal of tools from spin-glass theory ~\cite{mezard1987spin,nishimori2001statistical,parisi2020theory,charbonneau2023spin} have been developed and applied, not only to optimization problems, but also to information theory~\cite{sourlas1989spin, sourlas1994spin, mezard2009information}, machine learning~\cite{sarao2019afraid,agliari2020machine, bahri2020statistical}, and ecology and evolution~\cite{may1972will, diederich1989replicators, derrida1991evolution, bunin2017ecological, agarwala2019adaptive}. The techniques developed yield insights into the structure of ground states and metastable states of the energy landscape of disordered systems, as well as their dynamical properties. These techniques, including the replica method, cavity method, Kac-Rice formula, and dynamic mean field theory, have  only recently started to find applications and extensions to the realm of analog computing devices~\cite{yamamuraGeometricLandscapeAnnealing2024}. In particular, we still do not have an understanding of how the CIM energy landscape geometry and dynamics changes under annealing, when the connectivity of the system is non-random enough so as to bias low energy configurations towards a hidden, or planted configuration originating from structure in the connectivity. 

In this work, we leverage techniques from the statistical physics of disordered systems to understand the energy landscape geometry and the efficacy of the CIM in solving two prototypical NP-hard problems with such hidden, planted configurations. We first consider the case of the Sherrington-Kirkpatrick (SK) model with additional rank $1$ ferromagnetic coupling~\cite{sherrington1975solvable}, and we present a phase diagram for the properties of the global minima and most abundant local minima as the laser gain annealing parameter and the strength of the ferromagnetic coupling are varied. 
\begin{figure*}
    \centering 
    \includegraphics[width=\linewidth]{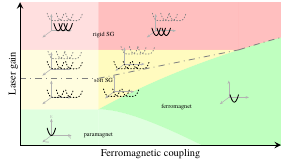}
    \caption{\textbf{Schematic phase diagram for global minima and most abundant local minima of the coherent Ising machine}. In each phase, we summarize the properties of global (black curves) and local (grey curves) minima by their positions in a three-dimensional coordinate system indicating their typical energy $E$, square distance from the origin $q$, and the net magnetization $m$. Additionally, we indicate whether either minima are soft (wide dashed curves) or rigid (sharp solid curves). At low laser gain and ferromagnetic coupling (Sec.~\ref{sec:Hessian_origin}), the system is in a paramagnetic phase, with a single rigid global minimum located at the origin (light green region). If the ferromagnetic coupling is large enough, further increase of the laser gain leads to a transition of the Baik-Ben Arous Peché type toward a ferromagnetic phase (dark green region, Sec.~\ref{sec:phase_boundary_para_to_ferro}). As the laser gain increases further, both the paramagnetic and the ferromagnetic phases become unstable toward a spin-glass (SG) phase (yellow and red regions). This overall SG phase can be subdivided into two phases, according to the behavior of the spectrum of the Hessian at global minima (analyzed in Sec.~\ref{sec:hessian_spin_glass}). For intermediate values of laser gain and ferromagnetic couplings, the lower edge of the Hessian at global minima touches the origin yielding a 'soft' spin-glass (light paramagnetic and dark ferromagnetic yellow regions). At larger values of the laser gain or ferromagnetic coupling, the spectrum of the Hessian at global minima is gapped away from the origin, yielding a rigid spin-glass (light paramagnetic and dark ferromagnetic red regions). Using a replicated Kac-Rice calculation (Sec~\ref{sec:Kac-Rice}), we further add to this picture by studying the properties of the most abundant local minima across the phase diagram. This yields a further subdivision of the soft spin-glass phase by a critical line (gray dash-dotted line) into two sub-phases. Above the critical line (but not below), local minima proliferate at an exponential rate with the size of the system. Interestingly, these minima are paramagnetic for {\it any} value of the ferromagnetic coupling, unlike the global minima, exist at higher energy than the global minima, and they are soft, with a gapless Hessian eigenspectrum. Furthermore, these exponentially many higher energy paramagnetic local minima persist into rigid spin-glass phase, and they retain their soft character there, despite the fact that global minima become rigid (light and dark red regions)  Numerical simulations and dynamical mean field theory (Sec.~\ref{sec:dynamics}) demonstrate that the annealed optimization dynamics of the CIM visits spin configurations that are much more similar to global minima than to the most abundant local minima. Importantly, annealing across the soft SG phase allows the CIM to evade higher energy local minima and reach better approximate solutions to the original combinatorial optimization problem, by exploiting soft modes of near global minima to descend. But when the annealed optimization dynamics enters the rigid SG phase, the dynamics finally terminates and the energy of the solution to the combinatorial optimization problem does not decrease further. \label{fig:cartoon}}
\end{figure*}

Our main finding for this model is schematically depicted in Fig.~\ref{fig:cartoon}. We unveil a rich phase diagram, for both global and local minima. At low value of the laser gain, the ground state of the CIM is either a paramagnet or a ferromagnet, depending on the strength of the ferromagnetic coupling. Further increase of the laser gain increases the ruggedness of the landscape, leading to a transition into a spin-glass phase. The latter can be classified into two distinct categories, depending on the features of the nonlinear excitations around global minima. Indeed, for intermediate values of laser gain and ferromagnetic coupling,  the spectrum of the Hessian at global minima is gapless, indicating many soft modes, and the support of the distribution of the soft spin population spans the whole real axis, with a nonzero value of its density at the origin. We call this phase a 'soft' spin-glass. Upon further increase of the laser gain, the spectrum of the Hessian at global minima develops a gap away from $0$, and the support of the distribution of the population of soft spins splits into two disconnected domains.  This corresponds to a rigid spin-glass phase, where in global minima, there is an absence of soft modes and all spins are strongly committed to finite nonzero values. We show that the soft-to-rigid transitions occurs at a value of the laser gain for which the effective energy of a mean-field model that we derive becomes non-convex.   

Furthermore, by a combination of dynamical mean field theory and numerical simulations, we discover that the two spin-glass phases have differential impacts on the efficacy of the annealed optimization dynamics. Indeed, the properties of the critical points visited during the annealing process are very similar to those of global minima. Thus in the soft spin-glass phase, linear excitations corresponding to soft modes with small Hessian eigenvalues are exploited by the CIM to flip the sign of uncommitted spins with values near $0$, allowing the CIM to reach approximate solutions of the QUBO problem with lower and lower energies as the annealing proceeds. On the other hand, in the rigid spin-glass phase, further annealing does not yield improved solutions. Here, no soft modes exist to be exploited, all the spins are committed to nonzero values,  and the CIM gets stuck in a single configuration. The analytical determination of the boundaries between the soft and rigid spin-glass phases and the importance of this soft to rigid phase transition in governing the efficacy of the CIM annealed optimization dynamics, is one of the main technical results of this work.

{We also push our analysis of the CIM energy landscape geometry beyond the structure of global minima. Indeed, we characterize, by means of a replicated Kac-Rice calculation, the properties of the most abundant local minima and critical points of the CIM as a function of laser gain and the ferromagnetic coupling. We identify a linear phase transition boundary within the soft spin-glass phase (dashed-dotted line in the yellow regions of~\ref{fig:cartoon}, separating a phase where local minima and critical points grow subexponentially with system size below the line, and a phase where they proliferate exponentially with system size above the line. Interestingly, the properties of typical critical points and local minima are significantly different from those visited by the dynamics, both in terms of average energy and magnetization. In particular, in the phase where their number grows at an exponential rate with the system size, the most abundant local minima  always have zero net magnetization with soft, gapless Hessian eigenspectra.  Overall, the picture that emerges from our analysis is that as the laser gain is annealed from low to high, the CIM descends through the energy landscape by exploiting the soft modes in the vicinity of global minima, thereby passing beneath an exponentially large cloud of proliferating higher energy local minima when they first appear. 

{Last, but not least, we study the global minima of the CIM with the Wishart planted ensemble~\cite{hamze2020wishart}, a model with a predefined ground state and tunable ruggedness.  The main results of our analysis of this model are presented in the phase diagram in Fig.~\ref{fig:wpe_cartoon}. At low laser gain, the CIM has a single global minimum at the origin, independently of the ruggedness of the original combinatorial optimization problem. If the parameter controlling the ruggedness of the combinatorial optimization problem is sufficiently large, increasing the laser gain leads initially to a ferromagnetic phase, where full recovery of the planted solution is possible. However, in this phase, spectral methods are equally effective in retrieving the planted solution. If the laser gain is increased further, the global minima of the CIM undergoes a spin-glass transition, no matter whether the planted solution in the  combinatorial optimization problem was originally in a hard-to-retrieve or easy-to-retrieve phase. In the easy-to-retrieve phase, spectral initialization of the CIM allows recovery of the planted solution, despite the spin-glass structure of the global minima. In the hard-to-retrieve phase, at large value of the laser gain, the annealing dynamics of the CIM obtains approximate solutions with a lower energy compared to a spectral method.}

Our paper is organized as follows. In Sec.~\ref{sec:model} we introduce the model of the CIM with a Sherrington-Kirkpatrick connectivity matrix. In Sec.~\ref{sec:hessian_and_results} we give a heuristic argument for the importance of soft modes in the annealing process and present the phase diagram for the global minima of the CIM. After a first look at the properties of global minima for small values of the annealing parameter in Sec.~\ref{sec:Hessian_origin}, we present a replica calculation of the resolvent of the Hessian at global minima, and illustrate how the lower edge of the spectrum can be determined in Sec.~\ref{sec:replica_resolvent}. In Sec.~\ref{sec:global_minima} we study the properties of the convex and spin-glass phases of the system. In Sec.~\ref{sec:Kac-Rice} we address the properties of the most abundant local minima of the system through a replicated Kac-Rice calculation, and identify a phase transition toward a phase where  most abundant minima with zero net magnetization proliferate. In Sec.~\ref{sec:dynamics} we address the dynamics of the model, showing that the properties of the ground state are a good proxy for the performance of the system. In Sec.~\ref{sec:wpe}, we move to the case of the Wishart planted ensemble, and through a replica calculation and numerical simulations of the annealing process we identify convex and spin-glass phases for the CIM applied to this problem. Finally we present conclusions and future directions in Sec.~\ref{sec:conclusion}.

\section{Model}\label{sec:model}
The combinatorial optimization problem of interest in this work is finding the ground state of the Ising energy function $E_\mathrm{Ising}$
\begin{equation}\label{eq:Hcomb}
    E_\mathrm{Ising}(\bs) = -\frac{1}{2}\bs\cdot\bJ_\text{Ising}\bs\,,
\end{equation}
defined over $N$ binary spin variables $\bs \in \{-1,1\}^N$. The connectivity matrix $\bJ_\text{Ising}$ depends on the specific problem at hand, but it is usually a large matrix with entries of different signs, for which the energy landscape of $E_\text{Ising}$ can be highly rugged. Instead of directly searching for the ground state of Eq.~\eqref{eq:Hcomb}, the CIM implements the dynamical evolution of a set of interacting degenerate optical parametric oscillators~\cite{wang2013coherent}. Each oscillator consists in an optical cavity, and all the cavities are supplied by a laser pump of uniform intensity across all the oscillators. The oscillators are allowed to exchange light among each other with an intensity and phase shift tuned by the experimentalist, and the intensity of the laser pump can be adiabatically increased throughout the process. The phase shift of each oscillator with respect to the reference phase of the laser pump can be measured during the dynamical evolution of the optical network, and for large enough values of the laser intensity, each oscillator collapses on either a in-phase state or an opposition-of-phase state. This binary configurations can then be mapped back to Ising configurations. We thus model the phase differences of the oscillators in the CIM as a set of $N$ soft spins $x_i$ that evolve dynamically to minimize an energy $E(\bx,a)$. The quantity $a$ denotes, up to a shift and a rescaling, the strength of the laser gain, and it is   adiabatically changed during the optimization process. The total energy $E(\bx,a)$, is given by
\begin{equation}\label{eq:E_CIM}
    E(\bx,a) \equiv \sum_{i=1}^N E_I(x_i,a) - \frac{1}{2}\bx\cdot\bJ_\text{Ising}\bx\,.
\end{equation}
The laser gain strength controls the shape of the single-site internal energy $E_I(x,a)$ defined as 
\begin{equation}\label{eq:E_I_single_site}
    E_I(x,a) \equiv \frac{1}{4}x^4 - \frac{a}{2}x^2\,.
\end{equation}
When $a<0$, $E_I(x,a)$ is a convex function of $x$, with a single minimum located at the origin. When $a>0$, the function $E_I$ is a double-well potential, with two minima located at $\pm\sqrt{a}$, separated by an energy barrier of height $\frac{1}{2}a^2$. For a very low and negative value of the laser gain, $a\ll0$, at low energy, the soft spins are localized around the origin, while for a large and positive value of the laser gain, $a\gg 0$, the soft spins are akin to binary variables, as the single site energy $E_I$ favors them to reside within one of the two wells. The second term on the right hand side of Eq.~\eqref{eq:E_CIM} represents the interaction among the different optical cavities, and we assume that the experimentalist has tuned the phase modulation of the different interactions to match the connectivity matrix $\bJ_{\text{\tiny{Ising}}}$ of the optimization problem of interest. The dynamics of this soft-spin model in presence of thermal fluctuations and for fixed value of the laser gain has been an object of study in the statistical physics of spin-glasses~\cite{kosterlitz1976spherical, sompolinsky1981dynamic, sompolinsky1982relaxational, cugliandolo1994out}. For a fixed value of $a$, the CIM dynamics  can be approximated by minimization of the energy through the gradient descent dynamics
\begin{equation}\label{eq:gradient_descent}
    \tau \frac{\dd}{\dd t}\bx(t) = -\bnabla E(\bx(t),a)\,,
\end{equation}
where $\tau$ is a microscopic timescale of the system and where $\bnabla = [\p_{x_1},\,\p_{x_2},\,\ldots,\,\p_{x_N}]^\mathrm{T}$ is the gradient operator in the space of the soft spin variables. The system is initialized from a random configuration at a low value of the laser gain $a\ll0$. The laser gain is then increased adiabatically, with a timescale much slower than the one required by the system to converge to a local minimum through the gradient descent dynamics in Eq.~\eqref{eq:gradient_descent}. The annealing of the laser gain proceeds further up to a final value of $a\gg0$, where the single-site energy $E_I$ given by Eq.~\eqref{eq:E_I_single_site} constrains the spins $x_i$ to localize around the positions $\pm \sqrt{a}$. At the end of the annealing process, the final configuration of the soft-spin system is transformed into an Ising spin configuration through the mapping $s_i = \sgn x_i$. The CIM exploits the  soft nature of the continuous spins $x_i$ to reach low energy configurations of the optimization problem given by Eq.~\eqref{eq:Hcomb}. 

We work with a connectivity matrix $\bJ_\text{\tiny{Ising}}$ for which the ground state of the corresponding combinatorial optimization problem has a partial or complete alignment with a preferential direction $\bt$. The latter is chosen, without loss of generality, to be $\bt\equiv[1,\ldots,1]^\mathrm{T}$. The first model we consider is given by $\bJ^\text{Ising}=\bJ^\text{\tiny{SK}}$, where $\bJ^\text{\tiny{SK}}$ is the connectivity matrix of a fully-connected spin-glass  with a finite ferromagnetic  coupling
 $J_0$, akin to the Sherrington-Kirkpatrick spin-glass~\cite{sherrington1975solvable}. It is obtained summing a rank-1 matrix and a random matrix $\bJ$,
\begin{equation}\label{eq:JSK}
    \bJ^{\text{\tiny{SK}}} = \frac{J_0}{N}\bt\otimes\bt + \frac{1}{\sqrt{N}}\bJ\,.
\end{equation}
The matrix $\bJ$ is drawn from the Gaussian orthogonal ensemble~\cite{potters2020first}: it is a symmetric matrix whose entries are i.i.d Gaussian variables with zero mean and variance $\langle J_{ij}^2\rangle_\bJ = 1+\delta_{ij}$. We denote by $\langle\ldots\rangle_{\bJ}$ the average over different realizations of the disordered couplings $\bJ$. The strength $J_0$ of the rank-1 perturbation favors ferromagnetic ordering, and it competes against the disordered interactions in the system, encoded by $\bJ$. For $J_0<1$, the ground state of the Ising energy in Eq.~\eqref{eq:Hcomb} with $\bJ^\text{Ising}=\bJ^\text{\tiny{SK}}$ is  a spin-glass with zero net magnetization, while for $J_0>1$, the ground state is a spin-glass with nonzero net magnetization,  partially aligned along the direction $\bt$~\cite{toulouse1980mean, nishimori2001statistical}. The resulting Ising model is a paradigmatic example in the theory of spin-glasses~\cite{mezard1987spin}. It presents a rugged energy landscape, with the number of minima close to the ground state growing exponentially with the size of the system. These low energy minima are organized in a complex hierarchical structure, whose description, achieved by means of the full replica symmetry breaking theory, is a landmark achievement in the theory of disordered systems~\cite{parisi1980sequence,talagrand2006parisi}. An efficient message-passing algorithm to find solutions arbitrarily close to the ground-state has been recently developed~\cite{montanari2025optimization}, but the Sherrington-Kirkpatrick spin-glass constitutes nevertheless an important target on which to derive and test our understanding of theoretical principles governing the performance and evolving energy landscape geometry of the CIM.  

In what follows, we denote by $E^\text{\tiny{SK}}(\bx,a)$ the energy of the CIM when the connectivity matrix is given by $\bJSK$, and we refer to this model as CIM-SK. Thus we focus on the energy $E^\text{\tiny{SK}}(\bx,a)$ given by 
\begin{equation}\label{eq:E_CIM_SK}
    E^\text{\tiny{SK}}(\bx,a) \equiv \sum_{i=1}^N E_I(x_i,a) - \frac{1}{2}\bx\cdot \bJ^\text{\tiny{SK}}\bx\,,
\end{equation}
with $\bJ^{\text{\tiny{SK}}}$ given by Eq.~\eqref{eq:JSK}. The main idea we put forward in this work is that the distribution of eigenvalues of the Hessian of  $E^\text{\tiny{SK}}$ around low energy minima is a good proxy for determining when the annealing process is effective. In the following Section, we provide a heuristic argument motivating this claim. 

\section{Heuristics and phase diagram for global minima}\label{sec:hessian_and_results}

Let us consider the annealing dynamics from a perspective complementary to the gradient descent given by Eq.~\eqref{eq:gradient_descent}, by looking at how its fixed points move as the laser gain increases. Consider a critical point $\bx_c$ of the CIM-SK, such that $\bnabla E^\text{\tiny{SK}}(\bx,a)\rvert_{\bx=\bx_c}=\mathbf{0}$. We can track the motion of the critical point as the laser gain is infinitesimally changed by taking the derivative of the stationary condition with respect to $a$, thus obtaining
\begin{equation}\label{eq:dynamics_cp}
    \bH^\text{\tiny{SK}}(\bx_c,a)\frac{\dd \bx_c}{\dd a} = \bx_c(a)\,,
\end{equation}
where we introduced the Hessian matrix of the CIM-SK energy $\bH^\text{\tiny{SK}}(\bx,a)$, defined as
\begin{equation}
    H^\text{\tiny{SK}}_{ij}(\bx,a) \equiv \frac{\p^2 E^\text{\tiny{SK}}(\bx,a)}{\p x_i \p x_j}\,\,.
\end{equation}
Equation~\eqref{eq:dynamics_cp} describes the motion of a critical point during the annealing process, assuming that critical points respond linearly to an infinitesimal change of the laser gain $a$. This is true if the Hessian $\bH^\text{\tiny{SK}}(\bx_c,a)$ is invertible. However, if the Hessian is not invertible, as is the case if its spectrum contains a zero eigenvalue, then Eq.~\eqref{eq:dynamics_cp} no longer holds, and we expect the critical point to jump discontinuously to a new configuration as the laser gain increases. Under gradient descent dynamics, the system then flows to a novel critical point. This sudden change might involve a change in the sign of some of the soft spins $x_i$, possibly allowing to reach a lower Ising energy.

The argument above suggests that an interesting proxy of the effectiveness of the annealing process is given by the distribution of eigenvalues of the Hessian at the critical points visited by the CIM during the annealing dynamics. In general, the relationship between the geometry of the energy landscape of a disordered system and its dynamics during a rapid quench or a slow annealing process is a broad, long-standing and challenging question, for which unambiguous answers have been provided mainly in pure spin-glass models on a sphere~\cite{cugliandolo1993analytical}, while recent work highlight the many subtleties occurring in more general cases~\cite{folena2020rethinking, lang2025numerical, citro2025strong}. To further characterize the performance of the CIM and its connection with the geometry of its changing landscape, we investigate the properties of the deepest energy minima of the model. These minima are typically unreachable on a reasonable timescale, but after their characterization we can ask whether they leave some trace at higher energy levels, which can affect the performance of the CIM.  

In the following, we analyze the properties of global minima and the spectrum of the associated Hessian as a function of the laser gain $a$ and the ferromagnetic alignment strength $J_0$. The main result of our analysis is presented in Fig.~\ref{fig:phase_diagram_global_minima_SK}.
\begin{figure*}
    \includegraphics[width=\columnwidth]{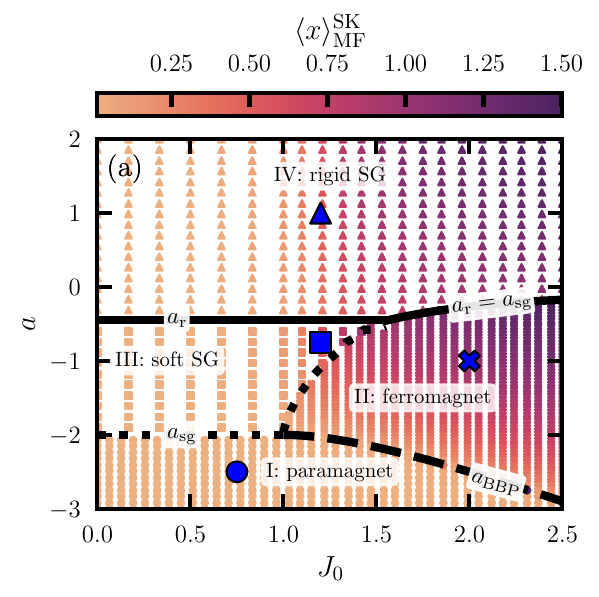}
    \includegraphics[width=\columnwidth]{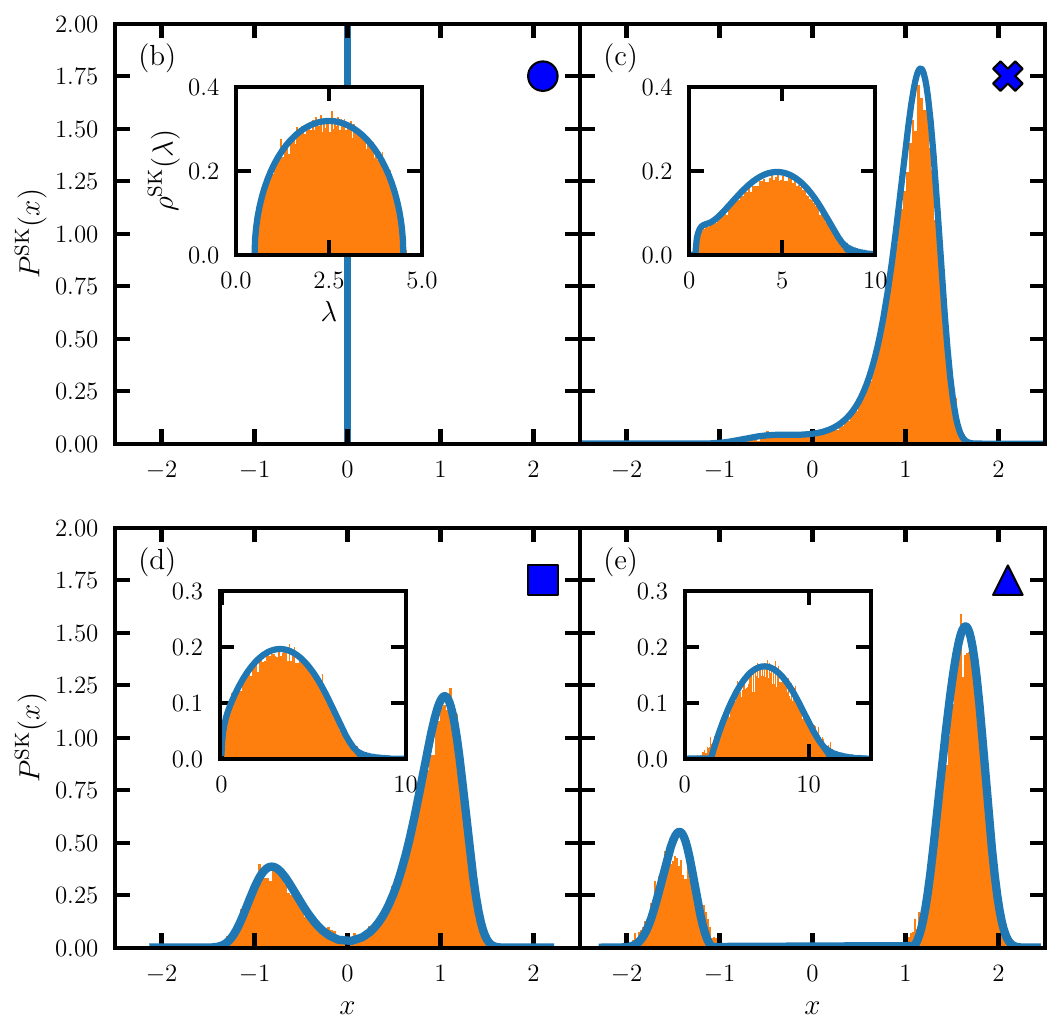}\caption{\textbf{Phase diagram of global minima of the CIM-SK with ferromagnetic bias} Panel (a): phase diagram in the $(J_0,\,a)$ plane: we identify a paramagnetic convex phase \Circled{I}, a ferromagnetic  phase \Circled{II}, a spin-glass phase \Circled{III} endowed with soft modes, and a rigid spin-glass phase \Circled{IV}. The color map in the plot is the mean value of the single spin distribution $P^\text{\tiny{SK}}(x)$ at global minima. The black lines identify different phase boundaries in the plane $(a,J_0)$. The dashed line $a_\text{\tiny{BBP}}(J_0)$ denotes the  boundary between the paramagnetic phase and the ferromagnetic phase. The dotted line $a_\text{\tiny{sg}}(J_0)$ is the  boundary between the replica symmetric phases and the spin-glass phase with soft modes at the global minima, and the solid line $a_\text{\tiny{r}}(J_0)$ is the boundary for the rigid spin-glass phase.  Panels (b-e): single-spin distribution $P^\text{\tiny{SK}}(x)$ and distribution of eigenvalues of the Hessian $\rho^\text{\tiny{SK}}(\lambda)$ (inset) at global minima in the different phases. The blue lines are the analytical results, while the orange histograms are lowest energy minima sampled from small, finite size systems. Panel (b): in the convex paramagnetic phase the soft spins are localized at the origin and the spectral distribution follows the Wigner semicircle law. Panel (c): in the ferromagnetic phase the distribution of the spin is skewed, with a finite probability density at the origin, and the spectrum of the Hessian is gapped. Panel (d): in the non-rigid spin-glass phase the single spin distribution is bimodal, with a nonzero probability density at the origin. The spectrum of the Hessian is gapless. Panel (e): in the rigid spin-glass phase, the single-spin probability density is nonzero in two disconnected domains along the $x$-axis, and the spectrum of the Hessian is gapped.\label{fig:phase_diagram_global_minima_SK} }
\end{figure*}
In panel (a), we reveal a phase diagram of the global minima of the system. The phase diagram exhibits four different regions, characterized by different properties of the distribution $P^\text{\tiny{SK}}(x)$ of single spin values, and the spectrum of the Hessian $\rho^\text{\tiny{SK}}(\lambda)$. In region \Circled{I},  there is a single, paramagnetic global minimum: the energy landscape is convex, and the spin distribution is concentrated around the origin. In region \Circled{II}, the system is a ferromagnet, with two global minima related by an inversion symmetry, and the global minimum of the system is ferromagnetic. The single spin distribution is skewed (panel (c)). In region \Circled{III} and \Circled{IV} the global minima have a spin-glass structure, with a rugged energy landscape, but the properties of the Hessian and of the single spin distribution at the global minima are different in the two phases. In phase \Circled{III}, the single spin distribution has a nonzero probability density at the origin, and the spectrum of the Hessian is gapless (panel (d)). We refer to this phase as a non-rigid spin-glass. We designate the boundary between the convex phase and the non-rigid spin-glass phase as $a$ increases by the curve $a_\text{\tiny{sg}}(J_0)$, the critical value of $a$ at which the transition occurs for a given $J_0$.  In phase \Circled{IV}, the spectrum of the Hessian is gapped away from the origin, and the single spin distribution has zero probability density at the origin (panel (e)). We refer to this phase as a rigid spin-glass. We denote by $a_\text{\tiny{r}}(J_0)$ the value of the laser gain at which this rigidity transition first occurs as $a$ increases. Together these results, discussed further below, begin to justify our schematic description introduced in Fig.~\ref{fig:cartoon} with regions \Circled{I}, \Circled{II}, \Circled{III}, and \Circled{IV}, corresponding to the light green, dark green, light and dark yellow, and light and dark red regions in Fig.~\ref{fig:cartoon} respectively. 

Interestingly, upon increasing the laser gain at fixed $J_0$, we see that two scenarios are possible, depending on the strength of the ferromagnetic coupling $J_0$. At lower values of $J_0$, we have $a_\text{\tiny{sg}}(J_0) <a_\text{\tiny{r}}(J_0)$: as $a$ increases, the system can transition from a convex phase, \Circled{I} or \Circled{II}, to a non-rigid spin-glass phase \Circled{III}, and then into the rigid spin-glass phase \Circled{IV}. Alternatively, for larger values of $J_0$, we can have $a_\text{\tiny{sg}}(J_0)=a_\text{\tiny{r}}(J_0)$: the system transitions from a convex phase \Circled{II} to a rigid-spin-glass phase \Circled{IV}. We next describe how we obtain these results.  

\section{Stability of global minimum at low laser gain}\label{sec:Hessian_origin}
Let us consider the case where the laser gain is  very low, $a\ll0$. In this case, the single-site energy $E_I$ in Eq.~\eqref{eq:E_I_single_site} is convex and dominates over the interaction term. We thus expect the system to have a single critical point at the origin, $\bnabla E^\text{\tiny{SK}}(\bx=\mathbf{0},a)=\mathbf{0}$, which is also the global minimum. The Hessian $ H^\text{\tiny{SK}}_{ij}(\bx,a)$ of the CIM-SK is
\begin{align}\label{eq:Hessian}
    \begin{split}
        H^\text{\tiny{SK}}_{ij}(\bx,a) &= \delta_{ij}\p^2_x E_I(x_i,a) - J^\text{\tiny{SK}}_{ij} \\
        &=\delta_{ij}(3x_i^2- a) - J^{\text{\tiny{SK}}}_{ij}
    \end{split}
\end{align}
The typical shape of the landscape in the vicinity of a configuration $\bx$ can be understood from the distribution of the eigenvalues $\rho^\text{\tiny{SK}}_\bx(\lambda)$ of $\bH^\text{\tiny{SK}}(\bx,a)$, obtained by averaging over realizations of the connectivity matrix $\bJ$,
\begin{equation}\label{eq:rho}
\rho^\text{\tiny{SK}}_\bx(\lambda) \equiv \lim_{N\to\infty}\frac{1}{N}\sum_{i=1}^N \left\langle\delta(\lambda_i(\bx) - \lambda)\right\rangle_\bJ\,,
\end{equation}
where $\{\lambda_i(\bx)\}_{i=1}^N$ is the set of eigenvalues  of $\bH^\text{\tiny{SK}}(\bx,a)$. When evaluated at $\bx=\mathbf{0}$,  The Hessian becomes
\begin{align}
    \begin{split}
        \bH^\text{\tiny{SK}}(\mathbf{0},a) &=  - a\mathbf{I} -\bJ^\text{\tiny{SK}}\,\\
        &= -a\mathbf{I} -\frac{1}{\sqrt{N}}\bJ - \frac{J_0}{N}\bt\otimes\bt\,.
    \end{split}
\end{align}
In the second equality we have decomposed the Hessian into a sum of a diagonal matrix, a random matrix from the Gaussian orthogonal ensemble and a rank-1 perturbation. The spectral properties of this class of random matrix have been widely studied. For $J_0=0$, the spectrum $\rho_\mathbf{0}^\text{\tiny{SK}}(\lambda)$ is given by Wigner semicircle law~\cite{wigner1967random} centered around $-a$,
\begin{equation}   
\rho^\text{\tiny{SK}}_\mathbf{0}(\lambda) = \frac{\sqrt{(2 -a- \lambda)(2+a+\lambda)}}{2\pi}\,,
\end{equation}
with support $\lambda \in [-2-a,2-a]$. For the origin to be a minimum, the lower edge of the support of the spectrum must be positive. This condition is violated for $a>-2$. Above this laser gain value, the origin is no longer a minimum of the system, and we expect a bifurcation in the energy landscape leading to new minima. When $J_0>0$, the ferromagnetic coupling can be thought of as a signal, while the random Gaussian couplings can be thought of as noise. The competition between these two elements, and how it affects the composite eigenstructure, has been studied by Baik, Ben Arous and P\'ech\'e (BBP)~\cite{baikPhaseTransitionLargest2005}. If $J_0<1$, the random couplings $\bJ$ overcome the ferromagnetic ones, and the spectrum $\rho_\mathbf{0}^\text{\tiny{SK}}(\lambda)$ remains identical to the $J_0=0$ case. Moreover, all eigenvectors of $\mathbf{J}$ are paramagnetic, or have $0$ overlap with $\bt$ in the large $N$ limit. However, if $J_0>1$, a phase transition occurs in which an isolated eigenvalue $\lambda_{\mathbf{0},\text{m}}$ detaches from the lower edge of the bulk spectrum $\rho^\text{\tiny{SK}}_\mathbf{0}$. Moreover, its associated eigenvector is ferromagnetic, or has a finite overlap with $\bt$. From the BBP theory, the isolated eigenvalue is
\begin{equation}
    \lambda_{\mathbf{0},\text{m}} = -J_0-J_0^{-1} - a\,.
\end{equation}
When $\lambda_{\mathbf{0},\text{m}}<0$, the origin of the CIM energy landscape becomes unstable. We thus predict the system to flow to a different minimum, which is partially aligned with the direction $\bt$. 

In summary, we obtain that the origin is a stable minimum of the CIM energy landscape,  as long as the lower edge of the support of its Hessian eigenspectrum $\rho_\mathbf{0}^\text{\tiny{SK}}(\lambda)$, which we denote by $\lambda_{\mathbf{0}, \text{m}}(J_0,a)$, is larger than $0$. $\lambda_{\mathbf{0},\text{m}}$ is given by
\begin{equation}\label{eq:lambda_m_BBP}  \lambda_{\mathbf{0},\text{m}}(J_0,a) = \begin{cases} -2-a\,&\quad \text{if }\, J_0\leq1 \\
    -J_0-J_0^{-1}-a\,&\quad\text{if }\,J_0>1\end{cases}\,.
\end{equation}
Equation~\eqref{eq:lambda_m_BBP}, together with the condition $\lambda_{\mathbf{0},\text{m}}>0$, identifies the region in the $(J_0,a)$ plane where the origin is a stable minimum of the system. 

Beyond this region, when the origin destabilizes, we expect the CIM to flow toward new minima. The properties of the minima with the lowest energy can be studied through the low temperature limit of the Boltzmann distribution
\begin{equation}
    P^\text{\tiny{SK}}_\text{\tiny{B}}(\bx) \equiv \frac{1}{Z^\text{\tiny{SK}}}\ee^{-\beta E^\text{\tiny{SK}}(\bx,a)}\,,
\end{equation}
where $\beta$ is the inverse temperature and $Z^\text{\tiny{SK}}\equiv \int\dd\bx\, \ee^{-\beta E^\text{\tiny{SK}}(\bx,a)}$ is the partition function of the system. Indeed, as $\beta \to \infty$, the probability density $P^\text{\tiny{SK}}_\text{\tiny{B}}(\bx)$ concentrates on minimum energy configurations. The typical properties of the shape of low energy minima are then studied using the low temperature limit of the Hessian eigenspectrum $\rho^\text{\tiny{SK}}(\lambda)$, defined as 
\begin{align}\label{eq:rhoSK}
    \begin{split}
        \rho^{\text{\tiny{SK}}}(\lambda) &\equiv \lim_{N\to \infty}\frac{1}{N}\int\dd\bx\, \left\langle P^\text{\tiny{SK}}_\text{\tiny{B}}(\bx)\sum_{i=1}^N \delta(\lambda_i(\bx) - \lambda)\right\rangle_\bJ\\
        &\equiv \lim_{N\to\infty}\left\langle \frac{1}{N}\sum_{i=1}^N \delta(\lambda_i(\bx) - \lambda)\right\rangle_{\bx,\,\bJ}^\text{\tiny{SK}}
    \end{split}
\end{align}
and taking the limit $\beta\to\infty$. The second equality defines the average over the Boltzmann distribution $\langle\ldots\rangle_\bx^{\text{\tiny{SK}}} \equiv \int\dd\bx\,\ldots P^\text{\tiny{SK}}_\text{\tiny{B}}(\bx)$. We note here that the properties of low energy configurations of the system can be alternatively, and equivalently, studied using a zero-temperature cavity method~\cite{mezard1987spin, cavagna2005cavity, talagrand2006parisi}. The spectrum $\rho^\text{\tiny{SK}}(\lambda)$ can be obtained from the resolvent $G^\text{\tiny{SK}}(z)$, defined as 
\begin{align}\label{eq:resolvent_def}
    \begin{split}
        G^\text{\tiny{SK}}(z) &\equiv \lim_{N\to\infty}\frac{1}{N}\left\langle\Tr\left[z-\bH^\text{\tiny{SK}}(\bx)\right]^{-1}\right\rangle_{\bx,\,\bJ}^\text{\tiny{SK}} \\
        &= \int \dd\lambda\, \frac{\rho^\text{\tiny{SK}}(\lambda)}{z-\lambda}\,.
    \end{split}
\end{align}
This formula can be inverted to express the spectrum $\rho(\lambda)$ as a function of the resolvent~\cite{potters2020first}
\begin{equation}
    \rho^\text{\tiny{SK}}(\lambda) = \lim_{\epsilon\to0} \frac{G^\text{\tiny{SK}}(\lambda-i\epsilon) - G^\text{\tiny{SK}}(\lambda+i\epsilon)}{2\pi i}\,,
\end{equation}
so that by computing the resolvent in the low temperature limit we can obtain the spectrum of the Hessian of global minima of the CIM machine. We next compute this resolvent.\\

\section{Computation of the resolvent of the Hessian}\label{sec:replica_resolvent}

We compute the resolvent following the approach of~\cite{edwards1976eigenvalue, urbani2022field}. We first note that from its definition in Eq.~\eqref{eq:resolvent_def}, the resolvent can be rewritten as\footnote{We use here the symbol $z$ as a shorthanded notation for $z+i0^+$, the limiting value of a complex number with real part $z$ and a vanishingly small positive complex part.}
\begin{align}\label{eq:Gz}
    \begin{split}
        G^\text{\tiny{SK}}(z) &= \lim_{N\to\infty}\frac{1}{N}\frac{\dd}{\dd z} \langle \Tr \log \left[z-\bH^\text{\tiny{SK}}(\bx,a)\right]\rangle_{\bx,\bJ}^\text{\tiny{SK}} \\
        &= \lim_{N\to\infty }\frac{1}{N}\frac{\dd}{\dd z}\lim_{l\to0}\p_l\langle\log Z^\text{\tiny{SK}}_l\rangle_\bJ\,,
    \end{split}
\end{align}
where we have introduced a modified partition function $Z_l^\text{\tiny{SK}}$,
\begin{align}
    \begin{split}
        Z_l^\text{\tiny{SK}} &\equiv \int \dd\bx\, \exp\Biggl[-\beta E^\text{\tiny{SK}}(\bx,a) \\
        &+ l \Tr\log[z - \bH^\text{\tiny{SK}}(\bx,a)]\Biggr]\,.
    \end{split}
\end{align}
To compute the average over the realizations of the disordered couplings, we resort to the replica trick:
\begin{equation} \label{eq:replicatrick}
    \langle\log Z^\text{\tiny{SK}}_l\rangle_\bJ = \lim_{n\to 0} \frac 1 n \log \langle(Z^\text{\tiny{SK}}_l)^n\rangle_\bJ\,.
\end{equation}
The limit $n\to0$ is approached by performing an analytic continuation from an integer $n$. For integer $n$, we can interpret $Z^n_l$ as the partition function of $n$ replicas of the original system, all with the same realization of the disordered couplings $\bJ$:
\begin{align}\label{eq:Znu_replicated}
    \begin{split}        \langle(Z_l^\text{\tiny{SK}})^n\rangle_\bJ &= \int_{\bx^\alpha } \,\Biggl\langle\exp\Bigg[\sum_{\alpha=1}^n \Big( -\beta E^\text{\tiny{SK}}(\bx^\alpha ,a) \\
        &+ l\Tr \log[z-\bH^\text{\tiny{SK}}(\bx^\alpha ,a)]\Big)\Bigg]\Biggr\rangle_\bJ\,,
    \end{split}
\end{align}
where $\int_{\bx^\alpha }\equiv \int \prod_{\alpha=1}^n \dd\bx^\alpha $ is an integral over the configurations of the $n$ different replicas. The term on the second line of Eq.~\eqref{eq:Znu_replicated} can be rewritten as 
\begin{align}\label{eq:log_z_rewriting}
    \begin{split}
        &l\sum_\alpha \Tr \log [z - \bH^\text{\tiny{SK}}(\bx^\alpha ,a)] \\&= 2 l\log\prod_\alpha \sqrt{\det[z-\bH^\text{\tiny{SK}}(\bx^\alpha ,a)]} \\
        &= \log\left\{\int_{\bphi^{\alpha}}\exp\left[ \frac{i}{2}\sum_\alpha\bphi^{\alpha}\cdot[z-\bH^\text{\tiny{SK}}(\bx^\alpha ,a)]\bphi^{\alpha}\right]\right\}^{-2l}\,,
    \end{split}
\end{align} 
where the symbol $\int_{\bphi^\alpha} \equiv \prod_{\alpha}\frac{\dd \bphi^\alpha}{\sqrt{2\pi i}}$ is a shorthand notation for the  integral over a set of $n$ bosonic fields $\{\bphi^\alpha\}_{\alpha=1}^N$ of dimension $N$.   
Substituting Eq.~\eqref{eq:log_z_rewriting} and Eq.~\eqref{eq:Znu_replicated} into the expression of the resolvent in Eq.~\eqref{eq:Gz} we obtain
\begin{widetext}
    \begin{align}\label{eq:Gz_double_replicas}
        \begin{split}
        &G^\text{\tiny{SK}}(z) = \lim_{n\to0}\lim_{l\to 0}\lim_{N\to\infty} \frac{1}{Nn}\p_l\frac{\dd}{\dd z}\log \int_{\bx^\alpha }\,\left\langle\ee^{-\beta \sum_{\alpha} E(\bx^\alpha,\,a )}\left(\int_{\bphi^\alpha} \exp\left[ \frac{i}{2}\sum_\alpha\bphi^{\alpha}\cdot[z-\bH^\text{\tiny{SK}}(\bx^\alpha ,a)]\bphi^{\alpha}\right]\right)^{-2l}\right\rangle_\bJ\\
        &= -\lim_{n\to0}\lim_{l\to 0}\lim_{N\to\infty} \frac{2}{Nn}\p_l\frac{\dd}{\dd z}\log \int_{\bx^\alpha }\,\left\langle\ee^{-\beta \sum_{\alpha} E(\bx^\alpha,\,a )}\left(\int_{\bphi^\alpha}\,\exp\left[ \frac{i}{2}\sum_\alpha\bphi^{\alpha}\cdot[z-\bH^\text{\tiny{SK}}(\bx^\alpha ,a)]\bphi^{\alpha}\right]\right)^{l}\right\rangle_\bJ\\
        &= -\lim_{n\to0}\lim_{l\to 0}\lim_{N\to \infty} \frac{2}{Nn}\p_l\frac{\dd}{\dd z}\log \int_{\bx^\alpha ,\,\bphi^{\alpha\rho}}\,\left\langle\exp\left[-\beta \sum_{\alpha=1}^n E(\bx^\alpha,\,a ) + \frac{i}{2}\sum_{\alpha=1}^n\sum_{\rho=1}^l \bphi^{\alpha\rho}\cdot [z\mathbf{1} - \bH^\text{\tiny{SK}}(\bx^\alpha ,a)]\bphi^{\alpha\rho}  \right]\right\rangle_\bJ\,.
    \end{split}
    \end{align}
\end{widetext}
In the first equality, we assumed as customary when resorting to the replica trick, that the limits $n,\,l\to0$ and $N\to\infty$ commute. In the second equality, we have treated $l$ as a dummy variable, rescaling it by a factor $2$ and changing its sign. In the third equality, we have expanded the integral over the bosonic variables as an integral over a new set of replicated fields $\bphi^{\alpha\rho}$ for $\alpha=1,\ldots,n$ and $\rho=1,\ldots,l$, with $n$ and $l$ integers whose limit to $0$ is eventually taken by means of an analytic continuation. We used a shorthanded notation for the integral over these replicated variables, $\int_{\bx^\alpha ,\,\bphi^{\alpha\rho}} \equiv \int \prod_{\alpha=1}^n\prod_{\rho=1}^m \frac{\dd\bx^\alpha  \dd\bphi^{\alpha\rho}}{\sqrt{2\pi i}}$.
The different replicated soft spins $\bx^\alpha $ and bosonic fields $\bphi^{\alpha\rho}$ share the same realizations of the quenched disordered couplings $\bJ$. This induces correlations among the different replicas. Such correlations are described by a set of order parameters $\bGamma$, that can be inserted in the expression of the resolvent after averaging over the disorder and using some manipulations detailed in Appendix~\ref{app:Gz_calculation}.
As a result, we obtain
\begin{align}\label{eq:Gz_after_average}
    \begin{split}
        G^\text{\tiny{SK}}(z) &= -\lim_{n\to 0}\lim_{l\to 0}\lim_{N\to \infty} \frac{2}{Nn}\\&\times\p_l\frac{\dd}{\dd z}\log \int \dd\bGamma\, \ee^{NS^\text{\tiny{SK}}[\bGamma]}\,. \\
    \end{split}
\end{align}
Where the set of parameters $\bGamma=\{\bQ, \bmm, \bP, \bT, \bM\}$ contains an $n\times n$ matrix $\bQ$, an $n$ dimensional vector $\bmm$, a $n\times n\times l$ tensor $\bT$ and a $n\times n\times l\times l$ tensor $\bP$. In the limit of large $N$, the integral in Eq.~\eqref{eq:Gz_after_average} is dominated by a saddle-point:
\begin{equation}\label{eq:GSK}
    G^\text{\tiny{SK}}(z) = -\lim_{n\to0}\frac{\dd}{\dd z}\lim_{l\to0}\frac{2}{n}\p_l \sup_{\bGamma} S^\text{\tiny{SK}}[\bGamma]\,.
\end{equation}
The extremization of the action $S$ yields a set of self-consistent equations that determine the order parameters $\bGamma$. These saddle point equations are evaluated in Appendix~\ref{app:saddle_point_G} within the assumption that there are no correlations among the replicated bosonic fields and the replicated soft spins. As we will see shortly, this ansatz amounts to neglecting correlations between the distribution of soft-spins in global minima and the Hessian matrix of global minima, across different realizations of the disorder. We show in Appendix~\ref{app:stability_diag_ansatz} that this approximation is stable against a small perturbation of the order parameters at the saddle point. We obtain the following self consistent equation for the resolvent,
\begin{equation}\label{eq:self_consistent_Gz}
    G^\text{\tiny{SK}}(z) = \int \dd x\,\frac{P^\text{\tiny{SK}}(x)}{z - \p^2_x E_I(x,a) - G^\text{\tiny{SK}}(z)}\,.
\end{equation}
This equation is equivalent to the Pastur self-consistent condition~\cite{pastur1972spectrum} for the resolvent of the sum of a Wigner matrix $\bJ$ and a diagonal matrix with independent, identically distributed elements equal to $\p^2_x E_I(x,a)$, where the value of $x$ for each diagonal entry is drawn i.i.d. from $P^\text{\tiny{SK}}(x)$, which is the mean field single-spin probability distribution, defined as
\begin{equation}
    P^\text{\tiny{SK}}(x) \equiv  \lim_{n\to0} \frac{1}{n} \sum_{\alpha=1}^n\langle \delta(x-x^\alpha ) \rangle_\text{\tiny{MF}}^\text{\tiny{SK}}\,.
\end{equation}
The average $\langle\ldots\rangle_\text{\tiny{MF}}^\text{\tiny{SK
}}$ is an average over a set of $n$ replicated single-site soft spins $x^\alpha $, and it reads
\begin{equation}
    \langle \ldots\rangle_\text{\tiny{MF}}^\text{\tiny{SK}} = \frac{1}{Z_\text{\tiny{MF}}^\text{\tiny{SK}}}\int \prod_{\alpha=1}^n\dd x^\alpha \,\ee^{-\beta E_\text{\tiny{MF}}^\text{\tiny{SK}}(\{x^\alpha \})}\,,
\end{equation}
where the mean field energy $E_\text{\tiny{MF}}^\text{\tiny{SK}}$ and the mean field partition function $Z_\text{\tiny{MF}}^\text{\tiny{SK}}$ are, respectively,\footnote{Note that, with a slight abuse of notation, we are using the symbol $\beta$ to denote both the inverse temperature and a replica index.}
\begin{align}
    \begin{split}
        E_\text{\tiny{MF}}^\text{\tiny{SK}} &= \sum_\alpha E_I(x^\alpha ,a) - \frac{\beta}{2}\sum_{\alpha,\beta}Q_{\alpha\beta}x^\alpha x^\beta  \\
        &- J_0\sum_\alpha m_\alpha x^\alpha  \\
        Z_\text{\tiny{MF}}^\text{\tiny{SK}} &= \int \prod_{\alpha=1}^n \dd x^\alpha \, \ee^{-\beta E_\text{\tiny{MF}}^\text{\tiny{SK}}[\{x^\alpha \}]}\,.
    \end{split}
\end{align}
The overlap matrix $Q_{\alpha\beta}$ and the magnetization vector $m_\alpha$ are determined by the saddle point equations
\begin{equation}\label{eq:sp_qm}
    Q_{\alpha\beta} = \langle x^\alpha  x^\beta\rangle_\text{\tiny{MF}}^\text{\tiny{SK}} \qquad m_\alpha = \langle x^\alpha  \rangle_\text{\tiny{MF}}^\text{\tiny{SK}}\,.
\end{equation}
The matrix $Q_{\alpha\beta}$ contains the correlations among different replicas, while the vector $m_\alpha$ expresses the average magnetization of each replica. Equations~\eqref{eq:self_consistent_Gz} to Eq.~\eqref{eq:sp_qm} contain, in principle, everything needed to determine the spectral distribution $\rho^\text{\tiny{SK}}(\lambda)$, once the single-spin distribution $P^\text{\tiny{SK}}$ is known. In order to determine the latter, an explicit structure of the overlap matrices has to be chosen. Before doing this, however, let us discuss how the position of the edge of the spectrum can be determined using the self-consistent equation for the resolvent. 

\subsection{Lower edge of the spectrum}

In this Section we discuss, following~\cite{tao2012topics, bouchbinder2021low}, how to determine the lower edge of the spectrum of the Hessian from the knowledge of the resolvent $G^\text{\tiny{SK}}(z)$ given in Eq.~\eqref{eq:self_consistent_Gz}. We introduce a function $g(z)\equiv  G^\text{\tiny{SK}}(z) - z$. Equation~\eqref{eq:self_consistent_Gz} can be inverted to express $z$ as a function of $g$, namely,
\begin{equation}\label{eq:z(g)}
    z = -\left[ g + 
    \int \dd x\,\frac{P^\text{\tiny{SK}}(x)}{g + \p^2_x E_I(x,a)}\right]\equiv z(g)\,.
\end{equation}
The support of the spectrum $\rho^\text{\tiny{SK}}(\lambda)$ is defined as the set of values $\lambda+i\epsilon$ for which $G^\text{\tiny{SK}}(\lambda+i\epsilon)$ has a nonzero imaginary part in the limit $\epsilon\to 0$. Outside this region, $G^\text{\tiny{SK}}(\lambda+i\epsilon)$ is real. Therefore, the support of $\rho^\text{\tiny{SK}}(\lambda)$ can be obtained as the complement of the image of $z(g)$ for real $g$. In particular, the lower edge of the spectrum is determined by the maximum attained by $z(g)$. From its definition, we see that $z(g)$ is defined in the interval  $g \in (-\infty,-E''_{I,\text{max}}] \cup [-E_{I,\text{min}}'',\infty)$, where $E''_{I,\text{max}}$, $E''_{I,\text{min}}$ are respectively the maximum and the minimum value attained by the function $\p_x^2 E_I(x,a)$, with $a$ fixed, in the region where $P^\text{\tiny{SK}}(x)\neq 0$. We expect the maximum of $z(g)$ to lie in the $[-E''_{I,\text{min}},\infty)$ branch. The lower edge $\lambda_\text{\tiny{min}}^\text{\tiny{SK}}$ of the spectrum of $\rho^\text{\tiny{SK}}(\lambda)$ is thus given by 
\begin{equation}\label{eq:lambda_min_from_g}
    \lambda^\text{\tiny{SK}}_\text{\tiny{min}} = z(g^*) \qquad g^* \equiv\argmax_{g \in [-E''_{I,\text{min}},\,\infty)} z(g)\,.
\end{equation}
In particular, if the function $z(g)$ reaches a maximum in the interval $(-E''_{I,\text{min}},\,\infty)$, then $g^*$ satisfies the condition
\begin{equation}\label{eq:g*}
    \frac{\dd z
    (g)}{\dd g}\Bigg\rvert_{g=g^*} = -1 + \int \dd x\,\frac{P^\text{\tiny{SK}}(x)}{(g^* + \p^2_x E_I(x,a))^2} = 0\,.
\end{equation}
The important interplay of this condition with the nature of the spin-glass phase of the CIM-SK will be discussed below. Once $P^\text{\tiny{SK}}(x)$ is determined by means of replica theory, we can determine where the lower edge of the spectrum of the Hessian lies. 
\begin{figure}          
    \includegraphics[width=\columnwidth]{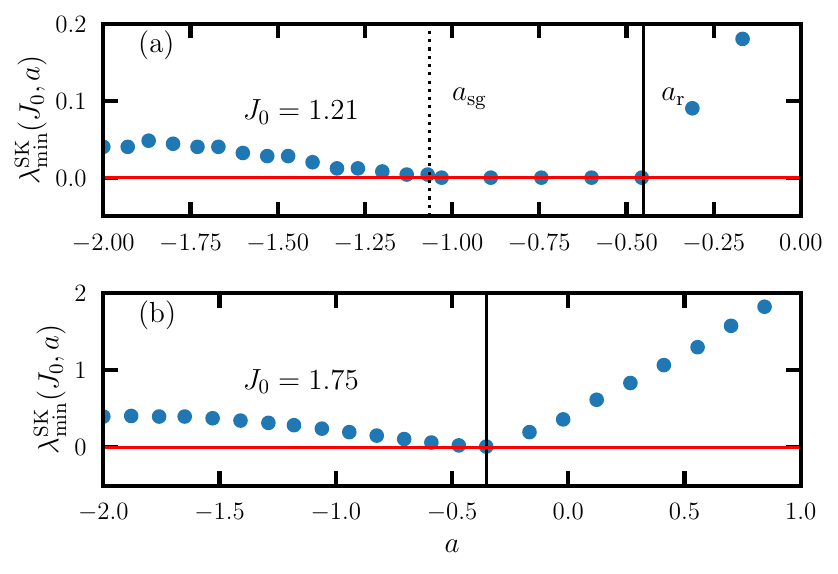}
    \caption{\textbf{Lower edge of the spectrum of the Hessian of global minima in the annealed landscape.} We plot $\lambda_\text{\tiny{min}}^\text{\tiny{SK}}(J_0,a)$ as a function of the laser gain $a$ for two different values of the mean connectivity $J_0$. Panel (a): in the replica-symmetric phase, the lower edge of the spectrum decreases as the laser gain increases, until it hits zero (horizontal solid red line) at the spin-glass transition value $a_{\mathrm{sg}}(J_0)$ (vertical dotted black line). Upon further increase of the laser gain in the spin-glass phase, the lower edge of the Hessian remains zero, as long as the laser gain is below the rigidity transition value $a_\mathrm{r}$ (vertical solid black line). For $a>a_\mathrm{r}$, the spectrum is gapped. In panel (b), the spin-glass and the rigidity transition occur at the same value of the laser gain, $a_\text{\tiny{r}}=a_\text{\tiny{sg}}$. The spectrum has a gap both above and below this value.\label{fig:lambda_m_J0_a}}  
\end{figure}
\section{The structure of global minima and their Hessian eigenspectra}\label{sec:global_minima}

\subsection{Convex phase}
To compute the probability density of the soft-spin population at global minina $P^\text{\tiny{SK}}(x)$ from the mean field calculation presented in Sec.~\ref{sec:replica_resolvent}, we need to make an ansatz about the structure of the overlap matrix $Q_{\alpha\beta}$ of the system. The simplest structure is a replica symmetric ansatz, invariant under permutations of the different replicas,  
\begin{align}\label{eq:RS_ansatz}
    \begin{split}
        Q_{\alpha\beta} &= q_d \delta_{\alpha\beta} + q_o(1-\delta_{\alpha\beta}) \\
        m_\alpha &= m\,.
    \end{split}
\end{align}
The diagonal part of $Q_{\alpha\beta}$ contains the overlap of a replica with itself, while its off-diagonal elements contain overlaps among different replicas. The difference $q_d-q_o$ is a measure of the width, in phase space, of the thermodynamic state considered. As the temperature decreases toward $0$, we expect the states occupied by the different replicas of the system to shrink toward the global minimum, which implies that $q_d-q_o\to0$. It is thus convenient to consider an $O(1)$ rescaled overlap difference, $\Delta \widetilde q\equiv \beta(q_d-q_o)$.\footnote{The scaling $q_d-q_o \propto \beta^{-1}$ is then self-consistently verified through the saddle point equations.}

A calculation detailed in Appendix~\ref{app:saddle_point_rs} shows that the saddle point equation for the order parameters $q_o$, $\Delta \widetilde q$ and $m$ are given by
\begin{align}\label{eq:sp_rs}
    \begin{split}
        \Delta \widetilde q &= \overline{\langle x^2\rangle_\text{\tiny{RS}}^\text{\tiny{SK}}-(\langle x \rangle_\text{\tiny{RS}}^\text{\tiny{SK}})^2}\\
        q_o &= \overline{\left(\langle x \rangle_\text{\tiny{RS}}^\text{\tiny{SK}}\right)^2}\\
        m &= \overline{\langle x\rangle_\text{\tiny{RS}}^\text{\tiny{SK}}}\,,
    \end{split}
\end{align}
where the replica-symmetric mean field average $\langle \ldots\rangle_\text{\tiny{RS}}^\text{\tiny{SK}}=\frac{1}{\ZRSSK(h)}\int \dd x\, \ldots\ee^{-\beta \ERSSK(x,h)}$ involves an effective partition function $\ZRSSK$ and an effective energy $\ERSSK$, which read, respectively,
\begin{equation}\label{eq:Z_RS_SK}
    \begin{split}
        Z^\text{\tiny{SK}}_\text{\tiny{eff}}(h) &\equiv \int \dd x\, \ee^{-\beta E_\text{\tiny{eff}}^\text{\tiny{SK}}(x,h)}\\
        E_\text{\tiny{eff}}^{\text{\tiny{SK}}}(x,h) &\equiv E_I(x, a+\Delta \widetilde q) -(\sqrt{q_o}h + J_0 m)x \\
        &=\frac{1}{4}x^4 -\frac{a+\Delta \widetilde q}{2}x^2 -(\sqrt{q_o}h + J_0 m)x\,,
    \end{split}
\end{equation}
and the average $\overline{\cdots} \equiv \int\frac{\dd h}{\sqrt{2\pi}}\ldots\ee^{-h^2/2}$ is an average over the realizations of a zero mean unit variance Gaussian field $h$. The single spin probability distribution in the convex phase $P^\text{\tiny{SK}}_\text{\tiny{RS}}(x)$ is instead given by
\begin{equation}\label{eq:P_rs_sk}
    P^\text{\tiny{SK}}_\text{\tiny{RS}}(x) = \overline{\ee^{-\beta E_\text{\tiny{eff}}^\text{\tiny{SK}}(x,h)}\left[\ZRSSK(h)\right]^{-1}}\,, 
\end{equation}
 As a result of the mean field analysis, the single spin distribution reduces to a Boltzmann distribution for a single-site soft spin system, with effective energy $\ERSSK$. The
rescaled overlap difference $\Delta \widetilde q$ acts as a shift in the laser gain of the single-site energy $E_I$, while the overlap among different replicas $q_0$ determines the coupling strength to the quenched random Gaussian field $h$. Finally, the mean connectivity $J_0$ determines the coupling strength of $x$ to the magnetization $m$ of the system. Note that here the magnetization $m$ is the mean of the  soft-spin distribution, and can therefore take values larger than $1$. 

In the low temperature limit and for a given realization of the random field $h$, the replica-symmetric partition function is dominated by the minimum $x^*(h)$ of the energy  $E^\text{\tiny{SK}}_\text{\tiny{eff}}$. The saddle-point equations then become 
\begin{align}\label{eq:self_consistent_RS}
    \begin{split}
        \Delta \widetilde q &= \overline{\left[\p_x^2 E_\text{\tiny{eff}}^\text{\tiny{SK}}\lvert_{x=x^*(h)}\right]^{-1}}, \\
        q_o &= \overline{x^*(h)^2}, \\
        m &=  \overline{x^*(h)}.
    \end{split}
\end{align}
This set of self consistent equations can be solved numerically through iterative methods for different values of the laser gain $a$ and the ferromagnetic strength $J_0$. The single spin probability distribution becomes
\begin{align}\label{eq:PSKRS_x}
    \begin{split}
    P^\text{\tiny{SK}}_\text{\tiny{RS}}(x) &= \overline{\delta(x-x^*(h))}\\
    &= \frac{1}{\sqrt{2\pi q_o}}\lvert \p^2_x E_I(x,a+\Delta \widetilde q)\rvert \ee^{-\frac{\left(h^*(x)\right)^2}{2}}\,,
    \end{split}
\end{align}
where the field $h^*(x)$ satisfies the stationary condition $\p_x \ERSSK(x,h)\rvert_{h=h^*(x)} = 0$, i.e., 
\begin{equation}\label{eq:stationary_h}
    \sqrt{q_o}h^*(x) + J_0 m = x\left[x^2 - (a+\Delta \widetilde q)\right].
\end{equation}
When $a+\Delta \widetilde q\leq 0$, the effective energy $\ERSSK(x,h)$ is a convex function of $x$ for {\it all} realizations of the random field $h$. The support of $P^\text{\tiny{SK}}_\text{\tiny{SK}}$ then extends over the whole real axis. If instead $a+\Delta \widetilde q>0$, the interval $(-\sqrt{a+\Delta \widetilde q},\,\sqrt{a+\Delta \widetilde q})$ is excluded from the support of $P^\text{\tiny{SK}}_\text{\tiny{RS}}(x)$. 

The saddle point equations are complemented with a study of the stability of the replica symmetric solution. This can be done expanding the action $S$ in Eq.~\eqref{eq:Gz_after_average} around the replica symmetric overlap matrix and magnetization. The replica symmetric solution remains an extremum of the action $S$ as long as the largest eigenvalue of the Hessian  remains negative. This eigenvalue, known as the replicon $\tilde\Lambda_\text{\tiny{R}}^\text{\tiny{SK}}$, has a known structure, and it evaluates to (details in App.~\ref{app:replicon})
\begin{equation}\label{eq:replicon}
    \tilde\Lambda^\text{\tiny{SK}}_\text{\tiny{R}} = -1 + \overline{[\p^2_x E_\text{\tiny{eff}}^\text{\tiny{SK}}(x,a)|_{x=x^*(h)}]^{-2}}\,.
\end{equation}
The replica symmetric solution is stable as long as $\tilde\Lambda^\text{\tiny{SK}}_\text{\tiny{R}}<0$. A positive replicon eigenvalue signals a nonlinear instability of the system under the perturbation generated by a small random external field applied to the soft spins. We remark that the sign that the replicon needs to have in order for a replica-symmetric solution to be stable depends on a sign convention in the action of Eq.~\eqref{eq:Gz_after_average}, and that the sign choice adopted in this work is different than the one often adopted in the spin-glass literature. By changing the integration variable in the average in Eq.~\eqref{eq:replicon} from the Gaussian field $h$ to the soft-spin configuration $x$ at the stationary point determined by Eq.~\eqref{eq:stationary_h}, and by using the single spin probability distribution in Eq.~\eqref{eq:PSKRS_x}, the expression for the replicon can be rewritten as 
\begin{equation}\label{eq:replicon_as_x}
     \tilde\Lambda^\text{\tiny{SK}}_\text{\tiny{R}} = -1 + \int \dd x\,\frac{P^\text{\tiny{SK}}_\text{\tiny{RS}}(x)}{\left(\p^2_x E_I(x,a) - \Delta \widetilde q \right)^2}\,.   
\end{equation}
 Using the saddle point equations and the stability condition, we can determine the phase boundaries shown in Fig.~\ref{fig:phase_diagram_global_minima_SK}. This is what we do next.\\

\tocless\subsubsection{Phase boundary of convex paramagnetic phase}\label{sec:phase_boundary_para_to_ferro}
 We first recover the results obtained in Sec.~\ref{sec:Hessian_origin} concerning the stability of the paramagnetic, convex phase. In the paramagnetic phase, we have $m=q_o=0$. The single spin distribution becomes a Dirac delta centered around the origin, $P^\text{\tiny{RS}}_\text{\tiny{SK}}(x) = \delta(x)$. The remaining order parameter $\Delta \widetilde q$ is determined from the saddle point equation
\begin{equation}\label{eq:Deltaq_para}
    \Delta \widetilde q = -\frac{1}{a + \Delta \widetilde q}\,.
\end{equation}
In App.~\ref{app:details_BBP}, we show that in order for the paramagnetic solution to be stable, we must have
\begin{equation}\label{eq:condition_i}
    \Delta \widetilde q \leq \min[1,J_0^{-1}]\,.
\end{equation}
Moreover, to ensure that $m=q_o=0$, the paramagnetic, replica symmetric energy $\ERSSK=\frac{1}{4}x^4 - \frac{1}{2}(a+\Delta \widetilde q)x^2$ must admit a minimum at $x=0$. This implies
\begin{equation}\label{eq:condition_ii}
    a + \Delta \widetilde q \leq 0\,.
\end{equation}
Combining together Eq.~\eqref{eq:condition_i} and Eq.~\eqref{eq:condition_ii} we obtain that the convex paramagnetic state is stable as long as
\begin{equation}\label{eq:stability_rs_minimum}
    a \leq \begin{cases} -2 &\quad \text{if } J_0\leq 1 \\
    -J_0 - \frac{1}{J_0} &\quad \text{if } J_0>1
    \,,\end{cases} 
\end{equation}
which is equivalent to the condition $\lambda_{\mathbf{0}
,\text{\tiny{m}}}>0$, with $\lambda_{\mathbf{0},\text{\tiny{m}}}$ given by Eq.~\eqref{eq:lambda_m_BBP}, which was obtained through random matrix theory. This analysis allows us to conclude that for $J_0<1$, the system has a transition into a spin-glass phase at $a_\text{\tiny{sg}}(J_0)=-2$. This is the horizontal dotted line in Fig.~\ref{fig:phase_diagram_global_minima_SK} (a). For $J_0>1$, the second case in Eq.~\eqref{eq:stability_rs_minimum} yields a transition line toward a ferromagnetic state. This is the dashed line $a_\text{\tiny{BBP}}(J_0)$ in Fig.~\ref{fig:phase_diagram_global_minima_SK} (a). We address the stability of this ferromagnetic phase next.        
\\\tocless\subsubsection{Stability of the ferromagnetic phase}

When the replica symmetric solution is ferromagnetic, its boundary with the spin-glass phase has to be computed numerically, solving the saddle point Equations~\eqref{eq:self_consistent_RS}. The replica symmetric solution can become unstable through two different pathways. In the first pathway, the replicon eigenvalue $\tilde\Lambda^\text{\tiny{SK}}_\text{\tiny{R}}$ changes sign continuously, similarly to what happens at the spin-glass transition from the paramagnetic phase. At the transition point, the effective energy $\ERSSK(x,h)$ is a convex function of $x$ for all the realizations of the quenched random field $h$.  In this case, as discussed more in detail in Sec.~\ref{sec:fRSB}, in the spin-glass phase the support of the single-spin distribution covers the whole real axis, and the lower edge of the Hessian is located at the origin. This corresponds to the soft spin-glass phase.  In the second pathway, the effective energy $\ERSSK(x,h)$ changes its convexity while the replicon eigenvalue is still negative. When this happens,  the Hessian of the replica-symmetric energy $E_\text{\tiny{RS}}^\text{\tiny{SK}}(x,h)$  vanishes when evaluated at its minimum $x_m(h)$ for some values of $h$, while the replicon eigenvalue in Eq.~\eqref{eq:replicon} jumps  discontinuously from a negative value to $\infty$, and so the convex phase becomes unstable.  We argue in Sec.~\ref{sec:fRSB} that the nonconvexity of the effective energy in the spin-glass phase implies that the support of the single-spin distribution splits into two disconnected domains, and that the spectrum of the Hessian at global minima is gapped away from the origin. These are the signature of the rigid spin-glass phase. The expression of $\ERSSK$ in Eq.~\eqref{eq:Z_RS_SK}, shows that the convex to non-convex transition occurs when
\begin{equation}
    a+\Delta \widetilde q = 0\,.
\end{equation}
This condition is used to determine the stability boundary between the ferromagnetic phase \Circled{II} and the rigid spin-glass phase \Circled{IV}. The above discussion concludes the identification of the boundaries of the convex phase of the CIM-SK model.\\  

\tocless\subsubsection{Hessian at the spin-glass transition}

At both spin-glass transitions, the spectral distribution of the eigenvalues of the Hessian for  global minima is gapless. To see this, let us consider separately the two pathways  to the instability of the replica-symmetric solution described in the preceding section.   In the first pathway, the replicon eigenvalue given by Eq.~\eqref{eq:replicon} approaches zero continuously at the transition. The condition $\tilde\Lambda^\text{\tiny{SK}}_\text{\tiny{R}}=0$ implies, from Eq.~\eqref{eq:replicon_as_x}, that
\begin{equation}\label{eq:replicon_vanishing_in_rs}
    -1 + \int \dd x\,\frac{P^\text{\tiny{SK}}_\text{\tiny{RS}}(x)}{\left(\p^2_x E_I(x,a) - \Delta \widetilde q \right)^2} = 0\,,
\end{equation}
where we made use of the definition of the replica symmetric mean field energy given by Eq.~\eqref{eq:Z_RS_SK}. Equation~\eqref{eq:replicon_vanishing_in_rs} can be compared with the extremization condition of $z(g)$ in Eq.~\eqref{eq:g*}, allowing us to conclude that $g^* = -\Delta \widetilde q$. Using Eq.~\eqref{eq:lambda_min_from_g}, the lower edge of the spectrum $\lambda_\text{\tiny min}^\text{\tiny{SK}}$ is thus given by
\begin{align}\label{eq:lambda_m_rs_to_sg_para}
    \begin{split}
    \lambda_\text{\tiny{min}}^\text{\tiny{SK}} &=z(-\Delta \widetilde q)\\ 
    &= \Delta \widetilde q - \int \dd x\,\frac{P^\text{\tiny{SK}}_\text{\tiny{RS}}(x)}{\p^2_x E_I(x,a) - \Delta \widetilde q} = 0\,.
    \end{split}
\end{align}
In the second equality, we made use of the saddle point equation for $\Delta \widetilde q$, given by Eq.~\eqref{eq:sp_rs}. Equation~\eqref{eq:lambda_m_rs_to_sg_para} is consistent with the random matrix analysis of the Hessian at the origin performed in Sec.~\ref{sec:Hessian_origin}. 

In the second pathway to instability of the convex phase, the replicon eigenvalue jumps discontinuously from a negative value to $\infty$ when $a + \Delta \widetilde q=0$. The fact that the replicon is negative implies, by comparing Eq.~\eqref{eq:replicon_as_x} against Eq.~\eqref{eq:g*}, that $z'(g)\rvert_{g=-\Delta \widetilde q}<0$. Therefore, since $z(g)$ is a concave function in the interval $[-E''_{I \text{min}},\infty)$, its maximum must lie in the interval $[-E''_{I,\text{min}},-\Delta \widetilde q)$. If the support of $P^{\text{\tiny{SK}}}_\text{\tiny{RS}}(x)$ spans the whole real axis, which, from Eq.~\eqref{eq:PSKRS_x}, is true for $a+\Delta \widetilde{q}\leq 0$, then $-E''_{I,\text{min}}=a$. The maximum of $z(g)$ must thus lie in the interval $[a,-\Delta \widetilde q)$. As the transition point is approached, this interval shrinks to the point $a$ along the real axis. At the transition, when $a=-\Delta \widetilde q$ the position of the lower edge of the spectrum is thus given by 
\begin{align}
    \begin{split}
        \lambda_\text{\tiny{min}}^\text{\tiny{SK}} &= z(g=-E''_{I,\text{min}}) = z(a) \\
        &= -a - \int \dd x\,\frac{P^{\text{\tiny{SK}}}_{\text{\tiny{RS}}}(x)}{\p^2_x E_I(x,a)+a}\\
        &= \Delta \widetilde q - \int \dd x\,\frac{P^{\text{\tiny{SK}}}_{\text{\tiny{RS}}}(x)}{\p^2_x E_I(x,a) - \Delta \widetilde q} = 0\,.
    \end{split}
\end{align}
In the third equality, we used the fact that $a+\Delta \widetilde q = 0$, while in the final passage we used again the saddle point equation for $\Delta \widetilde q$ given by Eq.~\eqref{eq:sp_rs}.

We thus see that when approaching the spin-glass transition from the replica symmetric phase, both pathways lead to the same fate for the lower edge of the spectrum of the Hessian. At the edge of the convex phase, soft modes arise around the global minima, leading to an instability that eventually brings the system into a spin-glass phase, which we now address. \\

\subsection{Spin-glass phase and rigidity transition}\label{sec:fRSB}

The glassy phase of the CIM-SK is described through the  full replica symmetry breaking solution~\cite{parisi1979infinite, mezard1987spin, nishimori2001statistical, parisi2020theory, charbonneau2023spin} (we will discuss in a moment the evidences for the instability of other types of spin-glass solutions). The phase space of the system shatters into many pure states, organized into an ultrametric structure. The inter-replica overlap $q_o$ is replaced by a continuous, monotonically increasing function $q(y)$, defined on the interval $[0,1]$, which is the  inverse of the cumulative distribution of overlaps among the different states of the system. The overlap difference $\Delta \widetilde q$ is now replaced by $\Delta \widetilde q = \beta (q_d - q(1))$. In Appendix~\ref{app:single-spin_frsb} we outline how the saddle point equations for $q(y)$, $\Delta \widetilde q$ and $m$ are obtained. These saddle point equations read
\begin{align}\label{eq:sp_frsb}
    \begin{split}
        q(y) &= \int \dd h\, P(y,h)\left[\p_h f(y,h)\right]^2 \\
        \Delta \widetilde q &= \int \dd h\, P(1,h) \p_h^2 f(1,h)  \\
        m &= \int \dd h\, P(1,h)\p_h f(1,h)\,.
    \end{split}
\end{align}
The function $f(y,h)$ describes a hierarchy of free energies of the system, computed across different levels of the ultrametric tree across which different states are arranged. The field $h$ encodes a form of quenched disorder stemming from the overlap among different pure states. While in the replica-symmetric solution this field is Gaussian, in the spin-glass phase a hierarchy of probability distributions $P(y,h)$ arises, and the distribution of internal fields $P(y=1,h)$ is no longer Gaussian. The function $f(y,h)$ and $P(y,h)$ satisfy the partial differential equations 
\begin{align}
    \begin{split}
        \p_y f(y,h) &= -\frac{1}{2}\frac{\dd q(y)}{\dd y}\left[ \beta y(\p_y f(y,h))^2 + \p_h^2 f(y,h)\right]\\
        \p_y P(y,h) &= \frac{1}{2}\frac{\dd q(y)}{\dd y} [\p_h^2 P(y,h) \\
        &- 2\beta \p_h\left(P(y,h)\p_h f(y,h)\right)]\,,
    \end{split}    
\end{align}
together with the boundary conditions
\begin{align}
    \begin{split}
        f(1,h) &= \beta^{-1}\log \int\dd x\,\ee^{-\beta \ERSSK(x,h)} \\
        P(0,h) &= \frac{1}{\sqrt{2\pi q(0)}}\ee^{-\frac{h^2}{2q(0)}}\,,
    \end{split}
\end{align}
where the single-spin, mean field energy $\ERSSK(x,h)$ is defined in Eq.~\eqref{eq:Z_RS_SK}. In the low-temperature limit, we show in App.~\ref{app:single-spin_frsb} that the single-spin distribution $P^\text{\tiny{SK}}(x)$ is obtained by evaluating the probability to realize a quenched field $h$ that minimizes the effective energy $\ERSSK(x,h)$, namely,
\begin{equation}\label{eq:Px_as_P1h}
    P_\text{\tiny{fRSB}}^\text{\tiny{SK}}(x) = \lvert \p^2_x E_I(x,a+\Delta \widetilde q)\rvert\,P(1,h^*(x))\,,
\end{equation}
where $h^*(x)$ is determined by the stationary condition in Eq.~\eqref{eq:stationary_h}. Using Eq.~\eqref{eq:Px_as_P1h} we can determine the resolvent and the spectrum of the Hessian in the spin-glass phase. The numerical integration of the full replica symmetry breaking equations at low temperature follows the methods developed in~\cite{schmidt2008method}, and it is discussed in Appendix~\ref{app:numerics_frsb}. The numerical method adopted here can also be used to investigate the nature of the spin-glass phase of the CIM. Indeed, the method approximates the full-replica symmetry breaking solution as a large, but finite, number of discrete steps of replica symmetry breaking. In Appendix~\ref{app:numerics_frsb}, Fig.~\ref{fig:qbetay_multistep}, we exploit this feature to show that, in the vicinity of the spin-glass transition line $a_\text{\tiny{sg}}(J_0)$ from the spin-glass phase, the one-step and two-steps replica symmetry breaking solutions yield different curves for the order parameter $q(y)$. This finding supports the claim that the entire spin-glass phase of the CIM is of the full-replica symmetry breaking type.

The full-replica symmetry breaking spin-glass phase is marginally stable in the space of all possible structures that the overlap matrix $\bQ$ can take. We report for completeness a derivation of this widely appreciated fact in Appendix~\ref{app:marginal_stability}, where we compute the replicon eigenvalue $\tilde\Lambda_\text{\tiny{R}}^\text{\tiny{SK}}$ and we show that
\begin{equation}\label{eq:marginal_stability_frsb}\tilde\Lambda_\text{\tiny{R}}^\text{\tiny{SK}} = -1 + \int \dd h\, P(1,h)\left[\p^2_h f(1,h)\right]^2 = 0\,. 
\end{equation}
Using the marginal stability of the spin-glass phase, and the possible change in convexity of the effective energy $\ERSSK$, we can study the properties of the linear excitations around the spin-glass global minima.\\

\tocless\subsubsection{Hessian in  spin-glass phase}\label{sec:hessian_spin_glass}

The behavior of the lower edge of the spectrum of the Hessian in the spin-glass phase is determined by an interplay between the shape of the effective single-site potential $\ERSSK(x,h)$ and the marginal stability of the spin-glass phase. If the effective single site energy $\ERSSK(x,h)$ is a convex function of $x$ for all the values of $h$, (i.e. when $a+\Delta \widetilde q<0$), soft modes appear in the spin-glass phase, and the lower edge of the spectrum of the Hessian for global minima touches the origin. If instead $\ERSSK(x,h)$
is a nonconvex function of $x$, as it is the case when $a+\Delta \widetilde q>0$, the spectrum of the Hessian is gapped away from the origin. 

When $a+\Delta \widetilde q<0$ we show in Appendix~\ref{app:marginal_stability} that the marginal stability condition in Eq.~\eqref{eq:marginal_stability_frsb} becomes
\begin{equation}\label{eq:marginal_stability_convex}
    \tilde\Lambda^\text{\tiny{SK}}_\text{\tiny{R}}= -1 + \int\dd x\, \frac{P_\text{\tiny{fRSB}}^\text{\tiny{SK}}(x)}{\left[\p^2_x E_I(x,a) -\Delta \widetilde q\right]^2} = 0\,.
\end{equation}
This condition implies that the maximum of $z(g)$, defined in Eq.~\eqref{eq:z(g)}, occurs for $g^*=-\Delta \widetilde q$. The marginal stability condition is, in fact, equivalent to the stationary condition $z'(g)\rvert_{g=-\Delta \widetilde q}=0$, as can be seen by comparing Eq.~\eqref{eq:marginal_stability_convex} with Eq.~\eqref{eq:g*}. We thus obtain for the lower edge of the spectrum
\begin{align}
    \begin{split}        \lambda_\text{\tiny{min}}^\text{\tiny{SK}} &= z(-\Delta \widetilde q) \\
        &= -\Delta \widetilde q + \int \dd x\,\frac{P_\text{\tiny{fRSB}}^\text{\tiny{SK}}(x)}{\p^2_x E_I(x,a)-\Delta \widetilde q} = 0\,,
    \end{split}
\end{align}
where the  third equality follows from the zero-temperature limit of the saddle point equation for $\Delta \widetilde q$ in Eq.~\eqref{eq:sp_frsb} (see also Eq.~\eqref{eq:app_sp_frsb_lowT_convex} in App.~\ref{app:low_T}, where the low temperature limit of the spin-glass phase is discussed in more detail). The spectrum of the Hessian is therefore gapless. 

If $a+\Delta \widetilde q >0$,  we show in Appendix~\ref{app:low_T} that the marginal stability condition is modified, and that the following inequality holds (see also Eq.~\eqref{eq:app_inequality_frsb}),
\begin{equation}\label{eq:marginal_stability_nonconvex}
     -1 + \int\dd x \frac{P_\text{\tiny{fRSB}}^\text{\tiny{SK}}(x)}{\left[\p^2_x E_I(x,a) -\Delta \widetilde q\right]^2} < 0\,.
\end{equation}
Combining this fact with Eq.~\eqref{eq:g*} implies that  $z'(g)\rvert_{g=-\Delta \widetilde q}<0$. On the other hand, the saddle point equations for $\Delta \widetilde q$ yield $z(g=-\Delta \widetilde q)=0$. Therefore, since $z(g)$ is a concave function in the domain $[-E''_{I \mathrm{min}}, \infty)$ (see Eq.~\eqref{eq:z(g)}), its maximum in this interval must be larger then zero, at least as long as $-E''_{I,\mathrm{min}} < -\Delta \widetilde q  $. But the latter condition is verified for $a+\Delta \widetilde q>0$. Indeed, when $a+\Delta \widetilde q>0$, the support of $P^{\text{\tiny{SK}}}_\text{\tiny{fRSB}}(x)$ is given by $(-\infty, -\sqrt{a+\Delta \widetilde q}) \cup (\sqrt{a+\Delta \widetilde q}, \infty)$, as can be seen from Eq.~\eqref{eq:Px_as_P1h} and Eq.~\eqref{eq:stationary_h}. $E''_{I,\mathrm{min}}$ is the minimum of the Hessian of the bare single-site energy $E_I(x,a)$ restricted to the support of $P^{\text{\tiny{SK}}}_\text{\tiny{fRSB}}$, and therefore it evaluates to 
\begin{align}
    \begin{split}
        -E''_{I,\mathrm{min}} &= \left[a - 3x^2\rvert_{x=\sqrt{a+\Delta \widetilde q}}\right]\\
        &= -2a - 3\Delta \widetilde q < -\Delta\widetilde{q}\,,
    \end{split}
\end{align}
where in the final equality we made use of the fact that $-a-\Delta \widetilde q<0$.  We thus conclude  that
\begin{align}
    \begin{split}
        \lambda_\text{\tiny{min}}^\text{\tiny{SK}} = z(g^*) > z(-\Delta \widetilde q) = 0\,,
    \end{split}
\end{align}
and that therefore the spectrum of the Hessian is gapped. Establishing the existence of gapped and gapless spin-glass phases in the CIM is one of the main results of this work. Our derivation highlights how this phenomenon stems from an interplay between a change of the shape of the effective single site energy and the nature of the spin-glass phase in the CIM-SK. 

We now turn to studying the properties of the most abundant, local minima of the evolving energy landscape of the CIM.\\ 

\section{Proliferation of  local minima and critical points with zero net magnetization}\label{sec:Kac-Rice}

In this Section we study the number and the properties of the most abundant minima of the CIM as the laser gain increases. This is done by means of a combination of the Kac-Rice formula and the replica approach~\cite{bray1980metastable, bray1981metastable, brayjpa1981metastable, bray1984weighted, ros2019complex, ros2022high, kent2023count, ros2025high}.\footnote{In the literature on disordered systems, the most abundant minima are also referred to as 'typical'. However, in this work, we will denote by 'typical' the points visited the gradient descent dynamics of the CIM, which is discussed in Sec.~\ref{sec:dynamics} and Sec.~\ref{sec:dmft}.}  We denote by $\mathcal{N}^\text{\tiny{SK}}(r,\bJ)$ the number of critical points with intensive index $r$, defined as the index of a critical point (number of negative eigenvalues of its Hessian) divided by $N$, in the coherent Ising machine for a given realization of the disorder $\bJ$. The complexity $\Sigma^\text{\tiny{SK}}(r,\bJ)$ is defined as the exponential rate at which $\mathcal{N}^\text{\tiny{SK}}(r,\bJ)$ grows as the size of the system increases:
\begin{equation}
    \mathcal{N}^\text{\tiny{SK}}(r,\bJ) \propto \ee^{N\Sigma^\text{\tiny{SK}}(r,\bJ)}\,, 
\end{equation}
where the proportionality factor accounts for subexponential contributions.
While the number of critical points fluctuates with different realizations of the disorder, we expect the complexity to concentrate around its average over different realizations of the connectivity $\bJ$. We are thus interested in computing the disorder-averaged complexity $\Sigma^\text{\tiny{SK}}(r)$, defined as
\begin{equation}
    \Sigma^\text{\tiny{SK}}(r) = \lim_{N\to \infty}\frac{1}{N}\langle \log \mathcal{N}^\text{\tiny{SK}}(r,\bJ)\rangle_\bJ\,.
\end{equation}
The number of critical points $\mathcal{N}^\text{\tiny{SK}}(r,\bJ)$ is counted using the Kac-Rice formula~\cite{azais2009level},
\begin{align}\label{eq:N_critical}
\begin{split}
    \mathcal{N}^\text{\tiny{SK}}(r,\bJ) &= \int \dd\bx\,\left(\prod_{i=1}^N\delta(\p_{i}E^\text{\tiny{SK}}(\bx,a))\right)\\&\times\lvert \det \bHSK(\bx,a)\rvert
    \delta(\mathcal{I}(\bx) -Nr)\,. 
\end{split}
\end{align}
The Dirac delta enforces $\bx$ to be a critical point of the CIM energy, while the quantity $\lvert \det \bHSK(\bx,a)\rvert$ is a phase space volume factor. The extensive index $\mathcal{I}(\bx)$ is the number of unstable directions around the critical point $\bx$. The number of critical points can thus be regarded as the partition function of a 'microcanonical' distribution, whose measure is given by the integrand in Eq.~\eqref{eq:N_critical}. Within this framework,  the complexity is thus analogous to the entropy of critical points in the system. An equivalent description, which we adopt here, employs a 'grand-canonical' ensemble. We introduce the grand-potential $\Omega^\text{\tiny{SK}}(\mu)$, defined as
\begin{align}\label{eq:Omega}
    \begin{split}
        \Omega^\text{\tiny{SK}}(\mu) &\equiv -\lim_{N\to\infty} \frac{1}{N} \Biggl\langle\log \int \dd\bx \left(\prod_{i=1}^N\delta(\p_{i}E(\bx))\right)\\
        &\times \lvert \det \bHSK(\bx,a)\rvert \ee^{\mu \mathcal{I}(\bx)}\Biggr\rangle_\bJ\\
        &\equiv -\lim_{n\to\infty}\frac{1}{N}\langle\log Z_\Omega\rangle_\bJ\,,
    \end{split}
\end{align}
where the last line defines the grand canonical partition function $Z^\text{\tiny{SK}}_\Omega$. The quantity $\mu$ is a chemical potential associated to the index of the critical points of the system. As $\mu\to-\infty$, the integrand in Eq.~\eqref{eq:Omega} concentrates around local minima of the energy landscape. The complexity is recovered from the grand-potential by means of a Legendre transform as
\begin{equation}\label{eq:Omega_and_Sigma}
    \Sigma^\text{\tiny{SK}}(r) = \inf_{\mu}\left[-\mu r - \Omega^\text{\tiny{SK}}(\mu) \right]\,.
\end{equation}
This extremization condition yields a relationship between the intensive index $r$ and the derivative of the grand-potential,
\begin{equation}\label{eq:r_from_Omega}
    r = -\frac{\p \Omega^\text{\tiny{SK}}(\mu)}{\p \mu}\,.
\end{equation}
In particular, when focusing on minima, we have $r=0$. Since the grand-potential is a convex function of the chemical potential we obtain, for the complexity of minima, that
\begin{equation}
    \Sigma(r=0) = -\lim_{\mu\to-\infty} \Omega^\text{\tiny{SK}}(\mu)\,.
\end{equation}
The evaluation of the grand-potential in the limit of large negative chemical potential allows us to compute the complexity of minima of the system. 

On the other hand, evaluating $\Omega^\text{\tiny{SK}}(\mu=0)$ gives access to the number of most abundant critical points, irrespective of the value of their intensive index $r$. This can be seen using the inverse of the Legendre transform in Eq.~\eqref{eq:Omega_and_Sigma}, evaluated at $\mu=0$, which yields
\begin{equation}
    -\Omega^\text{\tiny{SK}}(\mu=0) = \sup_r \Sigma^\text{\tiny{SK}}(r)\,.
\end{equation}
The index of the most abundant critical points can be computed using  the extremization condition in Eq.~\eqref{eq:r_from_Omega}, obtaining
\begin{equation}\label{eq:r*}
    r^* \equiv \argmax_r \Sigma^\text{\tiny{SK}}(r) = -\frac{\p \Omega^\text{\tiny{SK}}(\mu)}{\p\mu}\Biggr\rvert_{\mu=0}
\end{equation}
The discussion above highlights the two cases we are interested in: the complexity of the most abundant minima, and the complexity of the most abundant critical points, obtained respectively from the evaluation of the grand-potential in Eq.~\eqref{eq:Omega} for $\mu\to-\infty$ and $\mu=0$, respectively. 

The grand-potential is computed using the replica trick and averaging over the disorder. The details of this calculation are given in Appendix~\ref{app:computation_grand-potential}. The main difficulty presented by the computation arises from the presence of correlations between the gradient of the energy $\bnabla E^\text{\tiny{SK}}(\bx,a)$ and the determinant of the Hessian in Eq.~\eqref{eq:Omega} across different realizations of the disorder. Because of the presence of the single site energy $E_I$, these correlations are particularly difficult to analyze. Importantly, in order to make progress, we assume, as a working hypothesis, that these correlations are negligible (see also discussion after equation Eq.~\eqref{eq:approx_decoupling} in App.~\ref{app:computation_grand-potential}). The neglection of  these cross correlations is an uncontrolled approximation, and in Sec.~\ref{sec:first_look} we show \textit{a posteriori} that it yields a reasonable agreement for the properties of critical points with zero net magnetization on small size systems.  To enforce the stationary condition of the configurations contributing to the partition function in Eq.~\eqref{eq:Omega}, new order parameters, coupling different replicas, need to be introduced. These order parameters are then determined self-consistently by means of saddle point equations. The replica-symmetric solution reads
\begin{widetext}
\begin{equation}\label{eq:grand-potential}
    \begin{split}
        \Omega^\text{\tiny{SK}}(\mu) &= \frac{1}{2}\left[(\Delta A + A_o)^2 - A_o^2 + 2 A_d t_R + (\Delta C + C_o)(\Delta q + q_o) - C_o q_o \right] + vm - \overline{\overline{\log Z_{\text{\tiny{RS}},\,\Omega}(h_1,h_2)}}\\
        Z_{\text{\tiny{RS}},\,\Omega}(h_1,h_2) &= \int \dd x\, \frac{|\p^2_x \EOmegaeff(x,h_1)|}{\sqrt{2\pi \Delta q}}\exp\left[\frac{1}{2}\Delta C x^2 + x F(h_1,h_2) + \mu \mIMF(x) - \frac{(\Delta A x - \p_x \EOmegaeff(x,h_1))^2}{2\Delta q}\right]\\
        \EOmegaeff(x,h_1) &\equiv E_I(x,a+t_R) - (J_0 m  + \sqrt{q_o}h_1) x\\
        F(h_1,h_2) &\equiv v + \sqrt{C_o - \frac{A_o^2}{q_o}}h_2 + \frac{A_o}{\sqrt{q_o}}h_1\,.
    \end{split}
\end{equation}
The function $\mI_\text{\tiny{MF}}(x) \equiv \Theta\left(-\p_x^2 \EOmegaeff(x,h_1)\right)$, with $\Theta$ the Heaviside step function, is the index of a mean field effective energy $\EOmegaeff(x,h_1)$. The double overline $\overline{\overline{\cdots}}\equiv \int \frac{\dd h_1 \dd h_2}{2\pi}\,\ldots \ee^{-\frac 12 (h_1^2+ h_2^2)}$ denotes an average over the random Gaussian fields $h_1,\,h_2$. Note that the field $h_1$ stems from the overlap among different replicas of the system. Its physical origin is akin to the random Gaussian field appearing in the global minima calculation in Sec.~\ref{sec:global_minima}. The field $h_2$ stems instead from overlaps among a given replica and the Lagrange multipliers that enforce the replicated system to occupy a critical point. The $8$ order parameters $\Delta A, A_o, \Delta C, C_o, \Delta q, q_o, m, v$ are determined by the saddle point equations
\begin{align}\label{eq:saddle_point_grand-potential}
    \begin{split}
		\Delta A &= \frac{1}{2\Delta  q}\left( \overline{\overline{\langle x \p_x\EOmegaeff \rangle_{\scriptstyle\Omega}}}- \overline{\overline{\langle x \rangle_{\scriptstyle\Omega}\langle \p_x\EOmegaeff\rangle_{\scriptstyle\Omega}}}\right) - \frac{1}{2}t_R\\
        A_o &= -\frac{1}{\Delta q}\left(\Delta A q_o - \overline{\overline{\langle x \rangle_{\scriptstyle\Omega}\langle \p_x\EOmegaeff\rangle_{\scriptstyle\Omega}}} \right)\\
        \Delta q &= \overline{\overline{\langle x^2 \rangle_{\scriptstyle\Omega} - \langle x \rangle_\Omega^2}}\\ 
		  q_o &=\overline{\overline{\langle x \rangle_{\scriptstyle\Omega}^2}} \\
		\Delta C &= -\frac{1}{\Delta q} + \frac{\Delta A^2}{\Delta q}
        + \frac{1}{\Delta q^2}\left[-2\Delta A \left( \overline{\overline{\langle x \p_x\EOmegaeff\rangle_{\scriptstyle\Omega} - \langle x\rangle_{\scriptstyle\Omega}\langle \p_x\EOmegaeff\rangle_{\scriptstyle\Omega}}}\right) +  \overline{\overline{\langle (\p_x\EOmegaeff)^2 \rangle_{\scriptstyle\Omega} - \langle \p_x\EOmegaeff \rangle_{\scriptstyle\Omega}^2}}\right] \\
        C_o &= \frac{1}{\Delta q^2}\left(\Delta A^2 q_o - 2 \overline{\overline{\langle x \rangle_{\scriptstyle\Omega}\langle \p_x\EOmegaeff\rangle_{\scriptstyle\Omega}}} + \overline{\overline{\langle \p_x\EOmegaeff \rangle_{\scriptstyle\Omega}^2}}\right) \\ 
        v &= \frac{J_0}{\Delta q}\left( \overline{\overline{\langle \p_x \EOmegaeff(x,h_1)\rangle_{\scriptstyle\Omega}}} -m\Delta A \right)\\
        m &= \overline{\overline{\langle x \rangle_{\scriptstyle\Omega}}}\\
		t_R &= \overline{\overline{\left\langle\frac{1}{\lvert \p^2_x \EOmegaeff(x,h_1)\rvert}\right\rangle_{\scriptstyle\Omega}}}\,,
\end{split}
\end{align}
where $\langle\ldots\rangle_{\Omega}$ denotes an average over the single spin distribution $P^\text{\tiny{SK}}_{\Omega}(x,h_1,h_2)$, defined as
\begin{equation}\label{eq:P_Omega}
    P^\text{\tiny{SK}}_{\Omega}(x,h_1,h_2) \propto |\p^2_x \EOmegaeff(x,h_1)|\exp\left[\frac{1}{2}\Delta C x^2 + x F(h_1,h_2) + \mu \mIMF(x) - \frac{(\Delta A x - \p_x\EOmegaeff(x,h_1))^2}{2\Delta q}\right].
\end{equation}
Upon marginalizing over the random Gaussian fields $h_1$ and $h_2$, this expression yields the mean-field probability distribution of the population of soft spins at the critical points of the $N$ dimensional system, where each critical point is being weighted by an exponential factor $\ee^{\mu\mI}$, which thus skews the distribution toward appropriate value of the index $\mI$ through the chemical potential $\mu$. The order parameters $q_d$, $q_o$ are the overlap of a replica with itself and with another replica, respectively, in an assumed replica symmetric ansatz. The parameter $m$ is the magnetization of the system, and $v$ is a Lagrange multiplier that, when different from zero, enforces the magnetized configurations to be critical points. The order parameters $\Delta C$, $C_o$ contain the overlaps among different replicas of auxiliary fields that enforce the stationary condition in the Kac-Rice formula. Similarly, $\Delta A$ and $A_o$ involve the cross-overlaps among auxiliary fields and replicated soft spin variables. 

The single spin distribution at the most abundant critical points selected by the chemical potential $\mu$ can be rewritten as a weighted sum of Dirac delta distributions centered around the critical points of the effective mean field energy $\EOmegaeff$, defined in Eq.~\eqref{eq:grand-potential}. We show in Eq.~\eqref{eq:app_calculation_P_alternative} of App.~\ref{app:rewriting_P_Omega}  that, for any continuous test function $f(x)$, we have
\begin{equation}
    \langle f(x)\rangle_\Omega =\int \dd x\, f(x) P_\Omega(x,h_1,h_2) \propto \left\langle\sum_{y \in \Crt\left[\EOmegaeff(x,h_1+h_0)\right]} \ee^{\frac{1}{2}\left(\Delta C - \frac{\Delta A^2}{\Delta q} \right)y^2 + y F(h_1,h_2) + \mu \mIMF(y) +\frac{\Delta A y}{\Delta q}h_0}f(y) \right\rangle_{h_0}\,,
\end{equation}
\end{widetext}
where the brackets $\langle\ldots\rangle_{h_0}$ denote an average over the realizations of a new random field $h_0$, which follows a Gaussian distribution of zero mean and variance $\Delta q$. The order parameters $\Delta A$, $A_o$, $\Delta C$, $C_o$, and $v$ have the effect of reweighting each critical point. When these order parameters are zero, the average over the single-spin distribution reduces to
\begin{equation}
    \langle f(x)\rangle_\Omega\propto \left\langle \sum_{\scriptstyle y \in \Crt\left[\EOmegaeff(x,h_1+h_0)\right]} \ee^{ \mu \mIMF(y)}f(y)  \right\rangle_{h_0}\,,
\end{equation}
where we see that only the chemical potential controls the relative weight of minima, saddle points and maxima in the computation of the averages. 

The order parameters $\Delta A,\,\Delta C,\,A_o,\,C_o$, and $v$ play a special role with respect to the complexity of the energy landscape. In App.~\ref{app:convexity_implies_susy}, we show that these order parameters are self-consistently equal to $0$ when the effective energy $\EOmegaeff(x,h)$ is a convex function of $x$ for {\it all} the realizations of the field $h$. From Eq.~\eqref{eq:grand-potential}, we see that this condition is satisfied when $a+t_R<0$. In this phase, we show in App.~\ref{app:susy_implies_grand-potential_zero} that the grand-potential is $\OmegaSK(\mu)=0$ for all values of the chemical potential $\mu$. Thus, the number of most abundant critical points of any index grows at a sub-exponential rate with the size of the system. Moreover, from Eq.~\eqref{eq:r_from_Omega} we see that the intensive index of the most abundant critical points is $0$, which implies that the most abundant critical points are minima. The grand-potential in Eq.~\eqref{eq:Omega} could thus have been computed dropping the absolute value from the contribution coming from the determinant, which can then be computed by introducing fermionic and bosonic fields that share a supersymmetry~\cite{annibale2003role, annibale2003supersymmetric, crisanti2003complexity,aspelmeier2004complexity, crisanti2004spin, annibale2004coexistence, rizzo2005tap, cavagna2005cavity, muller2006marginal, kent2023count}. We thus refer to the phase where $\Delta A=A_o=\Delta C=C_o=v=0$ as the supersymmetric phase. The breakdown of the supersymmetric phase has important implications with respect to the properties of the landscape. Indeed, as we discuss in more detail below, the breakdown of supersymmetry implies a change in the nature of the critical points counted through the Kac-Rice formula~\cite{crisanti2003complexity, aspelmeier2004complexity, crisanti2004spin, rizzo2005tap, muller2006marginal}. In the supersymmetric phase, the critical points counted are stable minima. For marginally stable minima or unstable critical points to be properly counted by the Kac-Rice approach, the supersymmetry must breaks down. From a technical standpoint, allowing for supersymmetry breaking in the Kac-Rice formula is necessary because critical point with one or more marginally stable directions have zero determinant of the Hessian in Eq.~\eqref{eq:N_critical}, and would not contribute to the complexity if supersymmetry was not allowed to be broken. The additional supersymmetry-breaking order parameters are introduced precisely to circumvent this issue.

We emphasize that from the discussion above, we concluded that in the supersymmetric phase the most abundant critical point are stable minima, and they grow at a subexponential rate with the size of the system. The proliferation of marginally stable minima at an exponentially fast (in system size) rate occurs when the supersymmetric solution becomes unstable. Such an instability is thus akin to the one observed when counting metastable states at finite temperature in the binary Sherrington-Kirkpatrick spin-glass~\cite{crisanti2004spin}. 

The supersymmetric phase breaks down when the effective energy $\EOmegaeff(x,h)$ becomes a nonconvex function of $x$ for some values of $h$, i.e. when $a+t_R>0$. We then have $\Omega(0)\neq 0$. Thus the number of critical points increases at an exponential rate with the size of the system. Moreover, the most abundant critical points are no longer minima, as can be seen from Eq.~\eqref{eq:r*} which yields, within the replica-symmetric phase
\begin{equation}
    r^* = \overline{\overline{\langle \mIMF(\bx)\rangle_\Omega}}\Bigr\rvert_{\mu=0} >0\,,
\end{equation}

The discussion above implies that the line where $a+t_R=0$ in the $(a,J_0)$ plane determines the boundary separating a supersymmetric phase with a subexponential number of critical points, and a supersymmetry-broken phase where critical points proliferate at an exponential rate in the size of the system. We parametrize this phase boundary as $a_\Sigma(J_0)$. In the supersymmetry-broken phase the number of saddles is exponentially larger than the number of minima which are, as we discuss in the next Section, marginally stable.  

\subsection{Phase diagram of most abundant minima and critical points}\label{sec:grand-potential_phase_diagram}

The saddle point equations in Eq.~\eqref{eq:saddle_point_grand-potential} are solved numerically by iteration, in the cases $\mu\to-\infty$ and $\mu = 0$, for different values of $a$ and $J_0$. The results are represented in the phase diagram in Fig.~\ref{fig:critical_points_SK}. 
\begin{figure}
    \includegraphics[width=\columnwidth]{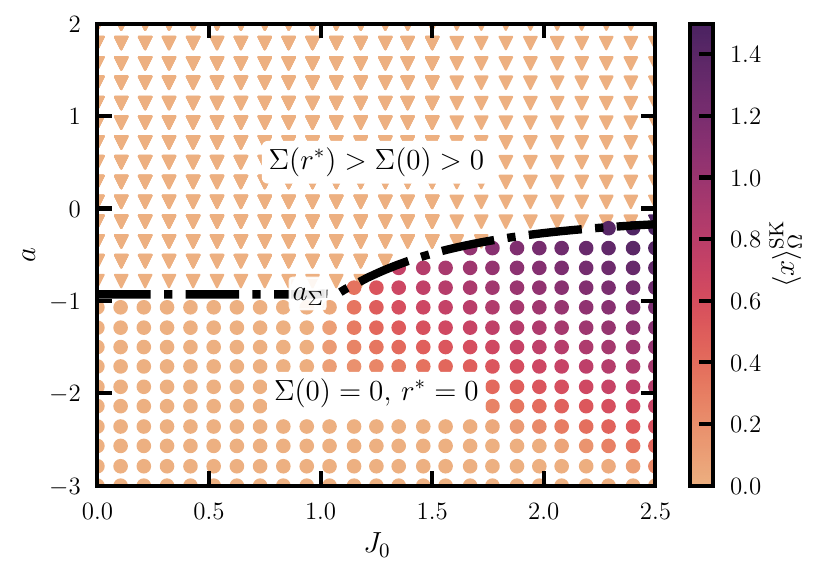}    \caption{\textbf{Proliferation of  critical points with zero net magnetization} Phase diagram of the most abundant minima and critical points of the CIM-SK, as obtained  from the replicated Kac-Rice calculation. The heat map is the mean of the soft spin variable $\langle x \rangle_\Omega^\text{\tiny{SK}}$, obtained for $\mu\to -\infty$. The dash-dotted line identifies a critical value $a_\Sigma(J_0)$ which separates a region where the complexity is zero from a region with an exponentially large (in the size of the system) number of  local minima and critical points with zero net magnetization. In this region, the index of the most abundant critical points, $r^*$, is larger than $0$, i.e., the most abundant critical points are saddles. The mean value of the soft spin distribution is zero both for the most abundant minima and for the most abundant critical points.\label{fig:SUSY_minima}}
\end{figure}
By checking the change in sign of $a+t_R$, we can draw the boundary separating the supersymmetric and the supersymmetry-broken phases. In the supersymmetric phase, the grand-potential is independent from $\mu$, so we show the results obtained for $\mu\to -\infty$. In this phase, we have $\Sigma(r)=0$ for all intensive indexes $r$. The mean value of the individual soft spin distribution at the critical points is $0$ for $J_0<1$, and grows monotonically as $J_0$ or $a$ are increased when $J_0>1$. For large values of $J_0$, the boundary of the supersymmetric phase overlaps with the boundary, displayed in Fig.~\ref{fig:phase_diagram_global_minima_SK}, between the ferromagnetic phase \Circled{II}  and the rigid spin-glass phase \Circled{IV}.

When supersymmetry is broken, the complexity of the most abundant critical points and of the minima is larger than $0$. The numerical solution of the saddle point equation shows that the mean value of the soft spin distribution in this phase is $0$ both for the most abundant critical point and for the case of minima. Thus at the supersymmetry-breaking transition, an exponentially large  `cloud' of  {\it unaligned} (i.e. with zero net magnetization) critical points proliferate across the energy landscape. Interestingly, this picture holds also for moderately large values of the ferromagnetic coupling $J_0$. The properties of the most abundant local minima for the coherent Ising machine in the absence of ferromagentic coupling have been studied in~\cite{yamamuraGeometricLandscapeAnnealing2024}. In particular, it has been shown that in the supersymmetry-broken phase the spectrum of the Hessian around these local minima is gapless. Therefore, we conclude that also in the presence of a ferromagentic coupling the Hessian at the most abundant local minima has a gapless spectrum, independent of the strength of $J_0$.

We further study the stability of the  solution in the supersymmetry-breaking phase with zero net magnetization against fluctuations of the mean spin value $m$ and of the associated Lagrange multiplier $v$. Our analysis, carried out in App.~\ref{app:stability_paramagnetic_susy}, shows that the $m=0$  solution is stable against these perturbation in the supersymmetry-broken phase. 

The determination of the phase boundary across which the complexity becomes different from $0$, and  critical points with zero net magnetization proliferate across the landscape, is also one of the main findings of this work. It justifies the dashed-dotted phase boundary in the yellow soft spin-glass region in Fig.~\ref{fig:cartoon}. We next complement this finding with an illustrative visualization of the properties of the critical points at large laser gain, and with an interpretation of a subset of the supersymmetry-breaking order parameters as reactivities of the grand-potential against certain specific perturbations of the single site energy of the coherent Ising machine.

\subsection{A portrait of critical points at large laser gain}\label{sec:first_look}

We can compare our theory to numerical experiments as follows. For a fixed value of $J_0$ and $a$, we solve the saddle point equations~\eqref{eq:saddle_point_grand-potential} numerically while scanning through different values of the chemical potential $\mu$. For each value of $\mu$, we compute theoretical predictions for the self-overlap $q_d(\mu)$, the mean value of the soft-spin distribution  $m(\mu)$, and the intensive energy $e(\mu)$ of the most abundant critical point with fixed intensive index $r(\mu)$. The intensive energy can be derived from the stationary condition of the gradient descent dynamics, Eq.~\eqref{eq:gradient_descent}, and it reads
\begin{equation}
    e(\mu) = \overline{\overline{\langle E_I(x,a)\rangle_\Omega}} - \frac{1}{2}\overline{\overline{\langle x \p_x E_I(x,a)\rangle_\Omega}}\,.
\end{equation}
We then construct parametric plots of how each of these predictions vary with the intensive index and we compare these theoretical predictions against experimental measurements obtained from direct sampling and enumeration of critical points in small size systems, following a procedure described in App.~\ref{app:numerical_sampling_critical_points}. We can compute pairs of properties of experimentally sampled critical points and display them in scatter plots, while comparing to theory. 

A theory-experiment comparison of this type is shown in Fig.~\ref{fig:critical_points_SK} for two different values of the ferromagnetic coupling $J_0$, at a large value of the laser gain $a$.
\begin{figure*}      
\includegraphics[width=\linewidth]{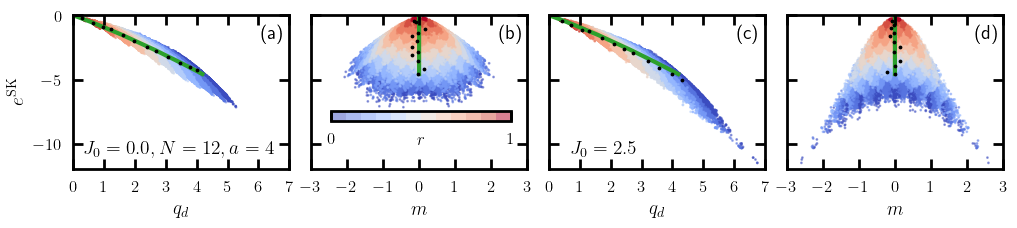}
    \caption{\textbf{The most abundant critical points  have zero net magnetization}. Scatter plots of critical points sampled from systems of small size $N=12$, at two different values of the ferromagnetic bias $J_0$, for values of the laser gain $a$ such that the ground state of the system is a spin-glass. The green line is the prediction from the Kac-Rice theory, the color of the points is the intensive index $r$ of the critical points, and the black dots are the properties of the most abundant critical points at fixed values of the intensive index $r$. Panels (a-b): scatter plots of the intensive energy as a function of the self-overlap $q_d \equiv \frac{1}{N} \sum_i \langle x_i^2\rangle_\bJ$ (mean squared distance of a soft-spin configuration from the origin), and of the intensive energy as a function of the mean of the soft spin distribution $m\equiv \frac{1}{N}\sum_i \langle x_i\rangle_{\bJ}$ for $J_0=0$. Panels (c-d): plots of the same quantities, but obtained for a system with large ferromagnetic coupling $J_0$. \label{fig:critical_points_SK}}
\end{figure*}
The numerical results show good agreement with the theoretical predictions, and a visual understanding of the properties of the critical points can be obtained. For both values of the ferromagnetic coupling, the intensive energy of the critical points decreases as their self-overlap increases. This observation demonstrates that the lower the energy of a critical point, the further away it is from the origin in the soft-spin configuration space. This trend is similar both without (Fig.~\ref{fig:critical_points_SK}a) and with (Fig.~\ref{fig:critical_points_SK}c) a strong ferromagnetic coupling $J_0$. The main difference in the two cases arises when considering parametric scatter plots of the intensive energy as a function of the magnetization. In the absence of ferromagnetic coupling (Fig.~\ref{fig:critical_points_SK}b), the cloud of critical points in finite size systems has the shape of a cone, with its spread increasing at lower energies and indices. In the presence of strong ferromagnetic coupling (Fig.~\ref{fig:critical_points_SK}d) the bottom of the cone bifurcates into an arrowhead, whose low energy tails consist of magnetized minima. However, the most abundant critical points for any given index  have zero net magnetization, both with and without a strong ferromagnetic coupling $J_0$. 

Overall, this theory-experiment comparison reveals a qualitative illustration of the salient properties of critical points in the supersymmetry broken phase, and shows that our replicated Kac-Rice calculation faithfully predicts the geometry of the landscape of the coherent Ising machine, at the level of most abundant critical points, even for modest system sizes.  A main summary is that at large laser gain $a$, for both weak and strong ferromagnetic coupling $J_0$, ferromagnetic minima occur at lower energy, but they are exponentially rarer than the most abundant minima, which  have zero net magnetization and occur at higher energies.

\subsection{Order parameters as reactivities of the grand-potential}\label{sec:reactivities}

We conclude this section by addressing how some supersymmetry-breaking order parameters appearing in the Kac-Rice saddle point equations~\eqref{eq:saddle_point_grand-potential} can be interpreted as derivatives of the grand-potential $\Omega^\text{\tiny{SK}}(\mu)$ with respect to perturbations of the single-site energy of the CIM. These perturbations are implemented by substituting the contribution from each single-site energy $E_I(x_i,a)$ in Eq.~\eqref{eq:E_CIM} with a modified version $E_I(x,a,\bepsilon)$, given by 
\begin{align}\label{eq:single_site_perturbed}
    \begin{split}
        E_I(x_i,a, \bepsilon) &= \frac 1 4 x_i^4 - \frac 1 2 (a + \epsilon_A) x_i^2 - \epsilon_v J_0 x_i \\
        &+ \sqrt{2 \epsilon_C}g_i x_i\,.
    \end{split}
\end{align}
The vector $\bepsilon \equiv [\epsilon_A,\, \epsilon_v,\,\epsilon_C]^\mathrm{T}$ contains the strength of the different types of perturbations performed on the CIM energy. They correspond respectively to changing the laser gain by an amount $\epsilon_A$, applying an external magnetic field of intensity $\epsilon_v J_0$, and applying a set of external random i.i.d Gaussian fields $g_i$, with mean $0$ and variance $1$. In App.~\ref{app:interpretation_SUSY} we detail how these perturbations modify the grand-potential, and obtain the relations
\begin{align}\label{eq:susceptibilities}
    \begin{split}
        \Delta A + A_o &= -\p_{\epsilon_{A}}\Omega^\text{\tiny{SK}}(\mu)\Big\rvert_{\bepsilon=\mathbf{0}} \\
        \Delta C &= -\p_{\epsilon_{C}}\Omega^\text{\tiny{SK}}(\mu)\Big\rvert_{\bepsilon=\mathbf{0}} \\
        v &= -\p_{\epsilon_{v}}\Omega^\text{\tiny{SK}}(\mu)\Big\rvert_{\bepsilon=\mathbf{0}}\,,
    \end{split}
\end{align}
which connect the order parameters to susceptibilities of the grand-potential $\Omega^\text{\tiny{SK}}$ under external perturbations of the single-site energy of the coherent Ising machine. In light of this analysis, in the supersymmetry-broken phase the number of critical points grows substantially upon external perturbation of the landscape. In Sec.~\ref{sec:grand-potential_phase_diagram}, we found that $v=0$ for the most abundant critical points even in the supersymmetry-broken phase. This implies that the landscape has a low reactivity to linear perturbations by means of an external magnetic field.   

The analysis carried out so far implies a fundamental discordance in the structure of the energy landscape at different energy levels: for large enough values of the ferromagnetic coupling $J_0$ and the laser gain $a$, global minima (and also rare low energy local minima) of the system are magnetized, but they are screened by exponentially many more higher energy most abundant local minima that  have zero net magnetization, and are thus unaligned with the ground state, even for large values of the ferromagnetic coupling. Given this landscape complexity, it is not clear how the dynamics of CIM energy minimization will then proceed, while the energy landscape geometry is annealed, by slowly increasing the laser gain. To address this important question, we now turn to directly analyzing the dynamics of the annealing process. 

\section{Annealed optimization dynamics of the CIM}\label{sec:dynamics}

\subsection{Numerical simulations}

We first simulate the landscape annealing dynamics of the CIM for a  system of $N=5\times 10^3$ spins. We focus on the case $J_0>1$, since the behavior of the annealing dynamics for $J_0=0$ has been explored in~\cite{yamamuraGeometricLandscapeAnnealing2024}, and for $J_0<1$ we saw that the global minima of the energy landscape  have zero net magnetization. For a given value of $J_0$, we first sample an instance of the connectivity matrix $\bJ$, and we quench the system at a fixed value of the laser gain $a$ in the convex ferromagnetic region of the phase diagram, i.e. phase \Circled{II} in Fig.~\ref{fig:phase_diagram_global_minima_SK}. This sets the initial condition at time $t=0$. We then simulate the annealing process by integrating numerically Eq.~\eqref{eq:gradient_descent} using Euler's method~\cite{suli2003introduction} with a time-step $\Delta t=0.15$. The laser gain $a=a(t)$ is slowly increased during the dynamics following the protocol
\begin{equation}
    a(t) = a(0) + \dot a t\,,
\end{equation}
with a fixed rate $\dot a$. As the dynamics proceeds, we track the minimum eigenvalue of the Hessian of the system $\lambda_\text{\tiny{min}}^\text{\tiny{SK}}(a(t))$, with the entries of the Hessian given by Eq.~\eqref{eq:Hessian}. We also track the intensive Ising energy of the system $e_\mathrm{Ising}(a(t)) = -\frac{1}{2N}\bs(t) \cdot \bJ\bs(t)$, with $s_i(t)=\sgn[x_i(t)]$. These measurements are then averaged over $50$ independent realizations of the quenched couplings $\bJ$. 
The effectiveness of the annealing process is compared against a spectral method, where the ground state configuration is approximated by $\bs_\text{\tiny spectral}$. The latter is obtained by looking for the real eigenvector with the largest eigenvalue of $\bJ$, and taking the sign of its entries. The approximated ground state is then used to obtain a spectral approximation to the energy $e_\text{\tiny spectral}^\text{\tiny SK}\equiv -\frac{1}{2 N}\bs_\text{\tiny spectral}\cdot \bJ\bs_\text{\tiny spectral}$. Results for an annealing rate $\tau \dot a=10^{-4}$ and two representative values of $J_0$ are shown in Fig.~\ref{fig:annealing_dynamics}.
\begin{figure}
    \includegraphics[width=\columnwidth]{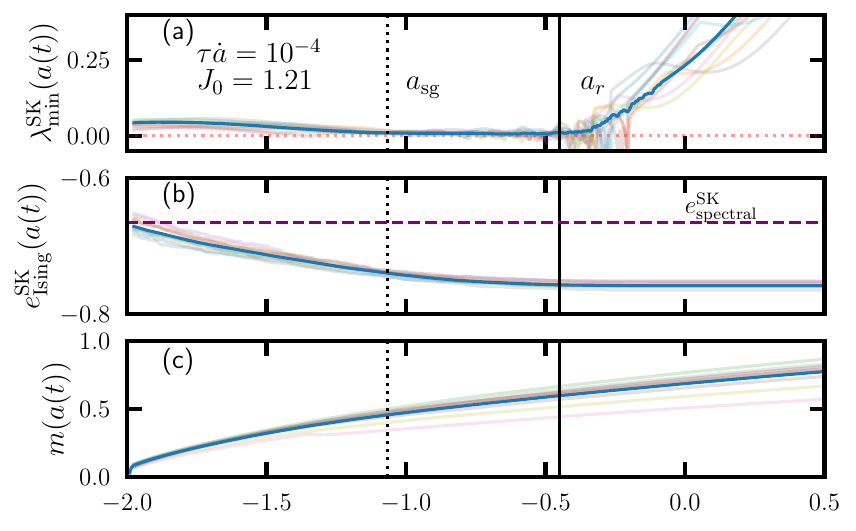}    \includegraphics[width=\columnwidth]{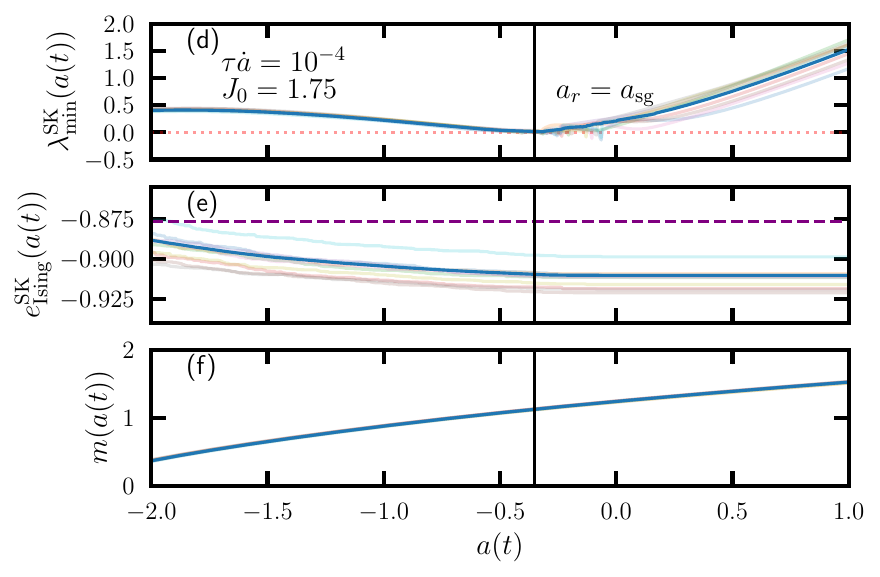}  \caption{\textbf{Annealing dynamics and rigidity transition} \label{fig:annealing_dynamics} Plots of the minimum value of the spectrum of the Hessian $\lambda^\text{\tiny{SK}}_\text{\tiny{min}}(a(t))$ and the Ising energy $e^\text{\tiny{SK}}_\text{\tiny{Ising}}(a(t))$ during the annealed dynamics of the coherent Ising machine for two different values of $J_0$. The solid blue curves represent an average over 50 independent runs, while the $10$ curves in the background are individual trajectories, included for illustrations. The horizontal dashed purple line denotes the average of the spectral approximation to the Ising energy across the different realizations of $\bJ$. The vertical dotted and solid ones denote the values of $a_\text{\tiny{sg}}$ and $a_\text{\tiny{r}}$ at which a spin-glass and a rigidity transition occur, respectively. Panel (a): parametric plot of $\lambda_\text{\tiny{min}}^\text{\tiny{SK}}(a(t))$ for $J_0=1.21$. The spin-glass transition and the rigidity transition take place at two different values of the laser gain, $a_\mathrm{sg}$ and $a_\mathrm{r}$, respectively. Panel (b): Parametric plot of the Ising energy $e_\text{\tiny Ising}^\text{\tiny SK}(a(t))$ for $J_0=1.21$. Panel (c): Parametric plot of the mean value of the soft spins $m(a(t))$ during the annealing process.  Panel (d): Parametric plot of $\lambda_\text{\tiny{min}}^\text{\tiny{SK}}(a(t))$ for $J_0=1.75$, where the rigidity transition and the spin-glass transition occur at the same value of the laser gain, $a_\text{\tiny r} = a_\mathrm{sg}$. Panel (e): Parametric plot of the Ising energy $e_\mathrm{Ising}^\text{\tiny{SK}}(a(t))$ for $J_0=1.75$. Panel (f): parametric plot of the mean value of the soft spin population $m(a(t))$ during the annealing process.}
\end{figure}

At small values of the laser gain, the annealing process takes place in the convex phase. The spectrum of the Hessian is gapped, and the gap monotonically decreases as the laser gain increases. Interestingly, the Ising energy of the system decreases too, as the system aligns more and more toward the ferromagnetic solution of the combinatorial optimization problem. This trend continues until the laser gain reaches the spin-glass transition value $a_\mathrm{sg}$. The dynamics of the system becomes qualitatively different depending on whether the laser gain value for the rigidity transition, $a_\text{r}$, is identical to or different from the spin-glass transition point $a_\mathrm{sg}$. 

If $a_\mathrm{sg} \neq a_\mathrm{r}$, as in Fig.~\ref{fig:annealing_dynamics}(a-b), the minimum value of the Hessian remains close to zero in the interval $[a_\mathrm{sg},\,a_\mathrm{r}]$. By looking at individual trajectories, represented as semitransparent lines in the figure, we observe that the lower edge of the spectrum can sometimes change sign. During these sign changes, the Ising energy of the system decreases in sudden steps. This phenomenon corresponds to the bifurcation of the current local stable minimum, where the system resides, into an unstable saddle. When such a bifurcation from stable minimum to unstable saddle occurs, the soft spins escape the saddle through its unstable direction(s), toward new configurations in which some of the spins may change their sign, yielding a lower Ising energy. In the average trajectory, these jumps are smoothed out in time, and the Ising energy gradually decreases as the laser gain increases. When $a \gtrapprox a_\mathrm{r}$, the spectral gap starts to increase with the laser gain, and the Ising energy stops decreasing,  reaching a plateau value. The system gets trapped in a minimum with a gapped spectrum of the Hessian, from which it cannot escape by an adiabatic increase of the laser gain.

In the second case, when $a_\mathrm{sg}= a_r$, as in Fig.~\ref{fig:annealing_dynamics} (c-d), the typical value of the lower edge of the spectrum of the Hessian touches the origin and then starts rising again for $a\approx a_\mathrm{sg}$, and the Ising energy immediately enters its plateau region. Some instability transitions are observed at the level of individual trajectories, but the average value of the edge of the spectrum increases monotonically with $a$. In this situation, annealing beyond the convex phase yields marginal reductions in the Ising energy reached by the system. Interestingly, however, in both cases considered the annealing process yields Ising configurations with lower energies compared to the spectral solution, even in the convex phase. 

Finally, we observe that for both values of $J_0$, the mean value of the population of soft spins during the annealing dynamics, $m(a(t))$, is different from $0$. The critical point in the vicinity of which the dynamics takes place are thus {\it not} one of the most abundant ones identified by means of the Kac-Rice calculation in Sec.~\ref{sec:Kac-Rice}. We observe that $m(a(t))$ grows monotonically with $a$. This can be understood by observing that as the laser gain increases, the separation between the two wells in the single site energy $E_I$, defined in Eq.~\eqref{eq:E_I_single_site}, increases too.  The dynamics thus takes place within a region of the phase space whose center is located further and further away from the origin as the laser gain increases, leading to larger $m$. 

The rigidity transition values we identified by means of our static calculation in Sec.~\ref{sec:fRSB} holds for the global minima of the system. These configurations may be different from the typical minima encountered by the annealed optimization dynamics. However,  the fact that the position of the lower edge of the spectrum of the Hessian increases with $a$ above the rigidity transition demonstrates that the properties of global minima are a good proxy of the properties of the typical minima explored by the CIM during landscape annealing. At the end of the annealing schedule displayed in Fig.~\ref{fig:annealing_dynamics}, we measure the energy of the coherent Ising machine per spin $e^\text{\tiny{SK}}(a(t))$. The final values obtained in  the cases investigated are $e^\text{\tiny{SK}}(a(t)=1) \approx -1.72 \pm0.01$ for $J_0=1.21$, and $e^\text{\tiny{SK}}(a(t)=1) \approx -2.16 \pm 0.02$ for $J_0=1.75$. These quantities are very close to the value of the ground state energy for the corresponding values of the laser gain and ferromagnetic coupling that can be computed from the analysis in Sec.~\ref{sec:global_minima}, which are respectively $e^\text{\tiny{SK}}_\text{\tiny{gs}} \approx -1.83$ and $e^\text{\tiny SK}_\text{\tiny gs} \approx -2.16$. 

Motivated by the above observations, we compare the properties of the typical soft-spin configurations visited during the annealing processes with the properties of global minima, most abundant local minima, and typical configurations reached through fast quenches at fixed laser gain $a$. Our results are shown in Fig.~\ref{fig:comparison_dynamics_vs_statics}.
\begin{figure*}
    \includegraphics[width=\linewidth]{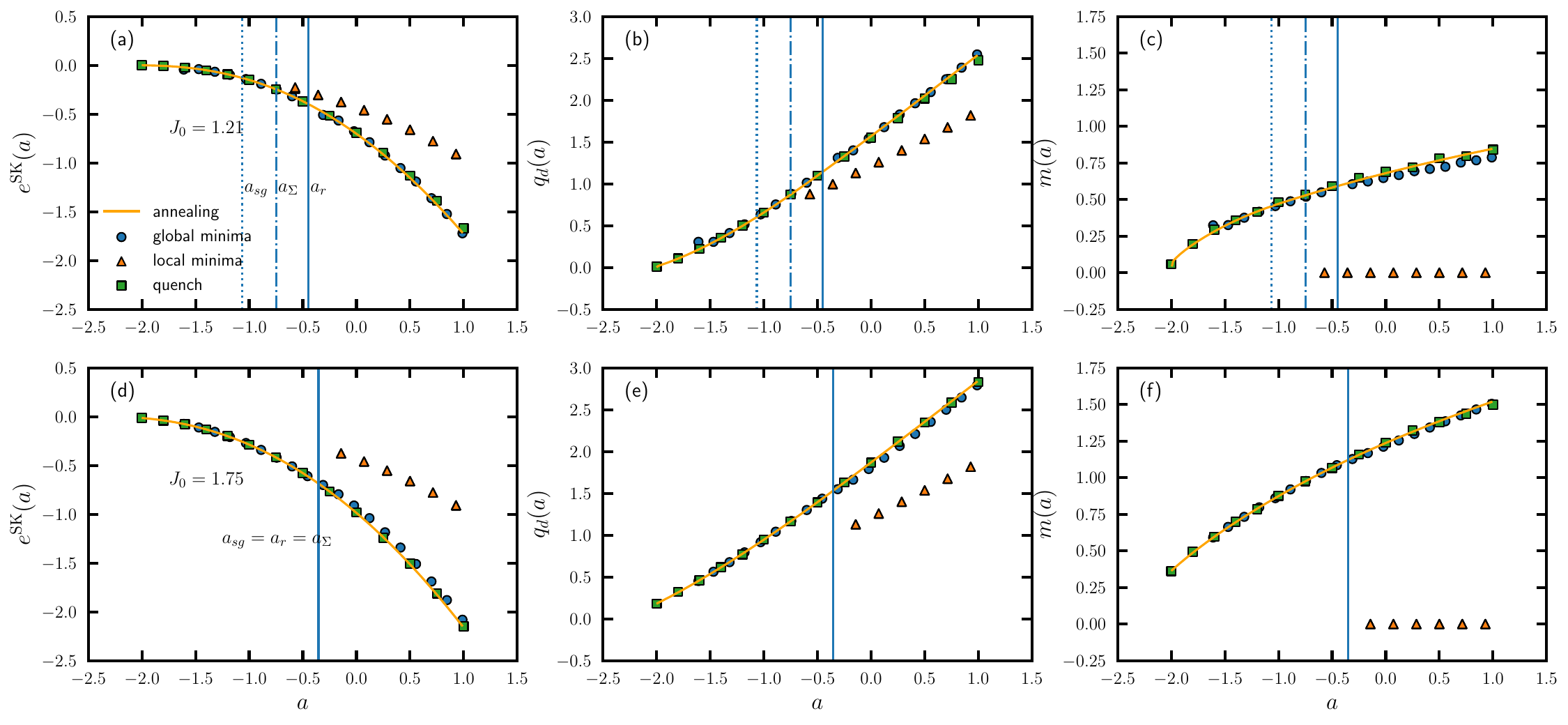}
    \caption{\label{fig:comparison_dynamics_vs_statics} \textbf{Configurations visited by quench and annealed dynamics are more similar to global minima than most abundant local minima}. The top and bottom row correspond to two different values of $J_0$, respectively. In each panel, we show the behavior of relevant average quantities as a function of the laser gain for: annealed dynamics (orange solid line); global minima (blue circles); local minima (red triangles); terminal state of a quenched dynamics at fixed $a$ (green squares). The average quantities of interest are: the CIM-energy per spin $e^{\text{SK}}(a)$ (panels (a) and (d)); the mean squared value of the soft spin population $q_d(a)$ (panels (b) and (e)); the mean value of the soft spin population $m(a)$ (panels (c) and (f)). The vertical blue lines signal the values of $a$ at which the spin-glass transition ($a_\text{\tiny{sg}}$, dotted line), the supersymmetry breaking transition ($a_\Sigma$, dot-dashed line), and the rigidity transition ($a_\text{\tiny{r}}, $ solid line) take place, respectively.  The orange solid line is obtained from numerical simulations of the annealing dynamics for a system of $N=5000$ spins and an annealing rate $\dot a \tau=10^{-4}$ (averaged over $50$ independent trajectories). The blue circles are obtained from the numerical integration of the full replica symmetry breaking solution for the properties of global minima. The red triangles are obtained from the numerical solution of the Kac-Rice calculation carried out in Sec.~\ref{sec:Kac-Rice} for the properties of the most abundant local minima. The green squares are obtained from the terminal configuration of a quench of the  dynamics at fixed $a$, with an initialization from the neighborhood of the eigenvector with the largest eigenvalue of the matrix $\bJ$ (averaged over $10$ independent trajectories). Overall the panels show from left to right that the annealed dynamics explores spin configurations more similar to global minima, with lower energy (left), higher distance from the origin (middle), and higher magnetization (right) than that of the most abundant local minima.}
\end{figure*}

We compare the CIM energy, the mean value of the soft spin population, and the self-overlap obtained during the annealed process. As soon as the complexity of local minima becomes nonzero, the properties of the most abundant minima become considerably different from the properties of the points visited by both quenched and annealed dynamics. The average CIM energy, the soft-spin population mean value, as well as the self-overlap $q_d$ at any given value of the laser gain in the annealing or the quench process are considerably closer to the values obtained for global minima rather than local minima. This difference becomes particularly striking when the mean-value of the soft-spin population $m(a)$ is considered, as this quantity is identically zero at all the values of the laser gain for the most abundant local minima, and increases with the laser gain both for global minima  and as the annealed dynamics unfolds. Interestingly both the annealed dynamics and quenches at fixed laser gain reach very similar configurations from the point of view of CIM energy, magnetization and mean squared value of the soft spin population. This result suggests that for intermediate values of the laser gain, the annealed optimization dynamics can successfully evade the exponential proliferation of most abundant higher energy local minima, and pierce down to much lower energy levels, to visit spin configurations more similar to lower energy global minima. 

To deepen our understanding of the dynamics of the soft spins across the rigidity transition, we develop a dynamical mean field theory (DMFT) which describes the annealing dynamics of the CIM in the thermodynamic limit, and we use it to describe how the probability distribution of the soft spins evolves during the annealing process.

\subsection{Dynamical mean field theory of the annealed optimization dynamics}\label{sec:dmft}

In the thermodynamic limit, the annealed landscape optimization dynamics of the CIM can be described using dynamical mean field theory (DMFT), involving an effective Langevin equation for the dynamics of a soft spin of the system during the annealing process. The derivation of the effective process using the dynamical cavity method~\cite{agoritsas2018out} is presented in App.~\ref{app:dmft}. Here we give an intuitive understanding of the terms involved. Let us consider the dynamics of a single tagged spin in the system. Since the number of interactions per spin grows linearly with system size, the interaction of the tagged spin with the rest of the system acts as a small perturbation to the dynamics of a system of $N-1$ spins, from which the tagged spin has been excluded (i.e. the cavity system of $N-1$ spins).  Therefore, the interactions among the tagged spin and the rest of the system are expanded using linear response theory, and decomposed into a random force, $\xi(t)$, and a response term that encodes the composite effect of the the perturbation that the tagged spin exerts on the cavity and the resultant back reaction of the cavity onto the tagged spin. Because of the ferromagnetic coupling $J_0$, the tagged spin is also subject to a magnetic field of strength $J_0 m(t)$, where $m(t)$ is the mean value of the soft-spin population as the dynamics takes place. In the thermodynamic limit, the random force $\xi(t)$ has Gaussian statistics due to the central limit theorem. The single-spin effective process $x(t)$  thus reads
\begin{equation}\label{eq:dmft}
    \begin{split}
        \tau \dot x(t) &= -\p_x E_I(x(t),a(t)) + J_0m(t) \\&
        + \int_0^t \dd\tau R(t,\tau)x(\tau) + \xi(t)\,.
    \end{split}
\end{equation}
The noise $\xi(t)$ is Gaussian with zero mean and temporal correlations defined as 
\begin{equation}\label{eq:noise_correlations}
    \langle\xi(t) \xi(t')\rangle_\bJ = C(t,t')\,.
\end{equation}
Because of the fully connected, mean field nature of the model, the knowledge of the statistics of the single spin dynamics allows us to self-consistently determine the noise autocorrelation function $C(t,t')$, the response function $R(t,t')$ and the magnetization $m(t)$, through the conditions
\begin{equation}\label{eq:self_consistent_dmft}
    \begin{split}
        R(t,t') &= \left\langle\frac{\delta x(t)}{\delta \xi(t')}\right\rangle_\bJ\\
        C(t,t') &= \left\langle x(t) x(t')\right\rangle_\bJ \\
        m(t) &= \left\langle x(t) \right\rangle_\bJ\,,
    \end{split}
\end{equation}
Equations~\eqref{eq:dmft},~\eqref{eq:noise_correlations} and Eq.~\eqref{eq:self_consistent_dmft} are a closed set of equations which constitute the DMFT of the CIM. They can be solved numerically for arbitrary initial conditions and annealing schedule through an iterative scheme described in Appendix~\ref{app:dmft_numerics}. Once these equations are solved, the time dependent single-spin distribution $P^\text{\tiny{SK}}(x,t)$ at time $t$ is obtained by running the dynamics given by Eq.~\eqref{eq:dmft} under different realizations of the noise $\xi(t)$. In Fig.~\ref{fig:dmft}, we show how the probability distribution of soft spins evolves at fixed $J_0=1.75$, as we anneal the system across the rigidity transition. 
\begin{figure*}
    \includegraphics[width=\linewidth]{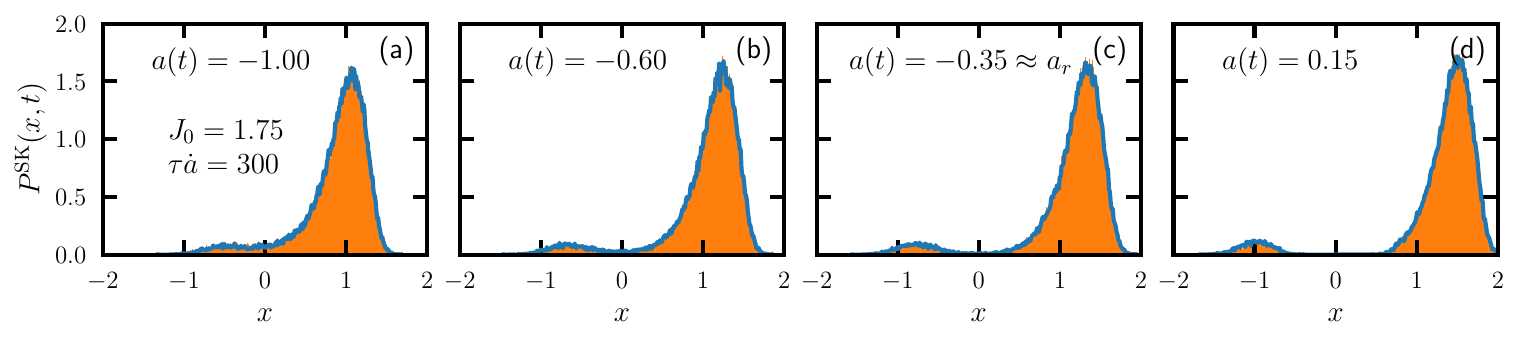}
    \caption{\textbf{Dynamical mean field theory of the annealing process of the coherent Ising machine.} Plots of the DMFT prediction for the single spin probability distribution $P(x,t)$ (blue lines) for the annealed dynamics of the CIM with $J_0=1.75$, for which $a_\mathrm{r} = a_\mathrm{sg}$.  The orange histograms are obtained from the numerical integration of Eq.~\eqref{eq:gradient_descent} for $50$  independent realizations of the disorder for a system of $N=2\,000$ spins. The different panels show snapshots of $P(x,t)$ at different stages of the annealing, depending on the value of the laser gain with respect to $a_\mathrm{r}$. The system is initialized at the global minimum for $a(0)=-1$ and is then gradually annealed with a constant rate $\tau \dot a$.  \label{fig:dmft} }
\end{figure*}
The initial condition $P(x,t=0)$ is taken to be the single spin distribution at the global minimum for $a(t=0)=-1$, which lies in the ferromagnetic, replica-symmetric region of the phase diagram in Fig.~\ref{fig:phase_diagram_global_minima_SK}.  In this initial condition, the distribution of spins is skewed toward the positive region of the $x$ axis, because of the influence of the ferromagnetic alignment, and the value of the probability density at the origin is finite. As the annealing dynamics proceeds, the value of $P(0,t)$ decreases, until it becomes zero approximately at the rigidity transition. For $a(t)\gtrapprox a_\mathrm{r}$, the single-spin probability distribution becomes gapped, and the width of the gap increases with $a$. This result supports, and rationalizes, the observation that above the rigidity transition the annealing process yields no improvement to the Ising energy of the system: as the laser gain increases, flipping a spin by a small increase of $a$ becomes harder and harder. Moreover, the dynamical mean field theory successfully predicts the evolution of the single-spin probability distribution for a moderate system size of $N=2\,000$ spins. 

In summary, we have developed a framework to identify the rigidity transition in the ground states of the CIM (Sec.~\ref{sec:fRSB}), and we have provided evidence that the line across which this transition occurs is a good indicator of the boundary below which the annealing process allows to reach configurations with lower Ising energy (Sec.~\ref{sec:dynamics} and Sec.~\ref{sec:dmft}). Moreover, the properties of the configurations visited during both quenched and annealed dynamics are much more similar to those of global minima rather than those of most abundant local minima. The strength of the ferromagnetic coupling $J_0$ determines whether the rigidity transition occurs within the spin-glass phase or at the boundary between the convex and the spin-glass phase.  

In the next section, we turn to a model where a specific ground state is engineered into a tunable, rugged energy landscape, and ask the question of how effective the CIM is in recovering such a planted solution.  

\section{Wishart Planted Ensemble}\label{sec:wpe}

\begin{figure}
    \includegraphics[width=\columnwidth]{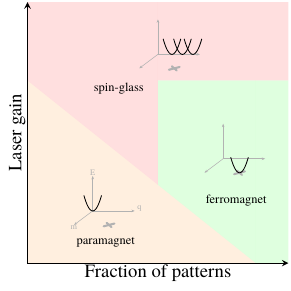}
    \caption{\textbf{Schematic phase diagram of global minima of the coherent Ising machine with Wishart planted ensemble.} Our mean field theory predicts the existence of three different regions, depending on the strength of the laser gain and fraction of patterns encoded in the connectivity matrix. At low laser gain, and below a critical threshold of the fraction of pattern, the system is a paramagnet, with the soft spin distribution concentrating at the origin, and the planted solution is not recovered. The boundary region of the paramagnetic phase is determined in Sec.~\ref{sec:wpe_large_laser_gain}. At intermediate value of the laser gain, if the fraction of patterns is above a critical threshold, the system is a ferromagnet (Sec.~\ref{sec:convex_ferro_wpe}), and full recovery of the planted solution is possible. At large value of the laser gain, the system is in a spin-glass phase, and the gradient descent dynamics from uninformed initial condition does not allow to recover the planted solution (Sec.~\ref{sec:wpe_dynamics}). \label{fig:wpe_cartoon}}
\end{figure}
The Wishart planted ensemble (WPE)~\cite{hamze2020wishart} is a fully connected, binary spin model with a known ground state, called the planted solution, and a tunable parameter that controls how hard it is to find the planted ground state through sampling methods such as parallel tempering~\cite{hukushima1996exchange}. The connectivity matrix $\bJWPE$ is tuned in such a way that the ground state of the Ising problem is given by the direction $\bt =[1,\ldots,1]^\mathrm{T}$, namely
\begin{equation}\label{eq:condition_WPE}
    \bt \in  \argmin_{\bs \in \{-1,1\}^N} \left(-\bs\cdot \bJWPE\bs\right) \,.  
\end{equation}
This is achieved by defining $\bJWPE$ as
\begin{align}\label{eq:def_WPE}
    \begin{split}
        \bJWPE &= \tilde{\mathbf{J}} - \text{diag}\, \tilde{\mathbf{J}}\\
        \tilde{\mathbf{J}} &=-\frac{1}{N}\sum_{\mu=1}^M \bw^\mu \otimes \bw^\mu\,.     
    \end{split}
\end{align}
The diagonal part of $\bJWPE$ is set to zero to hinder the recovery of the ground state using a spectral method like the one described in Sec.~\ref{sec:dmft}.  The matrix $\tilde{\mathbf{J}}$ is an $N\times N$ real-valued matrix, composed by a sum of outer products of $M$ random $N$-dimensional patterns $\{\bw^\mu\}_{\mu=1}^M$. The patterns $\bw^\mu$ are correlated random variables, defined as follows so all of them are orthogonal to $\bt$:
\begin{align}\label{eq:wmu}
    \begin{split}
        \bw^\mu &\equiv \bSigma^{1/2}\bz^\mu \\
        \bSigma^{1/2} &\equiv \sqrt{\frac{N}{N-1}}\left[\mathbf{1} - \frac{1}{N}\bt\otimes\bt^\mathrm{T}\right]\,,
    \end{split}
\end{align}
with $\{\bz^\mu\}_{\mu=1}^M$ a set of $M$ i.i.d. Gaussian vectors with zero mean and identity covariance. The matrix $\bSigma^{1/2}$ projects (and scales) any vector onto the orthogonal complement of $\bt$. The definition of $\bw^\mu$ implies 
\begin{equation}
    \bw^\mu \cdot\bt = 0\,,
\end{equation}
and this orthogonality condition implies $\tilde{\mathbf{J}}\bt=\mathbf{0}$. In this way, the connectivity matrix $\bJWPE$, defined in Eq.~\eqref{eq:def_WPE},  satisfies Eq.~\eqref{eq:condition_WPE}. In fact, for any binary spin configuration $\bs\in \{-1,1\}^N$, we have 
\begin{equation}
    -\bs\cdot\bJWPE\bs = \frac{1}{N}\sum_{\mu=1}^M|\bw^\mu\cdot \bs|^2 + \Tr \tilde{\bJ} \geq \Tr \tilde{\bJ}\,. 
\end{equation}
The last inequality becomes tight when $\bs=\bt$, which implies that $\bt$ is a ground state of the system. The Wishart planted ensemble can be seen as an anti-Hopfield model~\cite{nokura1998spin} with correlated patterns: the connectivity matrix $\bJWPE$ stores $M$ patterns, which are 
correlated in such a way as to be all orthogonal to the planted direction $\bt$. The minus sign in the definition of $\bJWPE$ ensures that the spin configurations that minimize the Ising energy $-\bs\cdot \bJWPE\bs$ are repelled by the stored patterns, instead of being attracted to them, as in the case of the traditional Hopfield model~\cite{hopfield1982neural, amit1985spin}. The relevant parameter determining the hardness of finding a ground state is the ratio between the dimensionality of the spin configuration, $N$, and the number of patterns $M$. We denote this ratio by $\alpha\equiv M/N$, with larger values of $\alpha$ indicating easier optimization problems, as we describe next.  

The phase diagram of this model with Ising spins and a finite temperature has been studied in~\cite{hamze2020wishart}, assuming a replica-symmetric structure. For $\alpha>1$, at high temperatures the global minimum of the free energy is paramagnetic. As the temperature is reduced below a critical value $T_c(\alpha)$ the system crosses a first order phase transition toward a state where a local  minimum completely unaligned with the planted solution coexists with the global, ferromagnetic minimum of the free energy. Upon further lowering the temperature below a value $T_m(\alpha)$, another transition occurs toward a state where the free energy is convex again, and the global minimum largely overlaps with the planted direction $\bt$. In this phase the ground state can be attained easily through Monte Carlo sampling methods. On the other hand, for $\alpha<1$, only the first order transition at $T_c(\alpha)$ is present. Below the critical temperature, the ferromagnetic state is the global minimum of the free energy, but its basin of attraction is very small, while the  unaligned state is a stable local minimum with a large basin of attraction. The metastability of the  unaligned state at all temperatures below $T_c(\alpha)$ for $\alpha<1$ hinders the retrieval of the planted solution. 

In the following, we study the properties of global minima of the CIM with the Wishart planted random connectivity matrix $\bJWPE$. The soft spins $\bx$ perform a gradient descent dynamics of the energy function $\EWPE$, defined as
\begin{equation}\label{eq:E_WPE}
    \EWPE(\bx) \equiv \sum_{i=1}^N E_I(x_i,a) - \frac{1}{2}\bx\cdot \bJWPE\bx\,,
\end{equation}
where the single site energy $E_I(x,a)$ is defined in Eq.~\eqref{eq:E_I_single_site}. We refer to this model as a CIM-WPE. We study the properties of the global minima of the CIM-WPE energy as the laser gain $a$ and the fraction of patterns $\alpha$ stored in $\bJWPE$ is varied. As is done for the CIM-SK, we first address the properties of the Hessian $H^\text{\tiny{WPE}}_{ij}(\bx) \equiv \frac{\p^2 \EWPE(\bx)}{\p x_i\p x_j}$. The probability density of its eigenvalues averaged over different realizations of the disorder, $\rho_\bx^\text{\tiny{WPE}}(\lambda)$, is defined as
\begin{equation}
\rho_\bx^\text{\tiny{WPE}}(\lambda) = \lim_{N\to\infty} \left\langle\frac{1}{N}\sum_{i=1}^N \delta(\lambda_i^\text{\tiny{WPE}}(\bx) - \lambda)\right\rangle_{\bz^\mu}\,,
\end{equation}
where $\{\lambda_i^\text{\tiny{WPE}}(\bx)\}_{i=1}^N$ it the set of eigenvalues of the Hessian matrix $\bHWPE(\bx)$, and $\langle\ldots\rangle_{\bz^\mu}$ is an average over the realization of the Gaussian random variables $\bz^\mu$, appearing in Eq.~\eqref{eq:wmu}. To evaluate the spectral distribution of global minima of the energy $\EWPE$, we can average $\rho_\bx^\text{\tiny{WPE}}$ over the Boltzmann distribution $P_\text{\tiny{B}}^\text{\tiny{WPE}}(\bx)$, defined as
\begin{equation}\label{eq:boltzmann_wpe}\rho_{\text{\tiny{B}}}^\text{\tiny{WPE}}(\bx) = \frac{\ee^{-\beta \EWPE(\bx)}}{\int \dd\bx'\, \ee^{-\beta \EWPE(\bx')}}\,,
\end{equation}
with $\beta$ the inverse temperature. In the limit $\beta\to\infty$, this distribution concentrates around global minima of $\EWPE$. We can then consider the spectral distribution $\rho^\text{\tiny{WPE}}(\lambda)$, defined as
\begin{equation}\label{eq:rholambda_wpe}
    \rho^\text{\tiny{WPE}}(\lambda) \equiv \lim_{N\to\infty} \left\langle \frac{1}{N}\sum_{i=1}^N \delta(\lambda_i^\text{\tiny{WPE}}(\bx) - \lambda)\right\rangle_{\bx, \bz^\mu}\,,
\end{equation}
where the average $\langle\ldots\rangle_\bx$ denotes an average of the soft spin configurations $\bx$ over the Boltzmann distribution for the CIM-WPE, $P_\text{\tiny{B}}^\text{\tiny{WPE}}$. The spectral distribution $\rho^\text{\tiny{WPE}}(\lambda)$ can be equivalently obtained from the knowledge of the resolvent $G^\text{\tiny{WPE}}(z)$, defined as
\begin{equation}\label{eq:G_WPE}
    G^\text{\tiny{WPE}}(z) \equiv \int \dd\lambda\, \frac{\rho^\text{\tiny{WPE}}(\lambda)}{z-\lambda}\,.
\end{equation}
Once $G^\text{\tiny{WPE}}(z)$
is known, it can be inverted to determine the spectral distribution $\rho^\text{\tiny{WPE}}$, through the same procedure described in Sec.~\ref{sec:replica_resolvent} for the CIM-SK.  To gather some intuition, we start by looking at low values of the laser gain $a$. 
\subsection{Low laser gain}\label{sec:wpe_large_laser_gain}

For low values of the laser gain $a\ll 0$, the single-site energy of the CIM, $E_I$ in Eq.~\eqref{eq:E_WPE}, dominates over the interaction term. As in the CIM-SK, we expect the origin to be the global minimum of the energy of the system. The Hessian $\bHWPE(\bx)$ evaluated at the origin is 
\begin{equation}\label{eq:Hessian_origin_WPE}
    \bHWPE(\mathbf{0}) = -\bJWPE - a\mathbf{1}\,.
\end{equation}
From the definition of $\bJWPE$ given by Eq.~\eqref{eq:def_WPE}, we see that an eigenvector of $\bHWPE(\mathbf{0})$ is given by $\bt/\sqrt{N}$. The associated eigenvalue is
\begin{equation}
    \frac{1}{N}\bt\cdot\bHWPE(\mathbf{0})\bt = \frac{1}{N}\Tr \tilde \bJ - a \approx -\alpha-a\,, 
\end{equation}
where in the first equality we used the definition of $\bJWPE$, given by Eq.~\eqref{eq:def_WPE}, and in the second we observed that in the large $N$ limit the quantity $\Tr \tilde\bJ$ concentrates around its mean value $-M$ by the central limit theorem. This eigenvalue becomes negative when
\begin{equation}\label{eq:instability_origin_wpe}
a>-\alpha\,,
\end{equation}
which yields a lower bound to the region in the $(\alpha,a)$ plane where the origin is no longer a stable minimum of the CIM-WPE energy. From Eq.~\eqref{eq:Hessian_origin_WPE}, we see that the spectrum of the Hessian around the origin follows the same distribution of the spectrum of the connectivity matrix $-\bJWPE$ with the support shifted by an amount $a$. The spectral distribution of $-\bJWPE$ has been computed in~\cite{hamze2020wishart}. It follows a Marchenko-Pastur law with a persistent Dirac delta distribution due to the presence of the planted solution $\bt$, and a shift of the support by an amount $\alpha$ since $\bJWPE$
 is constructed by subtracting from the matrix $\widetilde{\bJ}$ its diagonal elements. Overall, the spectral distribution around the origin is 
 \begin{equation}
     \rho_\mathbf{0}^\text{\tiny{WPE}}(\lambda) =\begin{cases} \alpha\delta(\lambda +a+\alpha) + \rho_\text{\tiny MP}(\lambda +a+\alpha) \quad \alpha\leq 1\\ 
     \frac{1}{N}\delta(\lambda + a+\alpha) + \rho_\text{\tiny{MP}}(\lambda+a+\alpha) \quad \alpha>1\end{cases}\,,
 \end{equation}
where $\rho_\text{\tiny{MP}}(\lambda)$ is the bulk of the Marchenko-Pastur law, which reads
\begin{align}
    \begin{split}
       \rho_\text{\tiny{MP}}(\lambda) &\equiv \frac{\sqrt{(\lambda - \lambda_-)(\lambda_+-\lambda)}}{2\pi \lambda}\\
       \lambda_\pm &= (1 \pm \sqrt{\alpha})^2\,, \quad \lambda \in [\lambda_-,\,\lambda_+]\,
    \end{split}
\end{align}
and, with a slight abuse of notation, we have included the persistent Dirac delta at the origin for $\alpha>1$, even though its weight vanishes in the thermodynamic limit. This mass is responsible for the instability of the origin as the laser gain increases above $-\alpha$. 

\subsection{Resolvent and single-spin distribution}

To determine the resolvent of the CIM-WPE in regions where the global minimum of the system is different from the origin, we assume that we can neglect the correlations between the matrix elements of $\bJWPE$ and the location of the global minimum of the CIM energy. This assumption is justified by the calculation performed above for the SK model. Within this approximation,  the Hessian $\bHWPE(\bx)$ is given by a random diagonal perturbation to $\bJWPE$. The resolvent of $\bJWPE$ is the resolvent of a Wishart matrix~\cite{potters2020first}, shifted by an amount $\alpha\mathbf{1}$ due to the subtraction of the diagonal element in Eq.~\eqref{eq:def_WPE}. The self consistent equation satisfied by the resolvent of $G^\text{\tiny{WPE}}(z)$ is then given by~\cite{tao2012topics} 
\begin{equation}\label{eq:G_pastur_wpe}
    G^\text{\tiny{WPE}}(z) = \int \dd x\,\frac{P^\text{\tiny{WPE}}(x)}{ z - \p^2_x E_I(x,a+\alpha) - \frac{\alpha}{1- G^\text{\tiny{WPE}}(z)}}\,, 
\end{equation}
where $P^\text{\tiny{WPE}}(x)$ is the probability distribution of a soft spin at the global minima of the CIM-WPE. It is obtained by computing the probability to find any spin in the system in a given configuration $x$ in the thermodynamic limit,
\begin{equation} \label{eq:Pwpe}
    \lim_{N\to\infty} \left\langle\frac{1}{N} \sum_{i=1}^N\delta(y_i-x )\right\rangle_\by^\text{\tiny{WPE}} =  P^\text{\tiny{WPE}}(x)\,.
\end{equation}
This computation is performed in Appendix ~\ref{app:wpe_replicas} using the replica trick. The final result is 
\begin{equation}\label{eq:P_WPE_MF}
    P^\text{\tiny{WPE}}(x) = \lim_{n\to 0} \frac{1}{n}\sum_{\alpha=1}^n\langle\delta(x-x^\alpha )\rangle_\text{\tiny{MF}}^\text{\tiny{WPE}}\,.
\end{equation}
The mean field average $\langle\ldots\rangle_\text{\tiny{MF}}^\text{\tiny{WPE}}$ is over a set of $n$ replicated soft spins, 
\begin{equation}
    \langle\ldots \rangle_\text{\tiny{MF}}^\text{\tiny{WPE}} = \frac{1}{Z_\text{\tiny{MF}}^\text{\tiny{WPE}}}\int \prod_{\alpha=1}^n\dd x^\alpha \,\ee^{-\beta \EWPEMF(\{x^\alpha \})}\,,
\end{equation}
with the limit $n\to 0 $ to be eventually taken by means of an analytic continuation. The mean field energy $\ERSWPE$ is\footnote{Note that, with a slight abuse of notation, we are using the symbol $\alpha$ to denote both the fraction of patterns stored in $\bJWPE$ and a replica index, and we are using the symbol $\beta$ to denote both the inverse temperature and a replica index.}
\begin{align}\label{eq:E_MF_WPE}
    \begin{split}
        E_\text{\tiny{MF}}^{\text{\tiny{WPE}}} &= \sum_{\beta=1}^n E_I(x^\alpha,a+\alpha ) - \frac{\alpha\beta}{2}\sum_{\alpha,\beta}r_{\alpha\beta}x^\alpha  x^\beta  \\
        &-  \alpha\sum_{\alpha=1}^n \widehat{m}_\alpha x^\alpha \,.
    \end{split}
\end{align}
and the mean field partition function $Z_\text{\tiny{MF}}^{\text{\tiny{WPE}}}$ is 
\begin{equation}
    Z_\text{\tiny{MF}}^\text{\tiny{WPE}} = \int \prod_{\alpha=1}^n \dd x^\alpha \, \ee^{-\beta E_\text{\tiny{MF}}^{\text{\tiny{WPE}}}[\{x^\alpha \}]}\,.
\end{equation}
We see from Eq.~\eqref{eq:E_MF_WPE} that the laser gain is shifted by an amount equal to the fraction of stored patterns $\alpha$.  The order parameters $r_{\alpha\beta}$, $\widehat{m}_\alpha$ arise as a consequence of the Wishart statistics of the connectivity matrix, and they are related to the matrix of overlaps $\bQ$ among the different replicas and a magnetization vector $\bmm$ encoding the overlap of each replica with the direction $\bt$. The saddle point equations determining the different order parameters are
\begin{align}\label{eq:saddle_point_wpe}
    \begin{split}
        \beta r_{\alpha\beta} &= -\frac{\p}{\p \beta Q_{\alpha\beta}} \Tr \log [\mathbf{1}_n + \beta \bQ - \beta \bmm \otimes \bmm] \\
        \widehat{m}_\alpha &= -\frac{1}{2}\frac{\p}{\p \beta m_\alpha} \Tr \log [\mathbf{1}_n + \beta \bQ - \beta \bmm \otimes \bmm]\\
        Q_{\alpha\beta} &= \langle x^\alpha  x^\beta \rangle_\text{\tiny{MF}}^\text{\tiny{WPE}}\\
        m_\alpha &= \langle x^\alpha \rangle_\text{\tiny{MF}}^\text{\tiny{WPE}}\,.
    \end{split}
\end{align}
These can be solved once an ansatz for the structure of the order parameters in replica space is chosen. In what follows, we study the replica-symmetric ansatz. 

\subsection{Convex phase}

The replica symmetric ansatz for the CIM-WPE is
\begin{align}
    \begin{split}
        r_{\alpha\beta} &= \delta_{\alpha\beta}r_d + (1-\delta_{\alpha\beta})r_o \\
        Q_{\alpha\beta} &= \delta_{\alpha\beta}q_d + (1-\delta_{\alpha\beta})q_o\\
        m_\alpha &= m \\
        \widehat{m}_\alpha &= \widehat m\,.
    \end{split}
\end{align}
Since we work in the low temperature limit $\beta\to\infty$, it is convenient to define the rescaled gaps
\begin{align}
    \begin{split}
        \Delta \widetilde r &\equiv \beta(r_d - r_o) \\
        \Delta \widetilde q &\equiv \beta(q_d - q_o)\,.
    \end{split}
\end{align}
In Appendix~\ref{app:wpe_rs} we compute the single-spin probability $P^\text{\tiny{WPE}}_\text{\tiny{RS}}(x)$ in the replica symmetric phase, starting from Eq.~\eqref{eq:P_WPE_MF}. The final result is
\begin{equation}\label{eq:PRSWPE}
    \PRSWPE(x) =\overline{\ee^{-\beta \ERSWPE(x,h)}\left[\int\dd x' \ee^{-\beta \ERSWPE(x',h)}\right]^{-1}}\,,
\end{equation}
where $\overline{\cdots} \equiv \int \dd h\, \ldots \frac{1}{\sqrt{2\pi}}\ee^{-h^2/2}$ is an average over the realizations of the Gaussian quenched disorder $h$. The replica-symmetric mean field energy $\ERSWPE(x,h)$ reads
\begin{align}\label{eq:E_WPE_RS}
    \begin{split}
        \ERSWPE(x,h) &\equiv E_I(x,a+\alpha+\alpha\Delta \widetilde r)  \\
        &- (\sqrt{\alpha r_o}h + \alpha \widehat{m})x\,.
    \end{split}
\end{align}
The laser gain in the mean field energy $\ERSWPE$ is shifted by an amount $\alpha+\alpha\Delta \widetilde r$, while the order parameter $\alpha\widehat m$ acts as a magnetic field induced by the presence of the correlated patterns orthogonal to $\bt$.  In the limit $\beta\to\infty$, the single spin probability distribution $\PRSWPE(x)$ concentrates, for each realization of the disorder $h$, around the minimum $x^*(h)$ of the mean field energy $\ERSWPE(x,h)$. The saddle point equations Eq.~\eqref{eq:saddle_point_wpe} then become
\begin{align}\label{eq:sp_wpe_rs}
    \begin{split}
        \Delta \widetilde q &= \overline{[\p^2_x \ERSWPE(x^*(h),h)]^{-1}}\\
        q_o &=  \overline{(x^*(h))^2}\\
        m &= \overline{x^*(h)}\\
        \Delta \widetilde r &= -\frac{1}{1+\Delta \widetilde q}\\
        r_o &= \frac{q_o - m^2}{(1+\Delta \widetilde q)^2}\\
        \widehat m &= \frac{m}{1+\Delta \widetilde q}\,.
    \end{split}
\end{align}
In App.~\ref{app:replicon_wpe}, we obtain a condition for the stability of the replica-symmetric solution based on the sign of the replicon $\widetilde\Lambda_\text{\tiny{R}}^\text{\tiny{WPE}}$, which reads
\begin{equation}\label{eq:wpe_replicon}
    \widetilde\Lambda_\text{\tiny{R}}^\text{\tiny{WPE}} \equiv -1 + \alpha \Delta \widetilde r^2 \overline{[\p^2_x \ERSWPE(x^*(h),h)]^{-2}} < 0\,. 
\end{equation}
The saddle point equations given by Eq.~\eqref{eq:sp_wpe_rs} can be solved numerically, while checking the stability condition in Eq.~\eqref{eq:wpe_replicon}, through an iterative method. Once the order parameters are known, we can obtain the Ising magnetization of the CIM-WPE at global minima from the single spin distribution $\PRSWPE(x)$,according to the formula  
\begin{equation}
    m_\text{Ising}= \int_0^{\infty}\dd x\, [2\PRSWPE(x)] - 1\,,
\end{equation}
where single spin distribution is evaluated in the low temperature limit $\beta \to\infty$. 

The resulting phase diagram is displayed in Fig.~\ref{fig:phase_diagram_global_minima_WPE}.
\begin{figure}
    \includegraphics[width=\columnwidth]{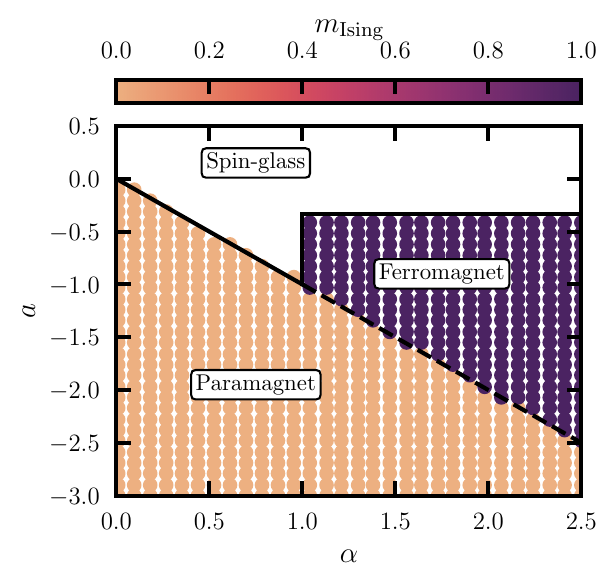}
    \caption{\textbf{Phase diagram of global minima for the coherent Ising machine with Wishart planted ensemble.} The color-code is the Ising magnetization of the global minimum of the CIM-WPE as the laser gain $a$ and the fraction of patterns $\alpha$ are changed. We identify a convex paramagnetic phase, where the soft spins concentrate at the origin; a ferromagnetic phase, where $m_\text{\tiny{Ising}}=1$ the planted solution $\bt$ is fully recovered, and a spin-glass phase (white region in the phase diagram). The phase diagram has been obtained by numerical integration of the saddle point equations in Eq.~\eqref{eq:saddle_point_wpe}, and the phase boundaries are obtained analytically, as discussed in Sec.~\ref{sec:convex_para_wpe} and Sec.~\ref{sec:convex_ferro_wpe}.\label{fig:phase_diagram_global_minima_WPE}}
\end{figure}
For $a<-\alpha$ the global minimum is paramagnetic. As anticipated in Sec.~\ref{sec:wpe_large_laser_gain}, the single site energy dominates over the interaction term of the CIM-WPE. As $a$ increases, two different scenarios occur depending on the fraction of patterns $\alpha$ stored into the connectivity matrix $\bJWPE$. If $\alpha<1$, the replicon eigenvalue $\widetilde\Lambda_\text{\tiny{R}}^\text{\tiny{WPE}}$ in Eq.~\eqref{eq:wpe_replicon} becomes positive for $a>-\alpha$. The replica symmetric solution is no longer stable, and the system enters a spin-glass phase. If instead $\alpha>1$, the paramagnetic minimum becomes unstable for $a>-\alpha$, and the global minimum of the system becomes ferromagnetic. The Ising magnetization $m_\text{\tiny{Ising}}$ jumps discontinuously from $0$ to $1$ upon crossing this boundary, meaning that the CIM fully retrieves the planted ground state $\bt$. Interestingly, even for $\alpha>1$, upon further increase of the laser gain, the system undergoes a spin-glass transition at $a=1/3$. The presence of a spin-glass phase suggests that for $\alpha>1$ the annealing process needs to be run from small values of the laser gain in order to successfully retrieve the ground state. We next discuss how these phases can be derived analytically from the saddle point equations in Eq.~\eqref{eq:sp_wpe_rs}, and comment on the form taken by the spectrum of the Hessian and the single-spin distribution.\\

\tocless\subsubsection{Convex paramagnetic phase}\label{sec:convex_para_wpe}

For $a<-\alpha$, we assume a paramagnetic ansatz $r_o=q_o =m=\widehat m=0$. The single-spin distribution $\PRSWPE(x)$ is then a Dirac delta distribution centered around the origin, 
\begin{equation}
    \PRSWPE(x) = \delta(x)\,.
\end{equation}
The saddle point equations for $\Delta \widetilde q$ and $\Delta \widetilde r$ then become 
\begin{align}\label{eq:sp_wpe_para}
    \begin{split}
        \Delta \widetilde q &= -\frac{1}{a+\alpha+\alpha \Delta \widetilde r}\\
        \Delta \widetilde r &= -\frac{1}{1+\Delta \widetilde q}\,,\\
    \end{split}
\end{align}
while the stability condition in Eq.~\eqref{eq:wpe_replicon} becomes
\begin{equation}
    -1 + \frac{\alpha \Delta \widetilde q^2}{(1+\Delta \widetilde q)^2} < 0\,,
\end{equation}
The order parameter $\Delta \widetilde q$ can be obtained by solving Eq.~\eqref{eq:sp_wpe_para}, and it reads
\begin{equation}\label{eq:Delta_q_para}
    \Delta \widetilde q = \frac{-1-a +\sqrt{(1-a)^2 - 4\alpha}}{a+\alpha}\,.
\end{equation}
As $a\to-\alpha$, the overlap gap $\Delta \widetilde q$ diverges, and the replicon eigenvalue tends to $\widetilde\Lambda_\text{\tiny{R}}^\text{\tiny{WPE}}\to -a+\alpha$. If $\alpha>1$, we then obtain that the replicon eigenvalue changes sign continuously as $a$ passes through the values $-\alpha$, and the paramagnetic phase becomes unstable. If $\alpha<1$, the mean field energy $\ERSWPE$ becomes nonconvex at the transition. The replicon eigenvalue jumps then discontinuously to $\infty$. The condition for the convexity of the effective single-site energy $\ERSWPE(x,h)$ is $a+\alpha+\alpha\Delta \widetilde r <0$, or, using the saddle point equations Eq.~\eqref{eq:sp_wpe_rs},
\begin{equation}\label{eq:convexity_condition}
    a + \frac{\alpha\Delta \widetilde q}{1+ \Delta \widetilde q} <0\,.
\end{equation}
Substituting Eq.~\eqref{eq:Delta_q_para} into the equation above, we obtain that the convex paramagnetic phase is no longer stable when $a>-\alpha$, and that a transition toward a spin-glass phase occurs for $\alpha<1$.\\

\tocless\subsubsection{Ferromagnetic minimum}\label{sec:convex_ferro_wpe}

For $\alpha>1$, the numerical solution of the saddle point equations shows that $r_o=0$, while $q_o,\,m$ and $\widehat m$ are different from zero. This type of solution implies that the single spin distribution is a Dirac delta centered around the point $x^*$ that minimizes the replica-symmetric mean field energy $\EWPERS(x)$ in Eq.~\eqref{eq:E_WPE_RS}, which no longer depends on the quenched random field $h$:
\begin{align}\label{eq:PRSWPE_ferro}
    \begin{split}
        &\PRSWPE(x) = \delta(x-x^*)\\
        &x^{*3} - (a + \alpha(\Delta \widetilde r + 1))x^* - \alpha \widehat m = 0\,. 
    \end{split}
\end{align}
Since $x^*\neq 0$, the Ising magnetization of the system becomes $m_\text{Ising}=1$. The coherent Ising machine fully recovers the planted ground state in this phase. 

The instability boundary of the ferromagnetic phase can be found analytically using our replica calculation. We first observe that, when $\alpha>1$ the replicon eigenvalue in Eq.~\eqref{eq:wpe_replicon} can jump discontinuously from a negative value to $\infty$ when the energy function $\ERSWPE$ in Eq.~\eqref{eq:E_WPE_RS} becomes non convex, i.e. when the stability condition in Eq.~\eqref{eq:convexity_condition} is violated. To determine the laser gain value for which the instability takes place, we need to find $x^*$. This can be done by observing that, if $r_o=0$, the saddle point equations Eq.~\eqref{eq:sp_wpe_rs} yields $\widehat m= \frac{x^*}{1+\Delta \widetilde q}$. Substituting this expression in the stationary condition in Eq.~\eqref{eq:PRSWPE_ferro} we obtain $x^*=\sqrt{a+\alpha}$, which is well-defined in the region $a>-\alpha$.The saddle point equation Eq.~\eqref{eq:sp_wpe_rs} for $\Delta \widetilde q$ becomes
\begin{equation}\label{eq:Deltaq_ferro}
    \Delta \widetilde q = \frac{1}{3 x^{*2} - a -\alpha\Delta \widetilde q/(1+\Delta \widetilde q)}\,. 
\end{equation}
Substituting the known value of $x^*$ allows to solve for $\Delta \widetilde q$, obtaining
\begin{equation}
    \Delta \widetilde q = \frac{1 - 2a - 3\alpha - \sqrt{1+4a+4a^2+2\alpha+ 12a\alpha+ 9\alpha^2}}{4(a+\alpha)}\,.
\end{equation}
Substituting this expression into the convexity condition in Eq.~\eqref{eq:convexity_condition} and solving for $a$ allows us to conclude that the ferromagnetic solution is stable when $\alpha>1$ and the laser gain $a$ obeys the bounds 
\begin{equation}
    -\alpha<a<-\frac{1}{3}\,.
\end{equation}
This is the purple region shown in Fig.~\ref{fig:phase_diagram_global_minima_WPE},
and we have concluded our discussion of the stability boundary of the ferromagnetic solution. 

The spectrum of the Hessian of the ferromagnetic minima in the replica-symmetric phase is given by a Marchenko-Pastur distribution shifted by an amount $2(a + \alpha)$. This can be seen by substituting the single spin probability distribution of Eq.~\eqref{eq:PRSWPE_ferro} into the expression for the resolvent given by Eq.~\eqref{eq:G_WPE}. The result is
\begin{align}
    \begin{split}
        G^{\text{\tiny{WPE}}}(z) &= \frac{1}{z-3x^{*2} - a-\alpha -\frac{\alpha}{1-G^\text{\tiny{WPE}}(z)} }\\
        &=\frac{1}{z-2(a+\alpha)-\frac{\alpha}{1-G^\text{\tiny{WPE}}(z)}}\,,
    \end{split}
\end{align}
which is the self-consistent equation for the resolvent of a Wishart matrix with parameter $\alpha$, perturbed by a constant diagonal matrix $2(a+\alpha)\mathbf{1}$. 

After discussing the properties of global minima of the CIM-WPE, we turn to a study annealing dynamics of the system and the degree to which the planted ground state can be recovered.\\  

\subsection{Dynamics}\label{sec:wpe_dynamics}
We perform numerical simulation of the annealing process of the CIM-WPE by integrating numerically the gradient descent dynamics 
\begin{equation}
    \tau \frac{\dd}{\dd t}\bx(t) = -\bnabla \EWPE(\bx)\,,
\end{equation}
for a system of $N=2000$  spins. For a fixed value of $\alpha$, we first initialize the system with a quench at $a=a(0)$ within the paramagnetic, replica symmetric phase. We then run the annealing dynamics, using the same protocol of Sec.~\ref{sec:dynamics}. During the annealing process, we track the position of the lower edge of the spectrum of the Hessian of the system $\lambda_\text{\tiny{min}}^{\text{\tiny{WPE}}}(t)$ in its current configuration, its Ising magnetization $m_\text{\tiny{Ising}}(t)$ and its Ising energy $e^\text{\tiny{WPE}}_\text{\tiny{Ising}}(t)$. The process is repeated for $50$ independent realizations of the disordered couplings $\bJWPE$. The results are displayed in Fig.~\ref{fig:annealing_dynamics_wpe}, 
\begin{figure}
    \includegraphics[width=\columnwidth]{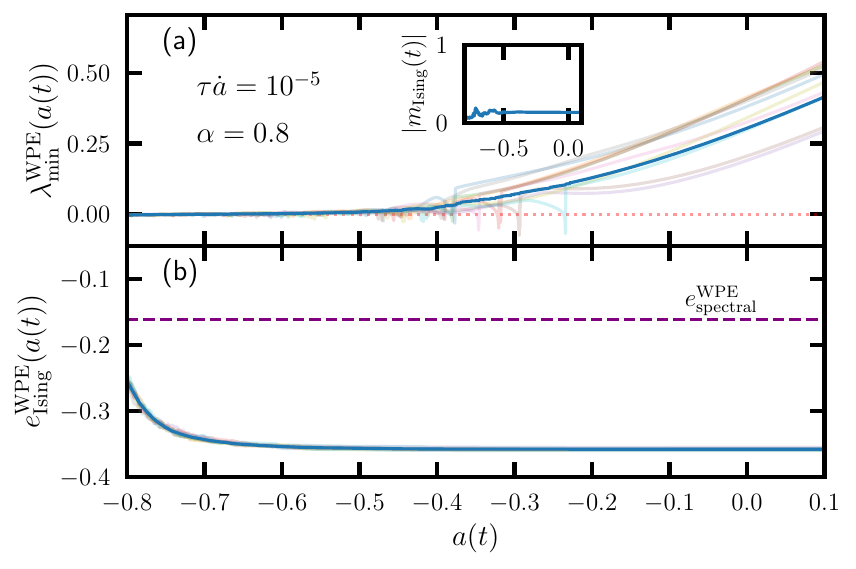}
    \caption{\textbf{Annealing dynamics of the coherent Ising machine in the Wishart planted ensemble} Panel (a): lower edge of the spectrum of the Hessian as a function of the laser gain $a(t)$. The solid blue line represent the average over different realizations of the dynamics and the connectivity matrix $\bJWPE$, while the semitransparent lines are random individual trajectories. The lower edge remains close to the value of zero (signaled by the red dotted line) at the beginning of the annealing process and starts increasing around $a(t)\approx 0.4$. In the inset, we show the Ising magnetization $m_\text{\tiny{Ising}}(t)$ as a function of the laser gain. Panel (b): Ising energy of the system during the annealing process. The dashed purple line is the estimate of the ground state energy obtained from a spectral method.  \label{fig:annealing_dynamics_wpe}}
\end{figure}
where we show results for the representative value of $\alpha=0.8$. The lower edge of the spectrum of the Hessian remains close to the origin in an interval of values above the spin-glass transition value $a_\text{sg}=-\alpha$. In the region where the lower edge of the Hessian remains close to the origin, the Ising energy of the system decreases through configurational changes similar to the ones witnessed for the Sherrington-Kirkpatrick model in Sec.~\ref{sec:dynamics}. The Ising magnetization of the system, shown in the inset of Fig.~\ref{fig:annealing_dynamics_wpe}(a) remains close to zero, indicating that the ground state of the system is far from being recovered.  The Ising energy found through the annealing dynamics is lower than the one obtained with a spectral method, represented by a purple dashed line in Fig.~\ref{fig:annealing_dynamics_wpe}(b). This analysis suggests that for $\alpha<1$, there is a paramagnetic spin-glass phase in the CIM-WPE, where the system finds lower energy solutions thanks to the annealing dynamics but struggles to align with the planted solution.

To investigate the spin-glass phase for $\alpha>1$, we simulate fast quenches from a random initial conditions for a density of stored patterns $\alpha=1.5$. The soft spins at initial times are independently sampled from a Gaussian distribution of standard deviation $0.1$. We perform quenches for two different values of the laser gain, at $a=-1.0$ and $a=0.0$, which correspond to the ferromagnetic and spin-glass states in the phase diagram in Fig.~\ref{fig:phase_diagram_global_minima_WPE}, respectively. The time evolution of the Ising energy and Ising magnetization during the quenches are shown in Fig.~\ref{fig:quench_dynamics_wpe}. 
\begin{figure}
    \centering
    \includegraphics[width=\columnwidth]{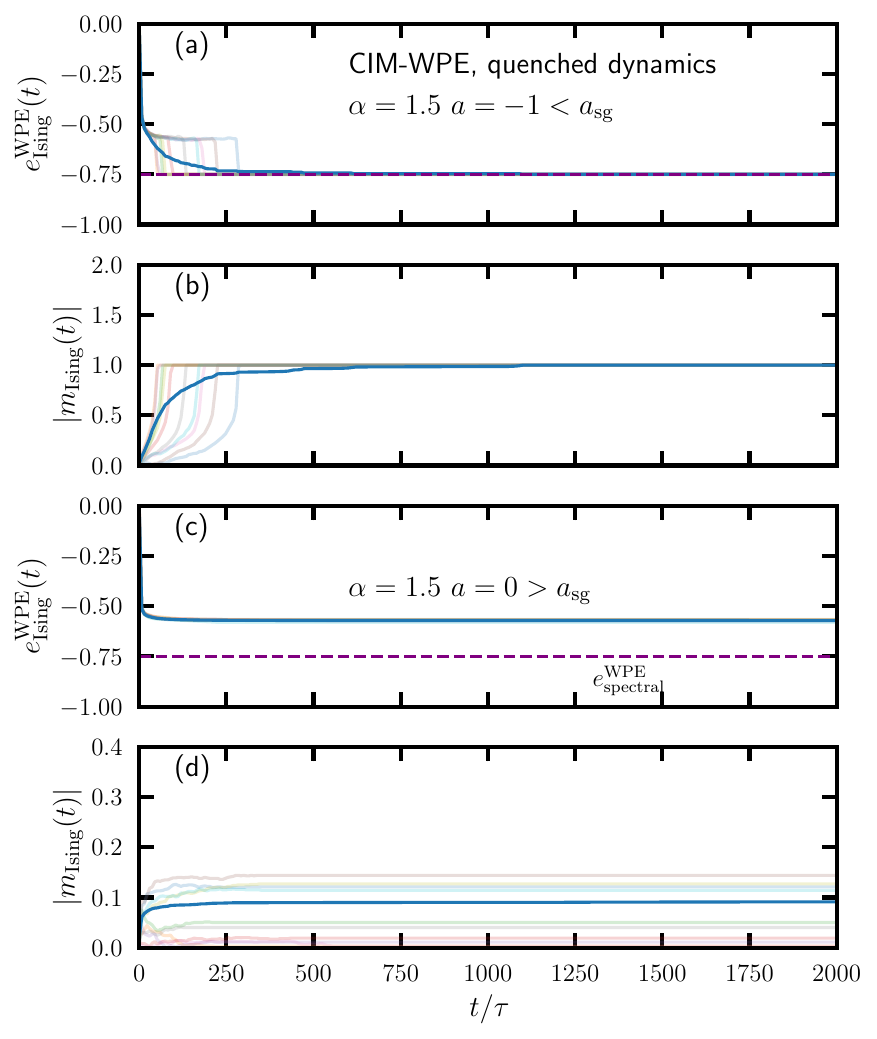}
    \caption{\textbf{Quench dynamics of the CIM-WPE in the convex and in the spin-glass phase.} Plots of the Ising energy $e_\text{Ising}(t)$ and of the Ising magnetization $m_\text{Ising}(t)$ as a function of time during a quench from random initial condition at fixed density of patterns $\alpha=1.5$, for two values of the laser gain. Panels (a-b): results for a quench in the convex ferromagnetic phase, $a<a_\text{\tiny{sg}}$. After an initial transient, the ground state is successfully  recovered. Panels (c-d): results for a quench in the spin-glass phase, $a>a_\text{sg}$. at long times, the Ising configuration of the system displays only a partial overlap with the ground state.}
    \label{fig:quench_dynamics_wpe}
\end{figure}
When $a<a_\text{\tiny{sg}}$, the Ising magnetization quickly converges to $1$, and the ground state is completely retrieved. The energy of the ground state matches that found using a spectral method, as expected when $\alpha>1$. When $a>a_\text{\tiny{sg}}$, the Ising magnetization at long times becomes considerably smaller, and the Ising energy recovered by the coherent Ising machine is greater than the ground-state energy. These results demonstrate the importance of initializing the annealing process from low values of the laser gain in the coherent Ising machine.

\section{Outlook}\label{sec:conclusion}

In this work, we employ a variety of techniques from the statistical physics of disordered systems, including the replica method, random matrix theory, Kac-Rice formulas, and dynamical mean field theory,  to gain insights into how the evolving geometry of a high dimensional energy landscape impacts the efficacy of the annealed optimization dynamics of the coherent Ising machine. We have characterized the low-energy minima of the coherent Ising machine when tackling the Sherrington Kirkpatrick model with ferromagnetic coupling, and found a region in the space of possible laser gain and ferromagnetic coupling where the global minima present soft modes. The appearance of these soft modes is tied with the nature of the spin-glass phase of the system and with the shape of the single site energy of the CIM. Our study of the annealed optimization dynamics shows that these soft modes are exploited to reach lower energies of the associated combinatorial optimization problem during the annealing process, yielding a mechanistic explanation for the efficacy of the CIM. Also, through a replicated Kac-Rice calculation, we showed that the annealing dynamics reaches states with a net nonzero magnetization in regions of the landscape where the most abundant critical points and minima are instead  paramagnetic.

Let us first discuss how we expect our results to be modified as more experimental details are taken into account. The soft-spin model of the CIM presented and analyzed here is constructed under the assumption that the photon decay rate of the laser pump is much larger then the one of the photons in the optical cavity, by neglecting sources of quantum and thermal noise, and by assuming that the connectivity matrix $\bJ_\text{\tiny{Ising}}$ modulating the feedback among the different optical cavities can be perfectly tuned. While the first assumption can be reasonably implemented in practice~\cite{yamamoto2020coherent}, the second and third assumptions cannot be perfectly realized. However, we expect thermal and quantum noise to dress up the structure of the landscape investigated here. In particular, we expect the rigidity transition, and its implication on the optimization dynamics, to be robust against the introduction of noise, which could on the other hand change the laser gain value at which the rigidity transition occurs. Concerning the possibility of errors in the calibration of the connectivity matrix, our analysis of the landscape reactivity in Sec.~\ref{sec:reactivities} suggests that the presence of this noise, at least at mild levels, should increase the number of critical points in the landscape.

Finally, we highlight here some venues of research to be explored building on our work. The first venue concerns the design and evaluation of novel Ising machine architectures. Soft modes appear as a crucial element in the CIM that determine the effectiveness of the annealing process. As these soft modes are tied to the shape of the single-site energies, it would be interesting to study the impact of higher order self-interactions on the appearance and lifetime of the soft modes, as well as the effect of heterogeneous modulations of the laser gain across the different units, mimicking mean field models of supercooled liquids~\cite{bouchbinder2021low, urbani2022field, folena2022marginal}. 

The second venue concerns the study of alternative dynamics for the CIM. Here, low-energy Ising configurations are found using a gradient descent dynamics. What are limits and possibilities when nongradient dynamics are employed~\cite{leleu2017combinatorial, leleu2019destabilization, ghimentiSamplingEfficiencyTransverse2023, xu2025combinatorial} is an interesting avenue of research that could be tackled using the dynamical mean field theory employed here. The latter could be combined with high-dimensional optimal control~\cite{urbani2021disordered}, to design annealing schedules subject to a fixed energy and time budget. 

A third avenue points at sharpening even further our understanding of the geometry of the annealed landscape and its relation with the dynamics. For instance, it would be interesting to extend our framework to allow the tracking of a tagged critical point as the laser gain increases, in the same spirit as what has been done to investigate temperature chaos in spin-glasses~\cite{rizzo2003chaos, chen2017temperature, subag2018free}. Furthermore, the  Kac-Rice calculation presented here addresses the properties of the most abundant critical points, independent of their energy. The construction of a full map of the critical points and a characterization of the linear excitations around them, conditioned on any possible energy level, is a challenging task, both from the numerical and the analytical standpoints, that future efforts could address.

\acknowledgments
We thank Aditya Mahadevan and Pierfrancesco Urbani for very insightful discussions. We are grateful to J\'er\'emie Klinger and Aditya Mahadevan for useful feedbacks on an early version of the manuscript.  We thank the Stanford Sherlock cluster for computing resources. F.G. acknowledges support from a postdoctoral fellowship of the Stanford's Leinweber Institute of Theoretical Physics. A.S. acknowledges funding from the National Science Foundation Graduate Research Fellowship, A.Y. acknowledges support from the Masason foundation. S.G. thanks the Simons foundation and a Schmidt foundation polymath award for support. This work was supported by the National Science Foundation, under the awards CCF-2423832 and CCF-1918549. During the finalization of this manuscript. we became aware of a parallel work~\cite{zhou2025phase} addressing the efficacy of the coherent Ising machine in the absence of ferromagnetic coupling, and how its performance can be improved by suitable modifications of the CIM energy landscape.

\section*{Data availability statement}

The data used to produce the figures in this paper are publicly available~\cite{github}.

\newpage
\appendix
\setcounter{tocdepth}{0}
\addtocontents{toc}{\protect\setcounter{tocdepth}{0}} 
\settocdepth{section}
\section{Derivation of Eq.~\eqref{eq:Gz_after_average}: \\ Computing the resolvent}\label{app:Gz_calculation}

We start by substituting the expression of the CIM-SK energy $E^\text{\tiny{SK}}(\bx,a)$, given by Eq.~\eqref{eq:E_CIM_SK} into the integral of Eq.~\eqref{eq:Gz_double_replicas}, which thus reads
\begin{widetext}
    \begin{align}
        \begin{split}
            \int_{\bx^\alpha ,\,\bphi^{\alpha\rho}}\, &\Biggl\langle\exp\Biggl[-\beta \sum_{\alpha=1}^n\sum_{i\leq j}^N\left( \delta_{ij}E_I(x_i^\alpha,a) -\frac{ J_{ij}}{\sqrt{N}} x_i^\alpha x_j^\alpha\right) +\frac{\beta J_0}{2N}\sum_{\alpha=1}^n\left(\sum_{i=1}^N x_i^\alpha\right)^2 \\
            &+ \frac{i}{2}\sum_{\alpha=1}^n\sum_{\rho=1}^l\sum_{i,j}^N \phi^{\alpha\rho}_i \left[(z - \p^2_x E_I(x_i^\alpha,a))\delta_{ij} + \frac{J_{ij}}{\sqrt{N}} + \frac{J_0}{2N} \right]\phi^{\alpha\rho}_j  \Biggr]\Biggr\rangle_\bJ\,,
        \end{split}
    \end{align}
where we recall that the quantity $z$ is a shorthanded notation for $z+i0^+$, the limiting value to the real axis of a complex variable with positive imaginary part. This choice ensures the convergence of the integrals over the bosonic variables $\phi^{\alpha\rho}_i$. The Gaussian average $\langle\ldots \rangle_\bJ$ over the realizations of the disorder is evaluated using the integral
\begin{equation}\label{eq:Gz_starting_point}
    \left\langle\ee^{\gamma J_{kl}}\right\rangle_\bJ = \int \prod_{i\leq j}\frac{\dd J_{ij}\,\ee^{-\frac{1}{2(1+\delta_{ij})} J_{ij}^2}}{\sqrt{2\pi}}  \ee^{ \gamma J_{kl}} = \ee^{\frac{\gamma^2}{2}(1+\delta_{kl})}\,,
\end{equation}
valid for any $\gamma \in \mathbb{C}$, and pair of indices $k,l$. In the large $N$ limit, the difference between the variance of diagonal and off-diagonal elements can be neglected, as its contribution in the exponential of Eq.~\eqref{eq:Gz_starting_point}
is of order $O(1)$, while we anticipate the leading contributions to grow linearly with $N$. Upon averaging over the disorder, the integral in Eq.~\eqref{eq:Gz_starting_point} becomes   \begin{align}\label{eq:integral_after_average_disorder}
        \begin{split}
            \int_{\bx^\alpha ,\,\bphi^{\alpha\rho}}\exp&\Biggl[-\beta \sum_\alpha \sum_{i=1}^N E_I(x_i^\alpha,a) + \frac{N\beta^2}{4}\sum_{\alpha,\beta}\left(\frac{\bx^\alpha  \cdot \bx^\beta}{N}\right)^2 + \frac{N\beta J_0}{2}\sum_\alpha\left(\frac{\bx^\alpha \cdot\bt}{N} \right)^2 + \frac{iN \beta}{2} \sum_{\alpha,\beta,\rho}\left(\frac{\bx^\alpha \cdot \bphi^{\beta\rho}}{N}\right)^2 \\
            &- \frac{N}{4}\sum_{\alpha,\beta,\rho,\sigma}\left(\frac{ \bphi^{\alpha\rho}\cdot \bphi^{\beta\sigma}}{N}\right)^2 +\frac{iNJ_0}{2}\sum_{\alpha,\rho}\left(\frac{\bphi^{\alpha\rho}\cdot\bt}{N}\right)^2+ \frac{i}{2}\sum_{\alpha,\rho}\sum_i \left(\phi^{\alpha\rho}_i\right)^2(z-\p^2_x E_I(x_i^\alpha,a))\Biggr]\,,  
        \end{split}
    \end{align}
\end{widetext}
where the indices $\alpha,\beta$ run from $1$ to $n$, while the indices $\rho$, $\sigma$ run from $1$ to $l$, and we recall that $\bt$ is a $N$-dimensional vector with all entries equal to one, $\bt \equiv [1,\,1,\,\ldots,\,1]^\mathrm{T}$. We note that, with a slight ambiguity, the symbol $\beta$ denotes both the inverse temperature and the index of the replicated soft spin. The quadratic terms can be linearized by means of a Hubbard-Stratonovich transformation 
\begin{equation}\label{eq:app_Hubbard_Stratonovich}
    \ee^{\frac{\gamma}{2N} x^2} = \int_{-\infty}^{\infty} \frac{\dd u}{\sqrt{2\pi/N\gamma}}\,\ee^{-\frac{N \gamma}{2}u^2 + \gamma u x}\,,
\end{equation}
with $\gamma$ a real positive quantity. This transformation introduces integrals over new variables which, as shown below, concentrate around the order parameters of the system in the thermodynamic limit. For instance, the term involving coupling among replicas of the soft spin system $\bx^\alpha $ is linearized by introducing an $n\times n$ matrix $\bQ$, thus becoming
\begin{align}\label{eq:Q_alpha_beta_HS}
    \begin{split}
    &\exp\left[ \frac{N\beta^2}{4}\sum_{\alpha,\beta}\left(\frac{\bx^\alpha \cdot\bx^\beta}{N}\right)^2 \right] = \int \prod_{\alpha\leq \beta}\frac{\dd Q_{\alpha\beta}}{\sqrt{2\pi/(N\beta^2)}} \\&\times \exp\left[-\frac{N\beta^2}{4}\sum_{\alpha,\beta}Q^2_{\alpha\beta} + \frac{\beta^2}{2}\sum_{\alpha,\beta}  Q_{\alpha\beta}\left(\bx^\alpha \cdot \bx^\beta\right)   \right]\,.
    \end{split}
\end{align}
Similar manipulations can be done for the other terms. This leads to the introduction of an $n\times n \times l$ tensor $\bT$, a $n\times l$ matrix $\bM$, an $n\times n\times l\times l$ tensor $\bP$, and an $n$-dimensional vector $\bmm$. We denote this set of new variables by $\bGamma = \{\bQ, \bmm, \bT, \bP, \bM\}$. From Eq.~\eqref{eq:Q_alpha_beta_HS}, we see that the Hubbard-Stratonovich transformation decouples spins and fields belonging to different sites within a single replica, at the expense of introducing couplings between spins at the same site of distinct replicas. The integral $\int_{\bx^\alpha ,\,\bphi^{\alpha\rho}}$ thus factorizes into a product of $N$ integrals over the $n\times l$ dimensional space of single-site, replicated soft spins $x^\alpha $ and fields $\phi^{\alpha\rho}$. Equation~\eqref{eq:integral_after_average_disorder} thus becomes
\begin{widetext}
    \begin{align}
        \begin{split}
                \int \dd\bGamma\,& \exp\Biggl[-\frac{N\beta^2}{4}\sum_{\alpha,\beta}Q_{\alpha\beta}^2 -\frac{N\beta J_0}{2}\sum_\alpha m_\alpha^2 - \frac{iN\beta}{2}\sum_{\alpha,\beta,\rho}T_{\alpha\beta\rho}^2 - \frac{N}{4}\sum_{\alpha,\beta,\rho,\sigma}P_{\alpha\beta\rho\sigma}^2 - \frac{iNJ_0}{2}\sum_{\alpha,\rho}M_{\alpha\rho}^2\Biggr] \\
                &\times \Biggl\{\int\prod_{\alpha=1}^n \prod_{\rho=1}^l\frac{\dd x^\alpha \dd \phi^{\alpha\rho}}{\sqrt{2\pi i}}\exp\Biggl[-\beta\sum_{i=1}^{N}E_I(x_i^\alpha) + \frac{\beta^2}{2}\sum_{\alpha,\beta}Q_{\alpha\beta}x^\alpha  x^\beta  + \beta J_0 \sum_{\alpha}m_\alpha x^\alpha  + i\beta \sum_{\alpha,\beta,\rho} T_{\alpha\beta\rho}x^\alpha \phi^{\beta\rho}\\
                &-\frac{i}{2}\sum_{\alpha,\beta,\rho,\sigma}\phi^{\alpha\rho}\phi^{\beta\sigma}P_{\alpha\beta\rho\sigma} + iJ_0 \sum_{\alpha,\rho}M_{\alpha\rho}\phi^{\alpha\rho} + \frac{i}{2}\sum_{\alpha,\rho} (\phi^{\alpha\rho})^2(z - \p^2_x E_I(x^\alpha, a ))\Biggr]\Biggr\}^N\\
                &\equiv \int \dd\bGamma\, \ee^{NS^\text{\tiny{SK}}[\bGamma]}\,.
            \end{split}
    \end{align}
We neglect the normalization factors in the differential $\dd\bGamma\equiv\prod_{\alpha\leq \beta}^n\prod_{\rho\leq \sigma}^l \dd Q_{\alpha\beta}\dd m_\alpha \dd T_{\alpha\beta\rho}\dd P_{\alpha\beta\rho\sigma}\dd M_{\alpha\rho}$, which contribute only with subleading order in $N$ to the exponential. The last line defines the action $S[\bGamma,z]$ as
\begin{align}\label{eq:Sz}
    \begin{split}
        S^\text{\tiny{SK}}[\bGamma,z] &\equiv -\frac{\beta^2}{4}\sum_{\alpha,\beta}Q_{\alpha\beta}^2 -\frac{\beta J_0}{2}\sum_\alpha m_\alpha^2 - \frac{i\beta}{2}\sum_{\alpha,\beta,\rho}T_{\alpha\beta\rho}^2 - \frac{1}{4}\sum_{\alpha,\beta,\rho,\sigma}P_{\alpha\beta\rho\sigma}^2 - \frac{iJ_0}{2}\sum_{\alpha,\rho}M_{\alpha\rho}^2 + \log Z_{x,\,\phi}^\text{\tiny{SK}}\\
        Z_{x,\,\phi}^\text{\tiny{SK}} &= \int\prod_\alpha \prod_\rho\frac{\dd x^\alpha \dd\phi^{\alpha\rho}}{\sqrt{2\pi i}}\exp\Biggl[-\beta \sum_{\alpha=1}^n E_I(x^\alpha,a) + \frac{\beta^2}{2}\sum_{\alpha,\beta}Q_{\alpha\beta}x^\alpha  x^\beta  + \beta J_0 \sum_{\alpha}m_\alpha x^\alpha  + i\beta \sum_{\alpha,\beta,\rho} T_{\alpha\beta\rho}x^\alpha \phi^{\beta\rho}\\
        &-\frac{i}{2}\sum_{\alpha,\beta,\rho,\sigma}\phi^{\alpha\rho}\phi^{\beta\sigma}P_{\alpha\beta\rho\sigma} + iJ_0 \sum_{\alpha,\rho}M_{\alpha\rho}\phi^{\alpha\rho} + \frac{i}{2}\sum_{\alpha,\rho} (\phi^{\alpha\rho})^2(z - \p^2_x E_I(x^\alpha,a ))\Biggr]\,\\
        &\equiv \int \prod_\alpha\prod_\rho \frac{\dd x^\alpha  \dd\phi^{\alpha\rho}}{\sqrt{2\pi i}}\,\ee^{-\beta E^\text{\tiny{SK}}_{x,\phi}[x^\alpha ,\phi^{\alpha\rho}]}
    \end{split}
\end{align}
\end{widetext}
$Z_{x,\,\phi}^\text{\tiny{SK}}$ is a mean field partition function of the system, computed in the space of the single-site replicated soft spins $x^\alpha $ and fields $\phi^{\alpha\rho}$. 

\section{Saddle point equations}\label{app:saddle_point_G}
In this section we evaluate the saddle point equations $\p_{\bGamma} S^\text{\tiny{SK}}[\bGamma,\,z]=\mathbf{0}$ for the action $S^\text{\tiny{SK}}[\bGamma,\,z]$ given by Eq.~\eqref{eq:Sz}. Taking the derivative with respect to the different order parameters we obtain
\begin{align}\label{eq:saddle_point_phix}
    \begin{split}
        Q_{\alpha\beta} &= \langle x^\alpha  x^\beta \rangle_{x,\,\phi}^\text{\tiny{SK}} \qquad m_\alpha = \langle x^\alpha  \rangle_{x,\,\phi}^\text{\tiny{SK}}\\
        T_{\alpha\beta\rho} &= \langle x^\alpha \phi^{\beta\rho}\rangle_{x,\,\phi}^\text{\tiny{SK}} \qquad M_{\alpha\rho} = \langle\phi^{\alpha\rho}\rangle_{x,\,\phi}^\text{\tiny{SK}}\\
        P_{\alpha\beta\rho\sigma} &= -i\langle \phi^{\alpha\rho}\phi^{\beta\sigma}\rangle_{x,\,\phi}^\text{\tiny{SK}}\,, 
    \end{split}
\end{align}
where the average $\langle\ldots\rangle_{x,\phi}$ is performed over the replicated soft spin variables and bosonic fields, and is defined as
\begin{equation}    \langle\ldots\rangle^\text{\tiny{SK}}_{x,\phi} \equiv \frac{1}{Z^\text{\tiny{SK}}_{x,\phi}}\int_{\bx^\alpha,\,\bphi^{\alpha\rho}}\,\ldots\ee^{-\beta E^\text{\tiny{SK}}_{x,\phi}[\{x^\alpha ,\phi^{\alpha\rho}\}]}\,.
\end{equation}
The saddle point equations are evaluated using the following ansatz for the order parameter encoding the correlations of the bosonic fields, 
\begin{equation}\label{eq:ansatz_P}
    P_{\alpha\beta\rho\sigma} = \Pi(z)\delta_{\alpha\beta}\delta_{\rho\sigma} \quad M_{\alpha\rho} = 0 \quad T_{\alpha\beta\rho} = 0\,.
\end{equation}
We have thus assumed that the replicated bosonic fields decouple from the replicated soft spins, and that there are no correlations among bosonic fields in different replicas.  Using this structure, the saddle point equations for the overlap matrix $Q_{\alpha\beta}$ and the magnetization $m_{\alpha}$ become
\begin{align}
    Q_{\alpha\beta} = \langle x^\alpha  x^\beta \rangle_\text{\tiny{MF}}^\text{\tiny{SK}} \quad m_\alpha = \langle x^\alpha  \rangle_\text{\tiny{MF}}^\text{\tiny{SK}}\,
\end{align}
while the saddle point equation for $\Pi(z)$ reads
\begin{equation}\label{eq:Pi_z}
    \Pi(z) = \left\langle\frac{1}{z - E_I''(x^\alpha, a ) - \Pi(z)}\right\rangle_\text{\tiny{MF}}^\text{\tiny{SK}}\,.
\end{equation}
The mean field average $\langle\ldots\rangle_\text{\tiny{MF}}^\text{\tiny{SK}}$ is an average over the replicated soft spin variables:
\begin{equation}
    \langle\ldots\rangle_\text{\tiny{MF}}^\text{\tiny{SK}} \equiv \frac{1}{Z_\text{\tiny{MF}}^\text{\tiny{SK}}}\int  \prod_\alpha\dd x^\alpha  \ee^{-\beta E_\text{\tiny{MF}}^\text{\tiny{SK}}[\{x^\alpha \}]}\,,
\end{equation}
and the mean field energy is defined as
\begin{align}\label{eq:E_MF_app}
    \begin{split}
        E_\text{\tiny{MF}}^\text{\tiny{SK}} &= \sum_\alpha E_I(x^\alpha ) - \frac{\beta}{2}\sum_{\alpha,\beta}Q_{\alpha\beta}x^\alpha x^\beta \\
        &- J_0\sum_\alpha m_\alpha x^\alpha  \\
        Z_\text{\tiny{MF}}^\text{\tiny{SK}} &= \int \prod_{\alpha=1}^n \dd x^\alpha \, \ee^{-\beta E_\text{\tiny{MF}}^\text{\tiny{SK}}[\{x^\alpha \}]}\,.
    \end{split}
\end{align}
We now have all the elements to compute the resolvent in Eq.~\eqref{eq:Gz_after_average}. Taking the derivative with respect to $z$ of the action we obtain
\begin{equation}
    \begin{split}
        G^\text{\tiny{SK}}(z) &=- \lim_{n\to0}\lim_{l\to0}\frac{i}{n}\p_l \sum_{\alpha,\beta,\rho,\sigma}\langle \phi^{\alpha\rho}\phi^{\beta\sigma}\rangle_{x,\phi} \\
        &= \lim_{n\to 0}\lim_{l \to 0}\frac{1}{n}\p_l \sum_{\alpha,\beta,\rho,\sigma}P_{\alpha\beta\rho\sigma}\\
        &= \lim_{n\to 0}\lim_{l\to 0}\frac{1}{n}\p_l n l\Pi(z)\\
        &= \lim_{n\to 0} \Pi(z)\,.
    \end{split}
\end{equation}
In the second line, we used the saddle point equation in Eq.~\eqref{eq:saddle_point_phix}, and in the third line we used our ansatz for the order parameter $P_{\alpha\beta\rho\sigma}$ given by Eq.~\eqref{eq:ansatz_P}. The resolvent can be thus identified with the quantity $\Pi(z)$. This implies that the resolvent satisfies a self-consistent equation
\begin{equation}\label{eq:Gz_self_consistent}
        G^\text{\tiny{SK}}(z) = \lim_{n\to 0}\left\langle \frac{1}{z-\p^2_x E_I(x^\alpha ,a)-G^\text{\tiny{SK}}(z)} \right\rangle_\text{\tiny{MF}}^\text{\tiny{SK}}\,,
\end{equation}
which can be rewritten, using the Dirac delta distribution, as
\begin{align}\label{eq:app_GSK}
    \begin{split}
        G^\text{\tiny{SK}}(z) &= \lim_{n\to0}\frac{1}{n}\sum_{\alpha=1}^n\int \dd x\,\frac{\langle \delta(x - x^\alpha )\rangle_\text{\tiny{MF}}^\text{\tiny{SK}}}{z - \p^2_x E_I(x) - G(z)}\\
        &= \int \dd x\,\frac{P^\text{\tiny{SK}}(x)}{z - \p^2_x E_I(x,a) - G^\text{\tiny{SK}}(z)}\,.
    \end{split}
\end{align}
The second line defines the single spin probability distribution $P^\text{\tiny{SK}}(x)$ as
\begin{equation}\label{eq:app_P_SK}
    P^\text{\tiny{SK}}(x) = \lim_{n\to 0} \frac 1 n \sum_{\alpha=1}^n \langle \delta(x - x^\alpha )\rangle_\text{\tiny{MF}}^\text{\tiny{SK}}\,.
\end{equation}
Equation~\eqref{eq:app_GSK} corresponds to Eq.~\eqref{eq:GSK} in the main text. \\
\subsection{Stability of the ansatz of Eq.~\eqref{eq:ansatz_P}\label{app:stability_diag_ansatz}}

The ansatz chosen in Eq.~\eqref{eq:ansatz_P} is valid as long as the action $S^\text{\tiny{SK}}[\bGamma,z]$ has a maximum at $\bGamma=\bGamma^*$, with $\bGamma^*$ the set of order parameters satisfying the saddle-point equations, Eq.~\eqref{eq:ansatz_P} and Eq.~\eqref{eq:Pi_z}. To check this condition, we expand the action around its saddle-point value:
\begin{align}\label{eq:expansion_S}
    \begin{split}
        S^\text{\tiny{SK}}[\bGamma^* &+\delta \bGamma,\,z] \approx S^\text{\tiny{SK}}[\bGamma^*,\,z] \\
        &+ \frac{1}{2}\sum_{\nu,\nu'} \frac{\p^2 S^\text{\tiny{SK}}[\bGamma,\,z]}{\p \Gamma_\nu \p \Gamma_{\nu'}}\Biggr\rvert_{\bGamma=\bGamma^*}\delta\Gamma_\nu\delta\Gamma_{\nu'}\,.
    \end{split}
\end{align}
The set of parameters $\bGamma^*$ maximizes the action as long as the eigenvalues of the Hessian matrix $\frac{\p^2 S^\text{\tiny{SK}}[\bGamma,\,z]}{\p \Gamma_\nu \p \Gamma_{\nu'}}\Biggr\rvert_{\bGamma=\bGamma^*}$ are all negative. Since within our ansatz in Eq.~\eqref{eq:ansatz_P} the bosonic fields and the soft spin variables are decoupled, we restrict our study on the subspace spanned the order parameters $P_{\alpha\beta\rho\sigma},\,T_{\alpha\beta\rho},\,M_{\alpha\rho}$. Within our ansatz, the only non zero entries of the Hessian matrix in Eq.~\eqref{eq:expansion_S} are
\begin{widetext}
\begin{align}\label{eq:PPS_PTM}
    \begin{split}
        \frac{\p^2 S^{\text{\tiny{SK}}}}{\p P_{\alpha\beta\rho\sigma} \p P_{\alpha'\beta'\rho'\sigma'}} &= -\frac{1}{2}\left[ \delta_{\alpha\alpha'}\delta_{\beta\beta'}\delta_{\rho\rho'}\delta_{\sigma\sigma'}+\delta_{\alpha\beta'}\delta_{\beta\alpha'}\delta_{\rho\sigma'}\delta_{\sigma\rho'}\right] \\&- \frac{1}{2}\left[\langle \phi^{\alpha\rho}\phi^{\beta\sigma}\phi^{\alpha'\rho'}\phi^{\beta'\sigma'}\rangle_{\text{\tiny{MF}}}^\text{\tiny{SK}} - \langle \phi^{\alpha\rho}\phi^{\beta\sigma}\rangle_{\text{\tiny{MF}}}^\text{\tiny{SK}}\langle\phi^{\alpha'\rho'}\phi^{\beta'\sigma'}\rangle_{\text{\tiny{MF}}}^\text{\tiny{SK}}\right]\\
        \frac{\p^2 S^\text{\tiny{SK}}}{\p T_{\alpha\beta\rho}\p T_{\alpha'\beta'\rho'}} &= i\delta_{\alpha\alpha'}\delta_{\beta\beta'}\delta_{\rho\rho'} - \beta^2 \langle x^\alpha x^{\alpha'} \phi^{\beta\rho}\phi^{\beta\rho'}\rangle_{\text{\tiny{MF}}}^\text{\tiny{SK}} = i\delta_{\alpha\alpha'}\delta_{\beta\beta'}\delta_{\rho\rho'} - \beta^2 Q_{\alpha\alpha'}\langle \phi^{\beta\rho}\phi^{\beta'\rho'}\rangle_\text{\tiny{MF}}^\text{\tiny{SK}} \\
        \frac{\p^2 S^\text{\tiny{SK}}}{\p M_{\alpha\rho}\p M_{\alpha'\rho'}} &= i\delta_{\alpha\alpha'}\delta_{\beta\beta'}\delta_{\rho\rho'} - J_0 \langle \phi^{\alpha\rho}\phi^{\alpha'\rho'} \rangle_{\text{\tiny{MF}}}^\text{\tiny{SK}}\,.
    \end{split}
\end{align}
\end{widetext}
The real parts of the terms involving partial derivatives with respect to $T_{\alpha\beta\rho}$ or $M_{\alpha\rho}$ are negative, and fluctuations within these subspace are thus irrelevant. We focus therefore on the block of the Hessian involving fluctuations of the tensor $P_{\alpha\beta\rho\sigma}$. The only nonzero entries in this block are
\begin{equation}\label{eq:nonzero_entries_S_PP}
    \begin{split}
        &\frac{\p^2 S^\text{\tiny{SK}}}{\p P_{\alpha\beta\rho\sigma}\p P_{\alpha\beta\rho\sigma}} = -\frac{1}{2}\left[1+\langle \left(\phi^{\alpha\rho}\right)\rangle^\text{\tiny{SK}}_\text{\tiny{MF}}\right]^2\\
        &\frac{\p^2 S^\text{\tiny{SK}}}{\p P_{\alpha\alpha\rho\rho}\p P_{\alpha\alpha\rho\rho}} =-1 \\
        &+ \int \dd x\,\frac{P^\text{\tiny{SK}}}{\left[z - \p^2_x E_I(x,a) - G^\text{\tiny{SK}}(z)\right]^2}=0\,,
    \end{split}
\end{equation}
where the $\alpha\neq\beta$ and $\rho\neq\sigma$. The term on the first line of Eq.~\eqref{eq:nonzero_entries_S_PP} has a negative definite real part, while the term on the second line is zero, as it can be verified by taking the derivative of the saddle point equation for the resolvent in Eq.~\eqref{eq:Gz_self_consistent} with respect to $G^\text{\tiny{SK}}$. Our ansatz is therefore marginally stable against longitudinal fluctuations in the space of the tensor $P_{\alpha\alpha\rho\rho}$. Motivated by past work in the analysis of the ground states of disordered systems~\cite{franz2015universal}, we assume that no instability develops when considering perturbations of higher order.

\section{Replica symmetric ansatz}\label{app:saddle_point_rs}

In this Section we show how to obtain the saddle point equations in the replica symmetric case, given by
\begin{align}\label{eq:app_rs_ansatz_sk}
    \begin{split}
        Q_{\alpha\beta} &= q_d \delta_{\alpha\beta} + q_o (1-\delta_{\alpha\beta})\\
        m_\alpha &= m\,.
    \end{split}
\end{align} 
Plugging Eq.~\eqref{eq:app_rs_ansatz_sk} into the mean-field partition function given by Eq.~\eqref{eq:E_MF_app}, we obtain
\begin{align}
    \begin{split}
        Z_\text{\tiny{MF}}^\text{\tiny{SK}} &= \int \prod_{\alpha=1}^n \dd x^\alpha \,\exp\Biggl[ -\beta\sum_\alpha E_I(x^\alpha,a ) \\
        &- \frac{\beta^2}{2}\Delta q \sum_{\alpha} (x^\alpha )^2 + \frac{\beta^2}{2}q_o\left(\sum_{\alpha} x^\alpha \right)^2 \\
        &+J_0m\sum_\alpha x^\alpha \Biggr]
    \end{split}
\end{align}
We now use the rescaled overlap gap $\Delta \widetilde q\equiv \beta(q_d -q_o)$ and perform a Hubbard-Stratonovich transformation, obtaining
\begin{align}\label{eq:app_Z_RS_SK}
    \begin{split}
        Z_\text{\tiny{MF}}^\text{\tiny{SK}} &= \int \prod_{\alpha=1}^n \dd x^\alpha \, \overline{\exp\Biggl[ -\beta\sum_\alpha E_I(x^\alpha,a)} \\
        &\overline{- \frac{\beta}{2}\Delta \widetilde q \sum_{\alpha} (x^\alpha )^2 - \beta (\sqrt{q_o}h - J_0 m)x^\alpha \Biggr]}\\
        &\equiv \overline{\left(\int \dd x\, \ee^{-\beta E_\text{\tiny{eff}}^\text{\tiny{SK}}(x,h)}\right)^n}\,\\
        &\equiv \overline{ \left(Z_\text{\tiny{RS}}^\text{\tiny{SK}}(h)\right)^n}\,,
    \end{split}
\end{align} 
where $\overline{\ldots}\equiv \int \frac{\dd h}{\sqrt{2\pi}}\ldots\ee^{-h^2/2}$ is an average over the realization of a random Gaussian field $h$, which acts as a quenched disorder in the system. In the second line, we have observed that the partition function factorizes among different replicas and we defined the replica symmetric energy function
\begin{equation}\label{eq:app_E_RS_SK}
    E_\text{\tiny{eff}}^\text{\tiny{SK}}(x,h) \equiv E_I(x,a+\Delta \widetilde q)  - (\sqrt{q_o}h + J_0 m)x\,.
\end{equation}
The last line of Eq.~\eqref{eq:app_Z_RS_SK} defines the replica symmetric partition function $Z_\text{\tiny{RS}}^\text{\tiny{SK}}(h)$ for a given realization of the Gaussian field $h$. We compute in a similar fashion the single spin distribution in Eq.~\eqref{eq:app_P_SK} within the replica symmetric phase,
\begin{align}
    \begin{split}
        P^\text{\tiny{SK}}(x) &= \lim_{n\to 0}\frac 1 n \sum_{\alpha=1}^n \int \prod_{\beta=1}^n \dd x^\beta\, \delta(x-x^\alpha)\\
        &\times \ee^{-\beta E^{\text{\tiny{SK}}}_\text{\tiny{RS}}[\{x^\beta\}]} \left[Z^\text{\tiny{SK}}_\text{\tiny{MF}}\right]^{-1}\\
        &= \lim_{n\to 0} \overline{\ee^{-\beta E{\text{\tiny{RS}}}^\text{\tiny{SK}}(x,z)} \left(Z^\text{\tiny{SK}}_\text{\tiny{RS}}(h)\right)^{n-1}} \\
        &= \overline{\ee^{-\beta E^\text{\tiny{SK}}_\text{\tiny{RS}}(x,z)}\left[Z^\text{\tiny{SK}}_\text{\tiny{RS}}(h)\right]^{-1}}\\
        &\equiv P^\text{\tiny{SK}}_\text{\tiny{RS}}(x)\,,
    \end{split}
\end{align}    
which is Eq.~\eqref{eq:P_rs_sk} of the main text.

\section{Stability of the replica symmetric solution}\label{app:replicon}

The replica symmetric solution is stable as long as the order parameters that satisfy the saddle point equations in Eq.~\eqref{eq:sp_rs} are an extremum of the action $S^\text{\tiny{SK}}$. This requirement breaks down when the Hessian of the action, introduced in Eq.~\eqref{eq:expansion_S}, acquires a positive eigenvalue along some direction related to a perturbation of the overlap matrix $Q_{\alpha\beta}$ and to the magnetization $m_\alpha$. The Hessian of the action involves the following components,  
\begin{equation}\label{eq:app_Hess_S_SK}
    \begin{split}
        &\frac{\p^2 S^\text{\tiny{SK}}}{\p Q_{\alpha\beta}\p Q_{\gamma\epsilon}} = -\frac{\beta^2}{2}(\delta_{\alpha\beta}\delta_{\gamma\epsilon} + \delta_{\alpha\gamma}\delta_{\beta\epsilon}) \\
        &+ \frac{\beta^4}{2}\left(\overline{\langle x^\alpha  x^\beta  x^\gamma x^\epsilon \rangle_\text{\tiny{RS}}^\text{\tiny{SK}}} - \overline{\langle x^\alpha  x^\beta  \rangle_\text{\tiny{RS}}^\text{\tiny{SK}}}\,\overline{\langle x^\gamma x^\epsilon \rangle_\text{\tiny{RS}}^\text{\tiny{SK}}}\right) \\ ˘
        &\frac{\p^2 S^\text{\tiny{SK}}}{\p Q_{\alpha\beta}\p m_{\gamma}} = \beta^3 J_0\left[ \overline{\langle x^\alpha  x^\beta  x^\gamma \rangle_\text{\tiny{RS}}^\text{\tiny{SK}}} - \overline{\langle x^\alpha  x^\beta  \rangle_\text{\tiny{RS}}^\text{\tiny{SK}}}\,\overline{\langle x^\gamma\rangle_\text{\tiny{RS}}^\text{\tiny{SK}}}\right]\\
        &\frac{\p^2 S^\text{\tiny{SK}}}{\p m_{\alpha}\p m_\beta} = -\beta J_0 \delta_{\alpha\beta} \\
        &+ (\beta J_0)^2 \left[\overline{\langle x^\alpha  x^\beta \rangle_\text{\tiny{RS}}^\text{\tiny{SK}}} -\overline{\langle x^\alpha \rangle_\text{\tiny{RS}}^\text{\tiny{SK}}}\,\overline{\langle x^\beta \rangle_\text{\tiny{RS}}^\text{\tiny{SK}}}\right]\,.
    \end{split}
\end{equation}
The first eigenvalue that changes sign at the instability transition is known as the replicon, $\Lambda_\text{\tiny{R}}^\text{\tiny{SK}}$. Its expression depends on the generic structure of the overlap matrix $Q_{\alpha\beta}$, rather than the specific model at hand, and it is given by~\cite{sherrington1978stability, de2006random}
\begin{align}\label{eq:app_replicon}
    \begin{split}           
    \Lambda_\text{\tiny{R}}^\text{\tiny{SK}}&= 2\frac{\p^2 S^\text{\tiny{SK}}}{\p Q_{\alpha\beta}\p Q_{\alpha\beta}} -4\frac{\p^2 S^\text{\tiny{SK}}}{\p Q_{\alpha\beta}\p Q_{\beta\gamma}} + 2\frac{\p^2 S^\text{\tiny{SK}}}{\p Q_{\alpha\beta}\p Q_{\gamma\epsilon}}\\
    &= \beta^2\left[ -1 + \beta^2\overline{\left(\left\langle x^2\right\rangle_\text{\tiny{MF}}^\text{\tiny{SK}} - \langle x\rangle_\text{\tiny{MF}}^\text{\tiny{SK}}{}^2  \right)^2}\right]\,. 
    \end{split}
\end{align}
The replica-symmetric saddle point is stable as long as $\Lambda_\text{\tiny{R}}^\text{\tiny{SK}}<0$.  In the low temperature limit $\beta \to \infty$, it is convenient to consider a rescaled version of the replicon eigenvalue $\tilde\Lambda_\text{\tiny{R}}^\text{\tiny{SK}} \equiv \beta^{-2}\Lambda_\text{\tiny{R}}^\text{\tiny{SK}}$. A saddle point evaluation of the replica-symmetric average in Eq.~\eqref{eq:app_replicon} yields the following stability condition for the replica symmetric phase,
\begin{equation}
    \tilde\Lambda_\text{\tiny{R}}^\text{\tiny{SK}} = -1 + \overline{[\p^2\ERSSK(x^*(h))]^{-2}} <0\,,
\end{equation}
where $x^*(h)$ is the minimum of the replica symmetric mean field energy $\ERSSK$ in Eq.~\eqref{eq:app_E_RS_SK} for a fixed realization of the quenched field $h$. We thus recover Eq.~\eqref{eq:replicon} of the main text. 

\section{Derivation of Eq.~\eqref{eq:condition_i}}\label{app:details_BBP}

In this Appendix we compute the stability boundary of the convex paramagnetic phase for the CIM-SK. Starting from the replica-symmetric saddle point equations Eq.~\eqref{eq:sp_rs}, we consider the case where $m=q_o=0$. The only relevant order parameter is $\Delta \widetilde q$, which can be computed from the saddle point equations as 
\begin{equation}\label{eq:app_sp_Deltaq_sk_para}
    \Delta \widetilde q = -\frac{1}{a+\Delta \widetilde q}\,.
\end{equation}

The stability of this paramagnetic phase is determined by the condition $\tilde\Lambda_\text{\tiny{R}}^\text{\tiny{SK}}<0$ in Eq.~\eqref{eq:replicon}, which for $m=0,\,q_o=0$ yields
\begin{equation}\label{eq:app_stability_paramagentic_SK}
\Delta \widetilde q^2 < 1 \implies \Delta \widetilde q<1\,.
\end{equation}
We also need to impose the paramagnetic state to be stable against perturbation of the magnetization, i.e. that $\frac{\p^2 S^\text{\tiny{SK}}}{\p m_\alpha \p m_\beta}<0$. From Eq.~\eqref{eq:app_Hess_S_SK} we obtain
\begin{equation}\label{eq:app_stability_paramagnetic}
    \Delta \widetilde q < J_0^{-1}\,.
\end{equation}
Combining Eq.~\eqref{eq:app_stability_paramagentic_SK} and Eq.~\eqref{eq:app_stability_paramagnetic} we obtain
\begin{equation}
    \Delta \widetilde q <\min[1,\,J_0^{-1}]\,.
\end{equation}
This is Eq.~\eqref{eq:condition_i} of the main text.

\section{Outline of the derivation of the full replica symmetry breaking equations}\label{app:single-spin_frsb}

In this Appendix we outline how to derive the full replica symmetry breaking equation for the CIM-SK, presented in Sec.~\ref{sec:fRSB}. When the replica-symmetric solution becomes unstable, a new ansatz for the overlap matrix $\bQ$ is required. Different replicas are no longer equivalent, and their overlaps organize in a hierarchical structure~\cite{parisi1979infinite, parisi1980sequence, mezard1987spin, nishimori2001statistical}. All the rows of $\bQ$ are the same up to a permutation of their elements. The description of the first row of $\bQ$ is thus enough to characterize the overlap matrix. The first row is divided in $k+2$ bands of different lengths. We label each band by an index $i=0,\ldots,k+1$. The band with index $k+1$ corresponds to the first diagonal entry of the matrix $\bQ$. For $i\leq k+1$, the $i$th band contains $m_{i}-m_{i+1}$ elements, all equal to a given overlap value $q_i$. The set $\{m_i\}_{i=0}^{k+1}$ is ordered in the following way
\begin{equation}
    1=m_{k+1}< m_{k} < \ldots< m_{1} < m_{0} = n\,.
\end{equation}
The overlap matrix becomes
\begin{align}
    \begin{split}
        Q_{\alpha\beta} &= q_d \delta_{\alpha\beta} \\
        &+ (1-\delta_{\alpha\beta})\left[ q_0 +\sum_{i=1}^k (q_i - q_{i-1})I_{\alpha\beta}^{m_i}\right]\,,
    \end{split}
\end{align}
where $I_{\alpha\beta}^{m_i}$ is an $n\times n$ covered along its diagonal by $n/m_i$ block matrices of size $m_i\times m_i$, with all the entries equal to $1$. When the limit $n\to0$ is taken by means of an analytic continuation, the order of the size of the bands is inverted, so that 
\begin{equation}\label{eq:ordering_m_i_nto0}
    0 = m_0 < m_1 <\ldots<m_k<m_{k+1} = 1\,.
\end{equation}Using this structure, the mean field partition function $Z_\text{\tiny{MF}}^\text{\tiny{SK}}$ becomes~\cite{nishimori2001statistical}
\begin{equation}\label{eq:app_logZ_frsb_discrete}
    \lim_{n\to 0} \frac{1}{n}\log Z_\text{\tiny{MF}}^\text{\tiny{SK}} = \beta \int \mathcal{D}h_0\, f_1(h_0 \sqrt{q_0})\,,
\end{equation}
where $\mathcal{D}h_0 \equiv \frac{\dd h_0}{\sqrt{2\pi}}\,\ee^{-\frac{h_0^2}{2}}$ is the Gaussian measure. The function $f_1(h)$ is the first member of a hierarchy of $k+1$ 'free energies', which satisfy a recursion relation,
\begin{equation}\label{eq:app_recursion_f}
    f_i(h) = \frac{1}{\beta m_i}\log \int \mathcal{D}h_i\,\ee^{\beta m_if_{i+1}(h +\sqrt{q_i - q_{i-1}}z_i)}\,,
\end{equation}
together with the boundary condition
\begin{equation}
    f_{k+1}(h) = \frac{1}{\beta}\log \int \dd x\, \ee^{-\beta E^\text{\tiny{SK}}_\text{\tiny{RS}}(x,h)}\,,
\end{equation}
from which we recognize $f_{k+1}$ as the free energy of a single soft spin system with energy $\ERSSK$. The field $h$ acts as a form of quenched disorder, but its distribution is no longer Gaussian, because of the hierarchical structure of the states of the system. 

The saddle point equations for the overlaps $q_i,\,q_d$ and the magnetization $m$ can be computed as derivatives of the logarithm of the partition function in Eq.~\eqref{eq:app_logZ_frsb_discrete}. We obtain
\begin{align}\label{eq:frsb_sp_discrete}
    \begin{split}
        q_i &= \int \dd h P_i(h)\left[\p_h f_{i+1}(h)\right]^2\\
        \Delta \widetilde q &= \int \dd h\,P_{k}(h)[\p^2_h f_{k+1}(h)]\\
        m &= \int \dd h\,P_{k}(
        h)\p_h f_{k+1}(h)\,.
    \end{split}
\end{align}
The function $P_i(h)$ is the distribution of internal fields $h$ acting at the level $i$ of the hierarchy. It is defined as the integrated response of the function $f_1$ to a change of the free energy at the level $i+1$ in the hierarchy,
\begin{equation}
    P_i(h) \equiv \int \mathcal{D}h_0 \frac{\delta f_1(h_0\sqrt{q_0})}{\delta f_{i+1}(h)}\,,
\end{equation}
where $\frac{\delta f_1(h_0)}{\delta f_{i+1}(h)}$ is a functional derivative. The distribution of internal fields satisfies a recursive relation too, namely,
\begin{align}\label{eq:app_recursion_P}
    \begin{split}
    P_i(h) &= \int\mathcal{D}h_i P_{i-1}(h-\sqrt{q_i-q_{i-1}}h_i)\\
    &\times \ee^{\beta m_i\left(f_{i+1}(h) - f_{i}(h-\sqrt{q_i-q_{i-1}}h_i)\right)}\,,
    \end{split}
\end{align}
with the boundary condition
\begin{equation}\label{eq:app_frsb_P0_discrete}
    P_0(h) = \frac{1}{\sqrt{2\pi q_0}}\ee^{-\frac{h^2}{2 q_0}}\,. 
\end{equation}
The single-spin distribution $P^{\text{\tiny{SK}}}$ becomes instead
\begin{equation}
    P^\text{\tiny{SK}}(x) = \int \dd h P_k(h)\frac{\ee^{-\beta \ERSSK(x,h)}}{\ZRSSK(h)}\,.
\end{equation}
The full replica symmetry breaking is obtained when the number of levels $k$ in the hierarchy is sent to infinity. The bands in the hierarchical matrix $\bQ$ form then a continuum, and the overlaps are described by a continuous function $q(y)$, with $y \in [0,1]$, which for finite $k$ is defined as
\begin{equation}
    q(y) = q_i \qquad m_i\leq y<m_{i+1}\,.
\end{equation}
The same continuum limit is taken for the hierarchical free energies $f_i(h)$ and for the internal field distribution $P_i(h)$, which become continuous functions $f(y,h)$ and $P(y,h)$ respectively. The recursive relations given by Eq.~\eqref{eq:app_recursion_f} and Eq.~\eqref{eq:app_recursion_P} become a backward and a forward Fokker-Planck equations, respectively,
\begin{align}\label{eq:app_FP_parisi}
    \begin{split}
        \dot f(y,h) &= -\frac{\dot q(y)}{2}\left[ f'' +  \beta y(  f')^2\right]\\
        \dot P(y,h) &=  \frac{\dot q(y)}{2}\left[  P'' - 2 \beta y (P f')'\right]\,,
    \end{split}
\end{align}
where we used a shorthanded notation for the derivatives, $\dot f(y,h) = \p_y f(y,h)$ and $f'(y,h) = \p_h f(y,h)$. The boundary conditions of the partial differential equations in Eq.~\eqref{eq:app_FP_parisi} are
\begin{align}\label{eq:app_frsb_bc_continuum}
    \begin{split}
        f(y=1,h) &=  \frac{1}{\beta}\log \int\dd x\, \ee^{-\beta E_\text{\tiny{eff}}^\text{\tiny{SK}}(x,h)}\\
        P(y=0,h) &= \frac{1}{\sqrt{2\pi q(0)}}\ee^{-\frac{h^2}{2q(0)}}\,.
    \end{split}
\end{align}
The saddle point equations in the continuum limit are
\begin{align}\label{eq:app_sp_frsb}
    \begin{split}
        q(y) &= \int \dd h\, P(y,h)[f'(y,h)]^2\\
        \Delta \widetilde q &= \int \dd h\,P(1,h)f''(1,h)\\
        m &= \int \dd h\,P(1,h)f'(1,h)\,,
    \end{split}
\end{align}
and the single spin distribution is
\begin{equation}\label{eq:app_P_single_spin_frbs}
    P_\text{\tiny{fRSB}}^\text{\tiny{SK}}(x) = \int \dd h\, P(1,h)\frac{\ee^{-\beta \ERSSK(x,h)}}{\ZRSSK(h)}\,.
\end{equation}

\subsection{Marginal stability}\label{app:marginal_stability}
The replicon in the full replica symmetry breaking phase reads, from its expression in Eq.~\eqref{eq:app_replicon},
\begin{equation}\label{eq:app_replicon_frsb}  
\tilde\Lambda_\text{\tiny{R}}^\text{\tiny{SK}} = -1+\int_{-\infty}^{\infty}\dd h\, P(1,h)[f''(1,h)]^2\,.
\end{equation}
We now show that $\tilde\Lambda^\text{\tiny{SK}}_\text{\tiny{R}}=0$ in the spin-glass phase. Taking the derivative with respect to $y$ in the saddle point equation for $q(y)$ Eq.~\eqref{eq:app_sp_frsb}, yields
\begin{align}
    \begin{split}
        \dot q(y) &= \int \dd h \dot P(y,h)[f'(y,h)]^2\\
        &+2\int \dd h\, P(y,h)\dot f'(y,h)f'(y,h)\,.
    \end{split}
\end{align}
Substituting Eq.~\eqref{eq:app_FP_parisi} into the expression above and integrating by parts several times we obtain
\begin{equation}
    \dot q(y) = \dot q(y)\int \dd h\, P(y,h)[f''(y,h)]^2\,,
\end{equation}
which implies, for $y=1$, 
\begin{equation}
    -1 + \int \dd h\, P(1,h)[f''(1,h)]^2 = 0\,.
\end{equation}
From the expression of the replicon in Eq.~\eqref{eq:app_replicon_frsb} we thus conclude that
\begin{equation}\label{eq:app_marginal_stability_frsb}
    \tilde\Lambda^\text{\tiny{SK}}_\text{\tiny{R}}  = 0
\end{equation}
in the spin-glass phase. 

\subsection{Low temperature solution}\label{app:low_T}
In the low temperature limit $\beta \to\infty$ the boundary condition for the function $f(y,h)$ is evaluated through a saddle point method, yielding
\begin{equation}\label{eq:app_f1_h_lowT}
    f(y=1,h) = -\ERSSK(x^*(h),h)\,,
\end{equation}
where $x^*(h)$ is the minimizer of the energy $\ERSSK(x,h)$ for a given realization of the internal field $h$. This minimizer satisfies the equation
\begin{equation}\label{eq:app_cubic_frsb}
    x^*{}(h)^3 - (a+\Delta \widetilde q)x^*(h) - J_0 m= h\,.
\end{equation}
If $a+\Delta \widetilde q\leq 0$, then Eq.~\eqref{eq:app_cubic_frsb} has a unique solution. The higher order derivatives of $f(y=1,h)$ can be obtained from Eq.~\eqref{eq:app_f1_h_lowT} as
\begin{align}\label{eq:app_derivatives_f1_h_lowT_convex}
    \begin{split}
        f'(y=1,h) &= x^*(h)\\
        f''(y=1,h) &= \p_h x^*(h) = \frac{1}{3x^*(h)^2 - a -\Delta \widetilde q}\,.
    \end{split}
\end{align}
The saddle point equations Eq.~\eqref{eq:app_sp_frsb} then become
\begin{align}\label{eq:app_sp_frsb_lowT_convex}
    \begin{split}
        q(y) &=\int \dd h\, P(1,h)[x^*(h)]^2\\
        \Delta \widetilde q &= \int \dd h\, \frac{P(1,h)}{3 x^*(h)^2 - a- \Delta \widetilde q}\\
        m &= \int \dd h\, P(1,h)x^*(h)\,,
    \end{split}
\end{align}
while the marginal stability condition is
\begin{equation}\label{eq:app_replicon_frsb_lowT_convex}
    \tilde\Lambda_\text{\tiny{R}}^\text{\tiny{SK}} = -1 + \int \dd h\, \frac{P(1,h)}{(3x^*(h)^2 -a - \Delta \widetilde q)^2} =0\,.
\end{equation}
Finally, a saddle-point evaluation of the single spin probability distribution in Eq.~\eqref{eq:app_P_single_spin_frbs} yields
\begin{align}\label{eq:app_P_as_Pfrsb_lowT}
    \begin{split}
        P_\text{\tiny{fRSB}}^\text{\tiny{SK}}(x) &\propto \lvert 3 x^2 -a-\Delta \widetilde q \rvert \\&\times P(1, h=x^3 -(a+\Delta \widetilde q)x + J_0 m )\,.
    \end{split}
\end{align}
For $a+\Delta \widetilde q>0$, the effective energy $\ERSSK$ is nonconvex. Equation~\eqref{eq:app_f1_h_lowT} is still valid, provided that $x^*(h)$ is the global minimum of $\ERSSK(x,h)$. However, the global minimum $x^*(h) $exhibits a discontinuous jump at $h=-J_0 m$. The integral in marginality condition given by Eq.~\eqref{eq:app_marginal_stability_frsb} must be carefully handled around the discontinuity as the zero-temperature limit is taken. As discussed in~\cite{folena2022marginal} for a related model of mean -field disordered systems, let us consider two distinct regimes, depending on the strength of the quenched disorder $h$ with respect to the temperature $T$. A first regime is obtained when $h+J_0m\gg T$, far away from the discontinuity, while a second one is given by $h+J_0m \sim T$.The first derivative of the free energy in Eq.~\eqref{eq:app_f1_h_lowT} is split into the two regimes as
\begin{widetext}
\begin{equation}\label{eq:app_fp1_frsb_lowT_nonconvex}
    f'(y=1,h) = \begin{cases} x^*(h) \quad \text{if } |h+J_0m|\gg T \\ 
    x^*(h) \tanh(\beta(h+J_0m)x^*(h)) \quad \text{if } |h+J_0 m|\sim T\,,\end{cases}
\end{equation}
The integral in the marginal stability condition in Eq.~\eqref{eq:app_marginal_stability_frsb} is then evaluated in the limit $T\to 0$ by separately considering the two regimes. The computation reads
\begin{align}\label{eq:long_calculation_replicon_term}
    \begin{split}
        &\int_{-\infty}^{\infty}\dd h\, P(1,h) [f''(1,h)]^2 = \int_{-\infty}^{\infty}\dd h\, P(1,h-J_0m) [f''(1,h-J_0m)]^2\\&= \int_{h\in(-\infty,\,-T)\cup (T,\infty)}\dd h\, P(1,h-J_0m)[f''(1,h-J_0m)]^2 + \int_{-T}^T\dd h\, P(1,h-J_0m)[f''(1,h-J_0 m)]^2\\
        &= \int_{-\infty}^{\infty} \dd h\, \frac{P(1,h)}{\p^2\ERSSK(x^*(h))^2 } 
        + \int_{-T}^T \dd h\, \beta^2 P(1,h-J_0 m)[x^*(h-J_0m)]^4(1-\tanh(\beta h)^2)^2\\
        &+\int_{-T}^T \dd h\, P(1,h-J_0 m)\frac{[x^*(h-J_0m)]^2(1-\tanh(\beta h)^2)^2(\beta h)^2}{\p^2 \ERSSK(x^*(h-J_0 m))}\\
        &= \int_{-\infty}^{\infty} \dd h\, \frac{P(1,h)}{\p^2\ERSSK(x^*(h))^2 } + \int_{-1}^1 \dd \tilde h\, \beta P(1,T\tilde h-J_0 m)[x^*(T\tilde h-J_0m)]^4(1-\tanh(\tilde h)^2)^2 \\
        &+\int_{-1}^1 \dd \tilde h\, T P(1,T\tilde h)\frac{[x^*(T\tilde h-J_0m)]^2(1-\tanh(\tilde h)^2)^2\tilde h^2}{\p^2 \ERSSK(x^*(T\tilde h-J_0 m))}\\
        &= \int_{-\infty}^{\infty} \dd h\, \frac{P(1,h)}{\p^2\ERSSK(x^*(h))^2 } + \int_{-1}^1 \dd \tilde h\,\tilde P(1,\tilde h)[x^*(J_0m)]^4(1-\tanh(\tilde h)^2)^2\\
        &+T^2\int_{-1}^1 \dd \tilde h\, \tilde P(1,\tilde h)\frac{[x^*(J_0m)]^2(1-\tanh(\tilde h)^2)^2\tilde h^2}{\p^2 \ERSSK(x^*(J_0 m))}\\
        &=\int_{-\infty}^{\infty} \dd h\, \frac{P(1,h)}{(3x^*(h)^2 - a -\Delta \widetilde q)^2} + \int_{-1}^1 \dd \tilde h\,\tilde P(1,\tilde h)(a+\Delta \widetilde q)^4(1-\tanh(\tilde h)^2)^2
    \end{split}
\end{align}
In the first equality, we shifted the integration variable by $J_0m$. In the second equality, we split the integration domain into two different regions, isolating a region around the origin of width $\sim T$. In the third equality, we explicitly computed the quantity $f''(y=1,h)$. In the fourth equality, a rescaled field was introduced $T \tilde h \equiv h$. In the fifth equality, we introduced a rescaled and shifted distribution of fields $T\tilde P(1, \tilde h) = P(1, Th + J_0 m)$. Finally, in the last equality, we substituted for the explicit expression of $x^*(h)$ and its derivatives. The marginal stability condition in Eq.~\eqref{eq:app_marginal_stability_frsb} thus reads
\begin{align}
    \begin{split}
    \tilde\Lambda_\text{\tiny{R}}^\text{\tiny{SK}} &= -1 + \int_{-\infty}^{\infty} \dd h\, \frac{P(1,h)}{(3x^*(h)^2 - a -\Delta \widetilde q)^2} + \int_{-1}^1 \dd \tilde h\,\tilde P(1,\tilde h)(a+\Delta \widetilde q)^4(1-\tanh(\tilde h)^2)^2 = 0\,.
    \end{split}
\end{align}
\end{widetext}
Notice that the third contribution to the replicon is positive. In order for the marginal stability condition of the full-replica symmetry breaking solution with respect to perturbations in the space of overlap matrices to be satisfied, the internal field distribution must decay to zero around $J_0m$ at least as fast as linearly, i.e. $P(1,h-J_0m) \sim |h|$ for $h\ll 1$. Moreover,  when $a+\Delta \widetilde q>0$ the marginality condition implies 
\begin{equation}
    -1 + \int \dd h\,\frac{P(1,h)}{(3x^*(h)^2 - a -\Delta \widetilde q)^2} < 0\,.
\end{equation}
This inequality can equivalently be expressed through a change of variable from the internal field distribution to the soft spin distribution as
\begin{equation}\label{eq:app_inequality_frsb}
    -1 +\int \dd x\, \frac{P_\text{\tiny{fRSB}}^\text{\tiny{SK}}(x)}{(3x^2 -a -\Delta \widetilde q)^2} < 0\,,
\end{equation}
which is Eq.~\eqref{eq:marginal_stability_nonconvex} in the main text.
\subsection{Numerical solution of full replica symmetry breaking equations}\label{app:numerics_frsb}

To solve the full replica-symmetry breaking equations at zero temperature, we adopt a method similar to the one developed in~\cite{schmidt2008method}. We discretize the continuous solution into $k$ steps. Since we are interested in the low temperature limit of the full replica-symmetry breaking equations, we adopt a rescaled size of the blocks $\widetilde m_i \equiv \beta m_i$, for $i=0,\ldots,k+1$. The rescaled variables $\widetilde m_i$ are ordered in the following way, which can be read from Eq.~\eqref{eq:ordering_m_i_nto0},
\begin{equation}
    0<\widetilde m_0<\widetilde m_1<\ldots<\widetilde m_k <\widetilde m_{k+1} = \beta \to \infty\,.
\end{equation}
Starting from an initial guess of the discretized Parisi function $q_i$, we iteratively solve the recursion relation for the functions $\tilde f_i(h) \equiv  \widetilde m_i f_i(h)$,
\begin{equation}\label{eq:app_frsb_parisi_numerics_recursion_fi}
    \tilde f_i(h) = \log \int \mathcal{D}w_i\,\ee^{\frac{\widetilde m_i}{\widetilde m_{i+1}} \tilde f_{i+1}(h + w_i)}\,,
\end{equation}
where $\mathcal{D}w_i\equiv \frac{\dd w_i}{\sqrt{2\pi(q_{i}-q_{i-1})}}\ee^{-\frac{w_i^2}{2(q_i-q_{i-1})}}$ is a Gaussian measure of variance $q_i - q_{i-1}$. The boundary condition of the recursion relation is
\begin{equation}
    \tilde{f}_{k+1}(h) = -\widetilde m_{k+1} \min_x \ERSSK(x,h)\,.
\end{equation}
Note that the use of the rescaled variable $\tilde f_i$ requires the computation of the ratio $\widetilde m_i/\widetilde m_{i+1}$, which is typically more well-behaved than the quantities $\widetilde m_i$. We also compute the derivatives $\tilde f_i'(h)$, using a recursion relation obtained from Eq.~\eqref{eq:app_frsb_parisi_numerics_recursion_fi}, 
\begin{align}
    \begin{split}
        \tilde{f}_i'(h) &= \frac{\widetilde m_i}{\widetilde m_{i+1}}\int \mathcal{D}w_i\, \tilde f'_i(h +w_i)\\&\times\ee^{\frac{\tilde{m_i}}{\widetilde m_{i+1}}\tilde f_{i+1}(h+w_i)-\tilde f_{i}(h)}\,,
    \end{split}
\end{align}
together with the boundary condition
\begin{equation}
    \tilde f'_{k+1}(h) = \widetilde m_{k+1} x^*(h)\,,
\end{equation}
with $x^*(h) \equiv \argmin_{x} \ERSSK(x,h)$. We then compute the probability distributions of the internal fields $P_i(h)$ using the recursion relations
\begin{align}
    \begin{split}
        P_i(h) &= \int \mathcal{D}w_i\, P_{i-1}(h-w_i)\\
        &\times \ee^{\frac{\widetilde m_i}{\widetilde m_{i+1}}\tilde f_{i+1}(h) - \tilde f_i(h-w_i)}\,,  
    \end{split}
\end{align}
with the boundary condition given by Eq.~\eqref{eq:app_frsb_P0_discrete}. If $q_0<0.01$, we replace the function $P_0$ by a Dirac delta centered at the origin. Once the distribution of internal fields are known, we compute the order parameters from the low temperature limit of the saddle point equations Eq.~\eqref{eq:frsb_sp_discrete}, which read
\begin{align}
    \begin{split}
        q_i &= \frac{1}{\widetilde m_i^2}\int \dd h\, P_i(h) [f'_{i+1}(h)]^2\\
        \Delta \widetilde q &= \int \dd h\, \frac{P_k(h)}{3x^*(h)^2 - a - \Delta \widetilde q}\\
        m &= \int \dd h\, P_k(h) x^*(h)\,.
    \end{split}
\end{align}
The position of the bands $\widetilde m_i$ is updated too, using the formula~\cite{schmidt2008method}
\begin{align}
    \begin{split}
    \widetilde m_{i} &= \frac{-B_i-\sqrt{B_i^2 - 4A_iC_i}}{2 A_i}\\
    A_i &= \frac{1}{4}(q_i^2- q_{i-1}^2)\\
    B_i &= -\frac{1}{\widetilde m_{i+1}(1-\delta_{ik}) + \delta_{ik}}\int \dd h\, P_{i-1}(h)\\&\times \int \mathcal{D}w_i \tilde f_i(h-w_i)\ee^{\frac{\widetilde m_i}{\widetilde m_{i+1}}\tilde f_{i+1}(h-w_i) - \tilde f_i(h-w_i)}\\
    C_i &= \int \dd h\,P_{i-1}(h) \tilde f_i(h)\,, 
    \end{split}
\end{align}
with the boundaries $\widetilde m_0 = 0$ and $\widetilde m_{k+1}\gg 1$ held fixed. The steps described can be iterated until the desired level of convergence is achieved. To build Fig.~\ref{fig:phase_diagram_global_minima_SK} (a), we iterated the numerical scheme $20$ times with $k=25$, which is enough to obtain convergence of the quantities $m$ and $a+\Delta \widetilde q$ with a degree of accuracy of $\sim 1\%$. At large values of the laser gain, the initial condition for the order parameters $q_i,\,\Delta \widetilde q, m$ was taken from results obtained at slightly larger laser gains. To obtain the curves for $P^\text{\tiny SK}(x)$, $\rho^\text{\tiny{SK}}(\lambda)$ and $\lambda_\text{\tiny m}(a,J_0)$ in Fig.~\ref{fig:phase_diagram_global_minima_SK}(b-e) and Fig.~\ref{fig:lambda_m_J0_a} the number of iterations was increased to $50$, and the number of steps to $k=35$. Moreover, instead of updating the quantity $\Delta \widetilde q$ during each iteration, we instead fixed a value of the effective laser gain $a_\text{\tiny eff} \equiv a+\Delta \widetilde q$. After convergence, we computed $\Delta \widetilde q$ and, consequently, $a$. We show in Fig.~\ref{fig:app_P&q_frsb_numerics} a representative result of the numerical iteration.
\begin{figure}
    \includegraphics[width=\columnwidth]{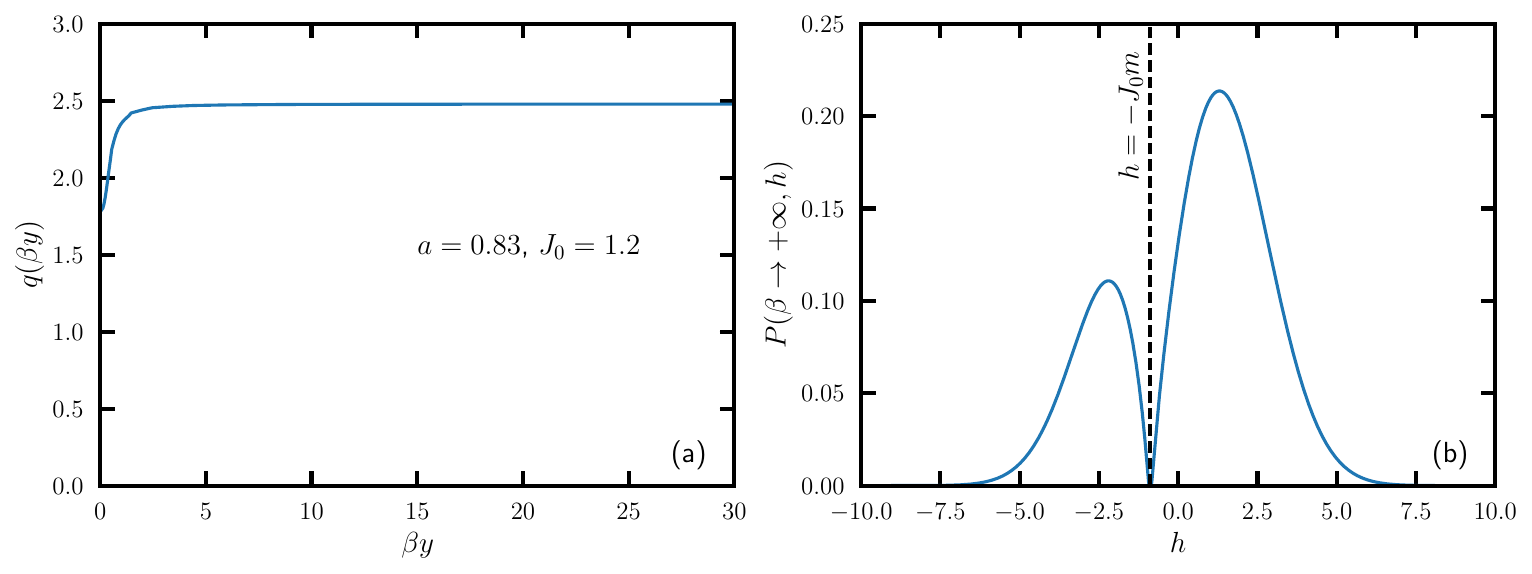}
    \caption{\textbf{Overlap function and distribution of internal fields in the spin-glass phase.} Panel (a): overlap function $q(\beta y)$ as a function of the rescaled quantity $\beta y$. Due to the presence of a spontaneous magnetization, the value of $q$ at the origin $q(0)$ is different from $0$. As $\beta y\to\infty$, the overlap function tends to a constant. Panel (b): distribution of internal fields $P(\beta y\to \infty,h)$. The distribution approaches $0$ linearly in $|h-J_0 m|$ for $h\approx -J_0 m$.   \label{fig:app_P&q_frsb_numerics}}
\end{figure}

The numerical procedure adopted here approximates the full replica-symmetry breaking solution as a large but finite number of replica symmetry breaking steps. We exploit this feature to show that in the vicinity of of the spin-glass transition line $a_{\text{\tiny{sg}}}(J_0)$ the one-step and the two-step replica symmetry breaking solutions are unstable, as their numerical solution yield different curves for the overlap function $q(\beta y)$. Numerical result supporting the need for a full replica symmetry breaking solution in the spin-glass are shown in Fig.~\ref{fig:qbetay_multistep}.
\begin{figure}
\includegraphics[width=\columnwidth]{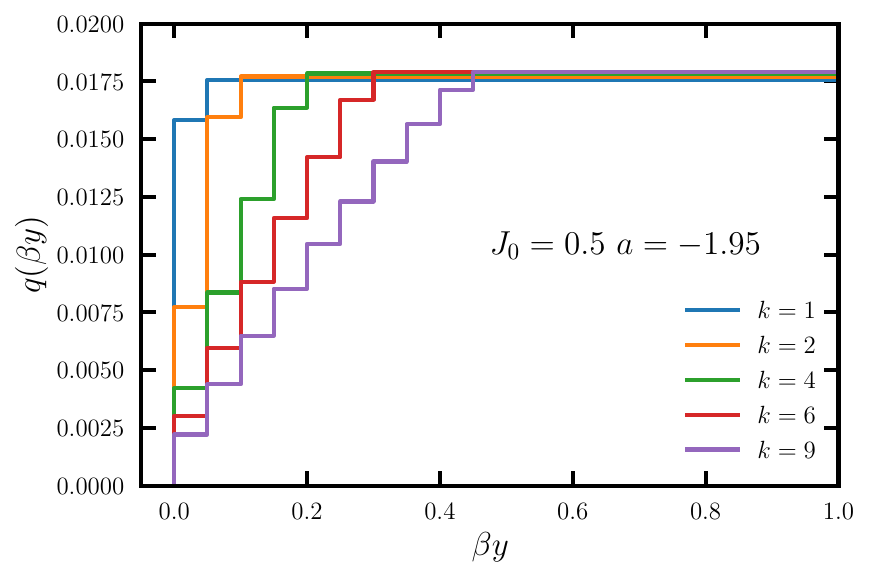}
\caption{\label{fig:qbetay_multistep}\textbf{Evidence for a full-replica symmetry breaking spin-glass in the CIM.} We show the overlap function $q(\beta y)$ at different steps of replica-symmetry breaking, for a value of the laser gain $a \lesssim a_\text{\tiny{sg}}(J_0)$ close to the spin-glass transition. The solution changes as we increase the number of steps in the replica-symmetry breaking solution.}
\end{figure}

\section{Replica calculation of the grand-potential}\label{app:computation_grand-potential}
This Appendix is devoted to the computation of the grand-potential in Eq.~\eqref{eq:Omega}, namely,
\begin{equation}
     \OmegaSK(\mu) = -\lim_{N\to \infty } \frac{1}{N} \langle \log \ZOmega\rangle_\bJ\,,
\end{equation}
where the grandcanonical partition function $\ZOmega$ is given by
\begin{equation}\label{eq:app_Zomega}
    \begin{split}
        \ZOmega &\equiv \int \dd\bx \left(\prod_{i=1}^N\delta(\p_{i}E^\text{\tiny{SK}}(\bx,a))\right)\\
        &\times \lvert \det \bHSK(\bx,a)\rvert \ee^{\mu \mathcal{I}(\bx)}\\
        &= \int_{\bu,\,\bx}\ee^{-\bu\cdot \bnabla E^\text{\tiny{SK}}(\bx,a) + \mu\mI(\bx)}\lvert \det \bHSK(\bx,a)\rvert\,. 
    \end{split}
\end{equation}
In the second line, we used an integral representation of the Dirac delta distribution. The integral $\int_{\bu,\,\bx} \equiv \int_{-i\infty}^{+i\infty}\frac{\dd\bu}{(2\pi i)^N}\,\int\dd\bx$ is over a $2N$ dimensional space of soft spin variables $\bx$ and auxiliary fields $\bu$. The average of the disorder is computed, as in Sec.~\ref{sec:replica_resolvent} through the replica trick
\begin{equation}\label{eq:app_Omega_replicated}
    \beta\OmegaSK(\mu) = -\lim_{n\to0}\lim_{N\to\infty}\frac{1}{Nn}\left(\langle (\ZOmega)^n\rangle_\bJ-1\right)\,.
\end{equation}
The disorder-averaged, replicated grandcanonical partition function $\langle \ZOmega\rangle_\bJ$ being 
\begin{widetext}
\begin{align}
    \begin{split}
        \langle(\ZOmega)^n\rangle_\bJ &=  \int_{\bx^\alpha,\, \bu^\alpha}\Biggl\langle\exp\left[ \sum_{\alpha=1}^n \left(\sum_{i=1}^N u_i^\alpha\left(-\p_x E_I(x_i^\alpha,a) + \frac{J_0}{N}\sum_{j=1}^N x^\alpha_j  + \frac{1}{\sqrt{N}}\sum_{j=1}^N J_{ij}x_j^\alpha \right)\right) + \mu\sum_{\alpha} \mI(\bx^\alpha)\right]\\&\times\prod_{\alpha }|\det \bHSK(\bx^\alpha,a)|\Biggr\rangle_\bJ\,.
    \end{split}
\end{align}
The integral $\int_{\bx^\alpha,\,\bu^\alpha}\equiv \int \prod_{\alpha=1}^n \frac{\dd\bx^\alpha \dd\bu^\alpha}{(2\pi i)^N}$ is over the space of $n$ replicated $2N$ dimensional systems of soft spins $\bx^\alpha$ and auxiliary fields $\bu^\alpha$. The absolute value of the determinant of $\bHSK(\bx^\alpha,a)$ is correlated with the auxiliary fields $\bu^\alpha$ and with the index $\mI(\bx^\alpha)$ of different replicas. A full treatment of the grand-potential requires taking these correlations into account, for instance by means of conditional averages, as done in~\cite{ros2019complex} for spiked tensor models. In our case, however, the analysis is hindered by the single-site term $\p_x E_I(x,a)$. To proceed further, we make an approximation by decoupling the gradient of the CIM-SK energy and the determinant of the Hessian among different realizations of the disorder, that is
\begin{align}\label{eq:app_Zomega_decoupling_gradients}
    \begin{split}
        \langle\ZOmega\rangle_\bJ &\approx \int_{\bx^\alpha,\, \bu^\alpha}\left\langle\exp\left[ \sum_{\alpha=1}^n \left(\sum_{i=1}^N u_i^\alpha\left(-\p_x E_I(x_i^\alpha,a) + \frac{J_0}{N}\sum_{j=1}^N x^\alpha_j  + \frac{1}{\sqrt{N}}\sum_{j=1}^N J_{ij}x_j^\alpha \right)\right)\right]\right\rangle_\bJ\\
    &\times\left\langle \prod_{\alpha }\ee^{\mu \mI(\bx^\alpha)}|\det \bHSK(\bx^\alpha,a)|\right\rangle_\bJ\,.
    \end{split}
\end{align}
Moreover, we neglect the correlations between different replicas when computing the index and the absolute value of the determinant of the Hessian, i.e.,
\begin{align}\label{eq:approx_decoupling}
    \begin{split}
        &\left\langle \prod_{\alpha }\ee^{\mu \mI(\bx^\alpha)}|\det \bHSK(\bx^\alpha,a)|\right\rangle_\bJ \\
        &\approx \prod_\alpha \left\langle \ee^{\mu \mI(\bx^\alpha)}\right\rangle_\bJ\left\langle|\det \bHSK(\bx^\alpha,a)| \right\rangle_\bJ\,.
    \end{split}
\end{align}
 We stress here that these approximations are uncontrolled, and that they are introduced as a working hypothesis, to make the calculation of the grandpotential tractable and to allow for further progress. As shown \textit{a posteriori} from numerical experiments in Sec.~\ref{sec:first_look}, these approximations lead to a reasonable agreement with numerical experiments, as long as no conditioning on the net magnetization or energy of the critical point is introduced. 

\subsection{First contribution}
We consider the first average over the disorder in Eq.~\eqref{eq:app_Zomega_decoupling_gradients}, which  can be computed in a similar way as done in App.~\ref{app:Gz_calculation}. We obtain
\begin{align}\label{eq:app_first_term_grandcanonical}
    \begin{split}
        &\left\langle\exp\left[ \sum_{\alpha=1}^n \left(\sum_{i=1}^N u_i^\alpha\left(-\p_x E_I(x_i^\alpha,a) + \frac{J_0}{N}\sum_{j=1}^N x^\alpha_j  + \frac{1}{\sqrt{N}}\sum_{j=1}^N J_{ij}x_j^\alpha \right)\right)\right]\right\rangle_\bJ \\&= \exp\left[ -\sum_{\alpha}\sum_i u_i^\alpha \p_x E_I(x_i^\alpha,a) + J_0\sum_\alpha\sum_{i}u_i^\alpha \left(\frac{1}{N}\sum_{j=1}^N x_j^\alpha\right) + \frac{1}{4N}\sum_{i,j}\left(\sum_\alpha u_i^\alpha x_j^\alpha+  u_j^\beta x_i^\beta\right)^2\right]\\
        &= \exp\left[ -\sum_{\alpha}\sum_i u_i^\alpha \p_x E_I(x_i^\alpha,a) + NJ_0\sum_\alpha \left(\frac{\bu^\alpha\cdot\bt}{N}\right)\left(\frac{\bx^\alpha\cdot \bt}{N}\right) + \frac{N}{2}\sum_{\alpha,\beta} \left[ \left(\frac{\bu^\alpha\cdot\bu^\beta}{N}\right)\left(\frac{\bx^\alpha\cdot\bx^\beta }{N}\right) + \left(\frac{\bu^\alpha\cdot \bx^\beta}{N}\right)^2\right]\right]\,,
    \end{split}
\end{align}
where we recall that $\{\bx^{\alpha},\,\bu^\alpha\}_{\alpha=1}^n$ is a set of $N$-dimensional vectors representing the soft spin and auxiliary field configurations of each replica. We then use the Hubbard-Stratonovich transform introduced in Eq.~\eqref{eq:app_Hubbard_Stratonovich} to linearize the quadratic term $\left(\frac{\bu^\alpha\cdot \bx^\beta}{N}\right)^2$, and we insert the following set of Dirac deltas, together with their integral representations, for each distinct pairs of replica indexes $\alpha,\,\beta$,  
\begin{align}
    \begin{split}
        \delta(N Q_{\alpha\beta}- \bx^\alpha\cdot\bx^\beta) &= \int_{-i\infty}^{i\infty} \frac{\dd \lambda_{\alpha\beta}}{2\pi i}\ee^{-N\lambda_{\alpha\beta}Q_{\alpha\beta} + \lambda_{\alpha\beta}(\bx^\alpha\cdot\bx^\beta)}\\
        \delta(N m_\alpha - \bx^\alpha\cdot \bt) &= \int_{-i\infty}^{i\infty} \frac{\dd v_\alpha}{2\pi i}\,\ee^{-Nv_\alpha m_\alpha + v_\alpha(\bx^\alpha\cdot\bt)}\,.
    \end{split}
\end{align}
Equation~\eqref{eq:app_first_term_grandcanonical} thus becomes
\begin{align}
    \begin{split}
    &\int\prod_{\alpha=1}^n \frac{\dd v_\alpha \dd m_\alpha}{2\pi i}\prod_{\alpha\leq \beta}\frac{\dd Q_{\alpha\beta}\dd\lambda_{\alpha\beta}}{2\pi i}\prod_{\alpha,\beta} \dd w_{\alpha\beta}\,\exp\left[-N\left( \bmm\cdot\bv + \Tr\left[\blambda\bQ + \frac{1}{2}\bw\bw^\mathrm{T}\right]\right)\right]\\
    &\times\exp\Biggl[ -\sum_{\alpha}\sum_i u_i^\alpha \p_x E_I(x_i^\alpha,a) + \sum_{\alpha,\beta} \lambda_{\alpha\beta}(\bx^\alpha\cdot \bx^\beta) + J_0 \sum_\alpha (\bu^\alpha\cdot\bt)m_\alpha + \sum_{\alpha} v_\alpha (\bx^\alpha\cdot \bt) + \frac 1 2\sum_{\alpha,\beta}Q_{\alpha\beta}(\bu^\alpha\cdot \bu^\beta)\\
    &-\sum_{\alpha,\beta}w_{\alpha\beta}(\bu^\alpha\cdot\bx^\beta)\Biggr]\,,
    \end{split}
\end{align}
where the $n\times n$ matrices $\blambda$, $\bQ$, $\bw$ and the $n$-dimensional vectors $\bmm$, $\bv$ have been introduced.\\   
\end{widetext}

\subsection{Second contribution}
The terms in the second average in Eq.~\eqref{eq:app_Zomega_decoupling_gradients} can be computed from the knowledge of $\langle \det \left[z - \bHSK(\bx,a)\right] \rangle_\bJ$, with $z$ a complex number close to the real axis. In fact, the index $\mI(\bx)$ can be obtained from the formula~\cite{cavagna2000index}
\begin{align}
    \begin{split}
        \mI(\bx) &= \frac{1}{2\pi i}\lim_{\epsilon\to 0} \bigl[\log \det [\bHSK(\bx,a) - i\epsilon] \\
        &- \log\det [\bHSK(\bx,a) + i\epsilon]\bigr]\,,
    \end{split}
\end{align}
which can be proved using Cauchy's residues theorem. The branch cut of the complex logarithm function is chosen so that the quantities $\log \lambda_i \pm i\epsilon$ are on the opposite side of the branch cut for any eigenvalue $\lambda_i$ of $\bHSK(\bx,a)$. For replica $\alpha$, we thus obtain
\begin{equation}
    \ee^{\mu \mI(\bx^\alpha)} = \lim_{\epsilon\to 0}\left(\frac{\det[\bHSK(\bx,a) - i\epsilon ]}{\det[\bHSK(\bx,a) + i\epsilon ]}\right)^{\mu/2\pi i}\,.
\end{equation}
The absolute value of the determinant can be instead expressed as
\begin{align}\label{eq:app_approximated_index_det}
    \begin{split}
        &|\det \bHSK(\bx^\alpha,a)| = \det[\bHSK(\bx^\alpha,a)]\ee^{\xi\pi i\mI(\bx^\alpha)}\\
        &= \lim_{\epsilon\to0} \det [\bHSK(\bx,a)]\left(\frac{\det[\bHSK(\bx,a) - i\epsilon ]}{\det[\bHSK(\bx,a) + i\epsilon ]}\right)^{\frac{\xi}{2}}\,,
    \end{split}
\end{align}
where $\xi$ can be taken to be either equal to $1$ or $-1$. This freedom in the choice of the sign will be exploited later on to simplify the calculations, as done in~\cite{yamamuraGeometricLandscapeAnnealing2024}. Finally, to proceed further, we assume that $\det [\bHSK(\bx,a)]$, and $\det [\bHSK(\bx,a) \pm i\epsilon]$ are independent random variables under different realizations of the disorder, i.e.,
\begin{align}
    \begin{split}
        &\langle \ee^{\mu \mI(\bx^\alpha)}\rangle_\bJ \approx \lim_{\epsilon\to 0}\left(\frac{\langle \det[\bHSK(\bx^\alpha,a) - i\epsilon ]\rangle_\bJ}{\langle \det[\bHSK(\bx^\alpha,a) + i\epsilon ]\rangle_\bJ}\right)^{\frac{\mu}{2\pi i}}\\
        &\langle|\det \bHSK(\bx^\alpha,a)| \rangle_\bJ \approx \langle \det \bHSK(\bx^\alpha,a)\rangle_\bJ \\
        &\times \lim_{\epsilon\to 0}\left(\frac{\langle \det[\bHSK(\bx^\alpha,a) - i\epsilon ]\rangle_\bJ}{\langle \det[\bHSK(\bx^\alpha,a) + i\epsilon ]\rangle_\bJ}\right)^{\frac{\xi}{2}}\,.
    \end{split}
\end{align}
Within this approximation, we thus see that we need to compute the average $\langle \det \bHSK(\bx^\alpha,a) \pm i\epsilon \rangle_\bJ$. We compute this next.\\ 

\tocless\subsubsection{Determinant of the Hessian}
We express the determinant $\det[ \bHSK(\bx^\alpha,a) \pm i\epsilon]$ by means of two sets of independent Grassmann variables $\psi_i^\alpha$, $\overline{\psi}_i^\alpha$~\cite{zinn2021quantum}, satisfying the anticommutation relations 
\begin{align}
    \begin{split}           \psi_i^\alpha\overline{\psi}_j^\alpha + \overline{\psi}_j^\alpha \psi_i^\alpha &=0\\  \psi_i^\alpha\psi_j^\alpha + \psi_j^\alpha \psi_i^\alpha &=0\\
    \overline{\psi}_i^\alpha \,\overline{\psi}_j^\alpha + \overline{\psi}_j^\alpha\, \overline{\psi}_i^\alpha &=0\,.
    \end{split}
\end{align}
We thus write
\begin{widetext}
\begin{align}\label{eq:app_calculation_detH}
    \begin{split}
        &\langle \det[\bHSK(\bx^\alpha,a) \pm i\epsilon] \rangle_\bJ = \int \prod_{i=1}^N \dd\psi_i^\alpha\dd\overline{\psi}_i^\alpha \left\langle \exp\left[\sum_{i,j} \overline{\psi}_i^\alpha\left(-\frac{J_{ij}}{\sqrt{N}} - \frac{J_0}{N}+ (\p^2_x E_I(x^\alpha_i,a) \pm i\epsilon)\delta_{ij}\right)\psi_j^\alpha\right] \right\rangle_\bJ \\
        &= \int_{\bpsi^\alpha,\,\overline{\bpsi}^\alpha}\, \exp\left[\frac{1}{2N}\sum_{i\leq j}\left(  \overline{\psi}^\alpha_i\psi^\alpha_j + \overline{\psi}^\alpha_j\psi^\alpha_i\right)^2 + \sum_{i,j} \overline{\psi}_i^\alpha \left(-\frac{J_0}{N} + ( \p^2_x E_I(x^\alpha_i,a) \pm i\epsilon)\delta_{ij}\right)\psi_j^\alpha \right] \\
        &= \int_{\bpsi^\alpha,\,\overline{\bpsi}^\alpha}\, \exp\left[-\frac{1}{2N}\left(\sum_i  \overline{\psi}^\alpha_i\psi^\alpha_i\right)^2 + \sum_{i,j} \overline{\psi}_i^\alpha \left(-\frac{J_0}{N} + ( \p^2_x E_I(x^\alpha_i,a) \pm i\epsilon)\delta_{ij}\right)\psi_j^\alpha \right] \\
        &= \int_{\bpsi^\alpha,\,\overline{\bpsi}^\alpha}\int_{-\infty}^{\infty} \frac{\dd t^\alpha}{\sqrt{2\pi/N}}\, 
        \exp\left[ -\frac{N}{2}(t^\alpha)^2 - i t^\alpha\left(\sum_i \overline{\psi}_i^\alpha\psi_i^\alpha\right)\right]\\&\times\exp\left[ \sum_{i,j} \overline{\psi}_i^\alpha \left(-\frac{J_0}{N} + (\p^2_x E_I(x^\alpha_i,a)\pm i\epsilon)\delta_{ij}\right)\psi_j^\alpha\right] \\
        &= \int_{-i\infty}^{+i\infty} \frac{i\dd t^\alpha}{\sqrt{2\pi/N}}\, \exp\left[ \frac{N}{2}(t^\alpha)^2 \right]\int_{\bpsi^\alpha,\,\overline{\bpsi}^\alpha}\exp\left[ \sum_{i,j} \overline{\psi}_i^\alpha \left(-\frac{J_0}{N} + (\p^2_x E_I(x_i^\alpha,a)\pm i\epsilon - t^\alpha)\delta_{ij}\right)\psi_j^\alpha\right]\\
        &= \int_{-i\infty}^{+i\infty} \frac{i\dd t^\alpha}{\sqrt{2\pi/N}}\,\exp\left[N\left( \frac 1 2 (t^\alpha)^2 + \frac{1}{N}\log \det \left[(\p^2_x E_I(x_i^\alpha,a) \pm i\epsilon -t^\alpha)\delta_{ij}-\frac{J_0}{N}\right]\right)\right]\\
        &\approx \int_{-i\infty}^{+i\infty} \dd t^\alpha \exp\left[ N\left(\frac 1 2 (t^\alpha)^2 + \frac{1}{N}\log \det\left[(\p^2_x E_I(x_i^\alpha,a) \pm i\epsilon - t^\alpha)\delta_{ij} \right] \right)\right]\\
       &= \int_{-i\infty}^{+i\infty} \dd t^\alpha \exp\left[ N\left(\frac 1 2 (t^\alpha)^2 + \frac{1}{N}\sum_{i=1}^N\log \left(\p^2_x E_I(x_i^\alpha,a) \pm i\epsilon - t^\alpha\right)\right)\right]\,.
    \end{split}
\end{align}
\end{widetext}
In the second line, we computed the average over the disorder and we used the shorthand notation for the integral $\int_{\bpsi^\alpha,\,\overline{\bpsi}^\alpha} = \int \prod_i \dd \psi_i^\alpha\dd \overline{\psi}_i^\alpha$. In the third line, we used the anticommutation properties of the Grassmann variables to compute the square
\begin{equation}
    \sum_{i\leq j}(\overline{\psi_i}^\alpha \psi_j^\alpha + \overline{\psi}_j^\alpha \psi_i^\alpha)^2 = -\left(\sum_i \overline{\psi}_i^\alpha\psi_i^\alpha\right)^2\,.
\end{equation}
In the fourth line, we used a Hubbard-Stratonovich transformation. In the fifth line, we changed the integration variable. In the sixth line, we expressed the integral over the Grassmann variable as the determinant of a matrix. In the seventh line, we neglected subleading contribution in $N$. Note that the ferromagnetic coupling is among these subleading contributions. This can be understood from the Weinstein–Aronszajn identity,
\begin{align}\label{eq:matrix_determinant_lemma}
    \begin{split}
        \det \left[\bA + \frac{J_0}{N}\bt\otimes\bt\right] &= \left(1 + \frac{J_0}{N}\bt\cdot  \bA^{-1}\bt \right)\det \bA\,, 
    \end{split}
\end{align}
where $\bt=[1,\,\ldots,1]^\mathrm{T}$ is a $N$-dimensional vector and $\bA$ is a real $N\times N$ invertible matrix with entries of $O(1)$ with respect to $N$. A determinant of this type contributes to the exponential in Eq.~\eqref{eq:app_calculation_detH} through 
\begin{align}
    \begin{split}
        \frac{1}{N}&\log \det \left[\bA + \frac{J_0}{N}\bt \otimes \bt\right] \\
        &= \frac{1}{N}\log\left(1+ \frac{J_0}{N}\Tr[\bA^{-1}]\right)+ \frac{1}{N}\log \det \bA \\
        &= \frac{1}{N}\log \det \bA + O(N^{-1})\,,
    \end{split}
\end{align}
we thus see that the ferromagnetic alignment $J_0$ contributes to the integral in Eq.~\eqref{eq:app_calculation_detH} through a term subleading in $N$. From the last line of Eq.~\eqref{eq:app_calculation_detH}, we anticipate that a saddle point evaluation of the integral is possible. Using this fact, and the approximations in Eq.~\eqref{eq:app_approximated_index_det}, we obtain
\begin{widetext}
\begin{align}\label{eq:app_emu_det}
    \begin{split}
        \left\langle \ee^{\mu\mI(\bx^\alpha)} |\det \bHSK(\bx^\alpha,a)|\right\rangle_\bJ &\approx \lim_{\epsilon \to 0} \int \dd t^\alpha_0\,\dd t^\alpha_+ \dd t^\alpha_- \exp\Biggl[N\left((t_0^\alpha)^2 + \left(\frac \xi 2 + \frac{\mu}{2\pi i}\right)\left((t_+^\alpha)^2 - (t_-^\alpha)^2\right)\right)\\
        &+ \left(\frac{\xi}{2} + \frac{\mu}{2\pi i}\right)\sum_i \log\frac{\p^2_x E_I(x_i^\alpha,a) -i\epsilon - t_+^\alpha}{\p_x^2 E_I(x_i^\alpha,a) + i\epsilon - t_-^\alpha } + \sum_i \log (\p^2_x E_I(x_i^\alpha,a) - t_0^\alpha) \Biggr]\,,
    \end{split}
\end{align}
where we neglected subleading contributions in the thermodynamic limit.
\\\subsection{Saddle point evaluation of the grand-potential}
We substitute Eq.~\eqref{eq:app_first_term_grandcanonical} and Eq.~\eqref{eq:app_emu_det} into the grandcanonical partition function of Eq.~\eqref{eq:app_Zomega}. We can now evaluate the grand-potential $\Omega(\mu)$ in the thermodynamic limit $N\to\infty$ through a saddle point integral, as done for the resolvent in App.~\ref{app:Gz_calculation}. We obtain 
\begin{align}\label{eq:app_grand-potential_first}
    \begin{split}
         \Omega^\text{\tiny{SK}}(\mu) &= \lim_{n\to 0} \frac{1}{n}\left[\Omega_0 - \log \lim_{\epsilon \to 0} \int \prod_{\alpha=1}^n \frac{\dd x^\alpha \dd u^\alpha}{2\pi i} \ee^{S_\Omega[\{x^\alpha,u^\alpha\}]} \right]\\
        \Omega_0 &= \sum_\alpha m_\alpha v_\alpha +  \sum_{\alpha,\beta}\left[\lambda_{\alpha\beta}Q_{\alpha\beta} + \frac{1}{2}w_{\alpha\beta}^2 \right] \\
        S_\Omega &= -\sum_\alpha u^\alpha \p_x E_I(x^\alpha,a) + \sum_{\alpha,\beta}\left[ \lambda_{\alpha\beta} x^\alpha x^\beta + \frac{1}{2}Q_{\alpha\beta}u^\alpha u^\beta - w_{\alpha\beta}u^\alpha x^\beta\right] + \sum_\alpha [J_0 m_\alpha u^\alpha + v_\alpha x^\alpha]\\
        &+ \frac 1 2\sum_\alpha\left((t_0^\alpha)^2 + \left(\frac \xi 2 + \frac{\mu}{2\pi i}\right)\left((t_+^\alpha)^2 - (t_-^\alpha)^2\right)\right)\\
        &+ \left(\frac{\xi}{2} + \frac{\mu}{2\pi i}\right)\sum_\alpha \log\frac{\p^2_x E_I(x^\alpha,a) -i\epsilon - t_+^\alpha}{\p_x^2 E_I(x^\alpha,a) + i\epsilon - t_-^\alpha } + \sum_\alpha \log (\p^2_x E_I(x^\alpha,a) - t_0^\alpha)\,.
    \end{split}
\end{align}
The order parameters satisfy the following set of saddle point equations
\begin{align}\label{eq:app_saddle_point_grand-potential}
    \begin{split}
        Q_{\alpha\beta} &= \langle x^\alpha x^\beta \rangle_\Omega \\ \lambda_{\alpha\beta} &= \frac 1 2\langle u^\alpha u^\beta\rangle_\Omega \\
        w_{\alpha\beta} &= -\langle x^\alpha u^\beta \rangle_\Omega \\ m_\alpha &= \langle x^\alpha \rangle_\Omega\\
        v_\alpha &= J_0 \langle u^\alpha \rangle_\Omega\\
        t_0^\alpha &= \left\langle \frac{1}{\p^2_x E_I(x^\alpha,a) - t_0^\alpha}\right\rangle_\Omega\\
        t_\pm^\alpha &= \left\langle\frac{1}{\p^2_x E_I(x^\alpha) \mp i\epsilon - t_\pm^\alpha} \right\rangle_\Omega\,,
    \end{split}
\end{align}
\end{widetext}
where the average $\langle\ldots\rangle_\Omega$ denotes an average over the grandcanonical mean field distribution for the replicated system of spins $x^\alpha$ and Lagrange multipliers $u^\alpha$, defined as,
\begin{equation}\label{eq:average_Omega}
    \langle\ldots \rangle_\Omega = \frac{\int \prod_\alpha \dd x^\alpha \dd u^\alpha\,\ldots \ee^{S_\Omega(\{x^\alpha,u^\alpha\})}}{\int \prod_\alpha \dd x^\alpha\dd u^\alpha\, \ee^{S_\Omega(\{x^\alpha,u^\alpha\})}  }\,.
\end{equation}
Note that the saddle point solution to the equation for $t_0^\alpha$ is degenerate, for if $t_0^\alpha$ is a solution, then its complex conjugate is a solution solution too. On the other hand, the presence of the term $\mp i\epsilon$ in the equation for $t_\pm^\alpha$ lifts this degeneracy. We thus write $t_\pm^\alpha \equiv t_R^\alpha \pm it_I^\alpha$, separating the real part of $t_\pm^\alpha$ from the imaginary part, while $t_0^\alpha = t_R^\alpha \pm i t_I^\alpha$. However, this degeneracy is irrelevant for the computation of the grand-potential $\beta \Omega^\text{\tiny{SK}}(\mu)$, because we have freedom on the choice of $\xi =\pm1$, as mentioned below Eq.~\eqref{eq:app_approximated_index_det}. In fact, it is possible to show that if we choose $t_0^\alpha = t_R \pm i t_I$ and $\xi = \mp 1$, then the expression of $S_\Omega$ in Eq.~\eqref{eq:app_grand-potential_first} does not depend on the choice of the stationary solution $t_0^\alpha$, and it becomes
\begin{widetext}
    \begin{align}\label{eq:EOmega_tR_tI}
        \begin{split}
             S_\Omega &= -\sum_\alpha u^\alpha \p_x E_I(x^\alpha,a) + \sum_{\alpha,\beta}\left[ \lambda_{\alpha\beta} x^\alpha x^\beta +\frac{1}{2}Q_{\alpha\beta}u^\alpha u^\beta - w_{\alpha\beta}u^\alpha x^\beta\right] + \sum_\alpha [J_0 m_\alpha u^\alpha + v_\alpha x^\alpha]\\
            &+ \sum_\alpha \left[\frac 1 2 \left((t_R^\alpha)^2 - (t_I^\alpha)^2\right) + \log\left[(\p_x^2 E_I(x^\alpha,a)-t_R^\alpha)^2 + (t_I^\alpha)^2\right]\right] \\
            &+\mu\sum_\alpha\left[\frac{t_R^\alpha t_I^\alpha}{\pi} +\frac{1}{2\pi i}\log \frac{\p_x^2 E_I(x^\alpha,a) - i(\epsilon + t_I^\alpha) - t_R^\alpha}{\p_x^2 E_I(x^\alpha,a) + i(\epsilon + t_I^\alpha) - t_R^\alpha}\right]\,.
        \end{split}
    \end{align}
The real and imaginary parts $t_R^\alpha$ and $t_I^\alpha$ satisfy the following saddle point equations, which are obtained from Eq.~\eqref{eq:app_saddle_point_grand-potential}, 
\begin{align}
    \begin{split}
        t_R^\alpha &= \lim_{\epsilon\to0}\left\langle \frac{\p_x^2E_I(x^\alpha,a) - t_R^\alpha}{(\p_x^2 E_I(x^\alpha,a) - t_R^\alpha)^2 + (t_I^\alpha + \epsilon)^2} \right\rangle_\Omega\\
        t_I^\alpha &= \lim_{\epsilon\to0}\left\langle\frac{\epsilon + t_I^\alpha}{(\p_x^2 E_I(x^\alpha,a) - t_R^\alpha)^2 + (t_I^\alpha + \epsilon)^2} \right\rangle_\Omega\,.
    \end{split}
\end{align}
From Eq.~\eqref{eq:app_saddle_point_grand-potential}, we see that $t_R^\alpha \pm i t_I^\alpha$  satisfy the same self consistent equations for the resolvent, evaluated at the origin, of a random matrix drawn from the Gaussian orthogonal ensemble, whose diagonal elements are perturbed by $\p_x^2 E_I(x^\alpha,a)$, with the $x^\alpha$ following the distribution defined in the average $\langle\ldots\rangle_\Omega$ in Eq.~\eqref{eq:average_Omega}. We now assume that $t_I^\alpha=0$ which, as explained in~\cite{yamamuraGeometricLandscapeAnnealing2024}, implies that the spectral density of the Hessian of the most abundant critical points of the system is zero at the origin. We checked through the numerical solution of the saddle point equation that this solution is robust when a nonzero value for the imaginary part is allowed. Within this approximation, the contribution proportional to $\mu$ to the action $S_\Omega$ in Eq.~\eqref{eq:EOmega_tR_tI} becomes, in the limit $\epsilon\to0$, 
\begin{align} 
    \begin{split}
        \lim_{\epsilon\to0}\mu\sum_\alpha\left[\frac{t_R^\alpha t_I^\alpha}{\pi} +\frac{1}{2\pi i}\log \frac{\p_x^2 E_I(x^\alpha,a) - i(\epsilon + t_I^\alpha) - t_R^\alpha}{\p_x^2 E_I(x^\alpha,a) + i(\epsilon + t_I^\alpha) - t_R^\alpha}\right]\ &=\mu\sum_\alpha \Theta(-(\p^2_x E_I(x^\alpha,a)- t_R^\alpha))\\ &\equiv \mu \sum_\alpha \mIMF(x^\alpha)\,,
    \end{split}
\end{align}
where $\Theta(x)$ is the Heaviside step function, and we have chosen the branch cut of the logarithm to lie on the positive side of the real axis. This choice has a physical justification: in this way, in the limit $\mu\to-\infty$ the energy $\beta E_\Omega$ selects for configurations $x^\alpha$ located at the minima of the energy function $\p^2_x E_I(x^\alpha,a+t_R^\alpha)$. The last equality defines the mean field index function $\mIMF(x)$.

We now present the final result for the replica calculation of the grandcanonical partition function. We first find convenient to make a change of variable, replacing the matrices $\lambda_{\alpha\beta}$ and $w_{\alpha\beta}$ with a new pair of matrices $A_{\alpha\beta}$ and $C_{\alpha\beta}$, defined as
\begin{align}
    \begin{split}
    A_{\alpha\beta} &\equiv -t_R^\alpha\delta_{\alpha\beta} - w_{\alpha\beta} \\
    C_{\alpha\beta} &\equiv  2\lambda_{\alpha\beta}\,.
    \end{split}
\end{align}
With this substitution and our ansatz $t_I^\alpha=0$, the expression of the grand-potential in Eq.~\eqref{eq:app_grand-potential_first} becomes
\begin{align}\label{eq:app_grand-potential_second}
    \begin{split}
        \Omega^\text{\tiny{SK}}(\mu) &= \lim_{n\to 0} \frac{1}{n}\left[\Omega_0 - \log \lim_{\epsilon \to 0} \int \prod_{\alpha=1}^n \frac{\dd x^\alpha \dd u^\alpha}{2\pi i} \ee^{S_\Omega[\{x^\alpha,u^\alpha\}]} \right]\\
        \Omega_0 &= \sum_\alpha m_\alpha v_\alpha +  \frac{1}{2}\sum_{\alpha,\beta}\left[C_{\alpha\beta}Q_{\alpha\beta} + A_{\alpha\beta}^2  \right] + \sum_\alpha A_{\alpha\alpha} t_R^\alpha\\
        S_\Omega &= -\sum_\alpha u^\alpha \p_x E_I(x^\alpha,a+t_R^\alpha) + \sum_{\alpha,\beta}\left[ \frac{1}{2}C_{\alpha\beta} x^\alpha x^\beta + \frac{1}{2}Q_{\alpha\beta}u^\alpha u^\beta + A_{\alpha\beta}u^\alpha x^\beta\right] + \sum_\alpha [J_0 m_\alpha u^\alpha + v_\alpha x^\alpha]\\
        &+ \sum_\alpha \log |\p_x^2 E_I(x^\alpha,\,a+t_R^\alpha)| + \mu\sum_\alpha \mIMF(x^\alpha)\,.
    \end{split}
\end{align}
The order parameters satisfy the following set of saddle point equations
\begin{align}\label{eq:app_sp_SUSY_Omega}
    \begin{split}
        Q_{\alpha\beta} &= \langle x^\alpha x^\beta \rangle_\Omega \\
        C_{\alpha\beta} &= \langle u^\alpha u^\beta\rangle_\Omega \\
        A_{\alpha\beta} &= \langle x^\alpha u^\beta \rangle_\Omega -t_R^\alpha\delta_{\alpha\beta}\\ m_\alpha &= \langle x^\alpha \rangle_\Omega\\
        v_\alpha &= J_0 \langle u^\alpha \rangle_\Omega\\ t_R^\alpha &= \left\langle\frac{1}{\lvert \p^2_x E_I(x^\alpha,a+t_R^\alpha)\rvert}\right\rangle_\Omega\,.
    \end{split}
\end{align}
These equations are evaluated in the replica-symmetric phase in the next Section. Eq.~\eqref{eq:app_grand-potential_second} is a two-group formulation of the grandpotential for a soft-spin-glass~\cite{parisi1995number}, for which supersymmetry breaking is signaled by nonzero order parameters in the matrices $A_{ab}$, $C_{ab}$, as we discuss in the next Section using a replica-symmetric ansatz.

\subsection{Replica symmetric ansatz}
We make the following ansatz for the structure of the order parameters in Eq.~\eqref{eq:app_sp_SUSY_Omega},
\begin{align}\label{eq:sp_rs_SUSY}
    \begin{split}
        A_{\alpha\beta} &= A_d\delta_{\alpha\beta} + A_o(1-\delta_{\alpha\beta}) = \Delta A\delta_{\alpha\beta} + A_o\\
        C_{\alpha\beta} &= \Delta C\delta_{\alpha\beta} + C_o\\
        Q_{\alpha\beta} &= \Delta q\delta_{\alpha\beta} + q_o\\
        m_\alpha &= m\\ v_\alpha &= v\\ t_R^\alpha &= t_R\,,
    \end{split}
\end{align}
and we compute the exponential weight
    \begin{align}\label{eq:app_calculation_exponentialweight_grandpotential_rs}
        \begin{split}
            \int &\prod_\alpha  \frac{\dd x^\alpha \dd u^\alpha}{2\pi i} \ee^{ S_\Omega[\{x^\alpha,\,u^\alpha\}]} = \int_{\bx^\alpha,\,\bu^\alpha}\left(\prod_\alpha |\p^2_x E_I(x^\alpha,a+t_R)|\right)\\
            &\times \exp \Biggl[ \frac{1}{2}\sum_\alpha\left[\Delta C (x^\alpha)^2 + 2\Delta A x^\alpha u^\alpha + \Delta q(u^\alpha)^2\right] + \sum_\alpha[x^\alpha v + J_0 m u^\alpha] \\
            &- \sum_\alpha u^\alpha \p_x E_I(x^\alpha, a+ t_R)+\mu\sum_\alpha \mIMF(x^\alpha)
            +\frac{1}{2}\left[ C_o \left(\sum_\alpha x^\alpha\right)^2 + 2A_o\left(\sum_\alpha x^\alpha\right)\left(\sum_\alpha u^\alpha\right) + q_o \left(\sum_\alpha u^\alpha\right)^2\right]\Biggr]\\
            &= \int_{\bx^\alpha,\,\bu^\alpha}\int \frac{\dd h_1 \dd h_2}{2\pi}\,\ee^{-\frac{h_1^2}{2} - \frac{h_2^2}{2}}\prod_{\alpha} |\p^2_x E_I(x^\alpha,a+t_R)| \exp\Biggl[\frac{1}{2}\left[\Delta C (x^\alpha)^2 + 2\Delta A x^\alpha u^\alpha + \Delta q(u^\alpha)^2\right] + \mu\mIMF(x^\alpha)\\
            &+\left(v + h_1\frac{A_o}{\sqrt{q_o}} + h_2 \sqrt{C_o - \frac{A_o^2}{q_o}}\right)x^\alpha - \left(\p_x E_I(x^\alpha,a+t_R)-J_0 m - h_1 \sqrt{q_o}\right)u^\alpha\Biggr]\\
            &\equiv  \int_{\bx^\alpha,\,\bu^\alpha}\int \frac{\dd h_1 \dd h_2}{2\pi}\,\ee^{-\frac{h_1^2}{2} - \frac{h_2^2}{2}}\prod_{\alpha}  |\p^2_x \EOmegaeff(x^\alpha,h_1)|\exp\Biggl[\frac{1}{2}\left[\Delta C (x^\alpha)^2 + 2\Delta A x^\alpha u^\alpha + \Delta q(u^\alpha)^2\right] + \mu\mIMF(x^\alpha)\\
            &+x^\alpha F(h_1,h_2)- \p_x \EOmegaeff(x^\alpha,h_1)u^\alpha\Biggr]\\
            &=
            \overline{\overline{\Biggl(\int \frac{\dd x \dd u}{2\pi}\, |\p^2_x \EOmegaeff(x,h_1)|\exp\Biggl[-\frac{\Delta q}{2}\left(u - i\frac{\Delta A x - \p_x\EOmegaeff(x,h_1)}{\Delta q}\right)^2}} \\
            &\overline{\overline{+ \frac{1}{2}\Delta C x^2 + x F(h_1,h_2) + \mu \mIMF(x) - \frac{(\Delta A x -\p_x\EOmegaeff(x,h_1))^2}{2\Delta q}\Biggr]\Biggr)^n}}\\
            &= \overline{\overline{\left(\int \dd x\,,\frac{ |\p^2_x \EOmegaeff(x,h_1)|}{\sqrt{2\pi\Delta q}} \exp\left[\frac{1}{2}\Delta C x^2 + x F(h_1,h_2) + \mu \mIMF(x) - \frac{(\Delta A x -\p_x\EOmegaeff(x,h_1))^2}{2\Delta q}\right]\right)^n}}\,,
        \end{split}
    \end{align}
\end{widetext}
where in the second equality, we have used a Hubbard-Stratonovich transformation. In the third equality, we have defined an effective energy $\EOmegaeff(x,h)$ and a field $F(h_1,h_2)$ as
\begin{align}
    \begin{split}
        \EOmegaeff(x,h) &\equiv E_I(x,a+t_R) - J_0 m x - \sqrt{q_o}h x\\
        F(h_1,h_2) &\equiv v+ h_1\frac{A_o}{\sqrt{q_o}} + h_2 \sqrt{C_o - \frac{A_o^2}{q_o}}\,.
    \end{split}
    \label{eq:Eomegaeff}
\end{align}
Finally in the last equality, we have rotated the integration domain of the fields $u^\alpha$ in the complex plane, defined the Gaussian average $\overline{\overline{\cdots}}\equiv \int \frac{\dd h_1\dd h_2}{2\pi}\ee^{-\frac{h_1^2}{2} - \frac{h_2^2}{2}}$ and observed that the integrals over the different replicated fields factorize. Note that within the replica symmetric ansatz, the distribution of the auxiliary field  $u$ is Gaussian, and it can be integrated out. The last line defines, up to a normalization constant, the single spin probability distribution $P^\text{\tiny{SK}}_\Omega(x,h_1,h_2)$ as
\begin{widetext}
    \begin{equation}
        P_\Omega(x,h_1,h_2) \propto |\p^2_x \EOmegaeff(x,h_1)|\exp\left[\frac{1}{2}\Delta C x^2 + x F(h_1,h_2) + \mu \mIMF(x) - \frac{(\Delta A x - \p_x \EOmegaeff(x,h_1))^2}{2\Delta q}\right]\,.
    \end{equation}
Within the replica-symmetric ansatz, the saddle point equations in Eq.~\eqref{eq:app_sp_SUSY_Omega} become
\begin{align}
    \begin{split}
        \Delta A &= \overline{\overline{\langle xu \rangle_\Omega - \langle x\rangle_\Omega\langle u \rangle_\Omega}} - t_R\\
        A_o &= \overline{\overline{\langle x\rangle_\Omega\langle u \rangle_\Omega}}\\
        \Delta C &= \overline{\overline{\langle u^2\rangle_\Omega - \langle u \rangle_\Omega^2}}\\
        C_o &= \overline{\overline{\langle u \rangle_\Omega^2}} \\
        \Delta q &= \overline{\overline{\langle x^2 \rangle_\Omega - \langle x\rangle_\Omega^2}}\\
        q_o &= \overline{\overline{\langle x \rangle_\Omega^2}}\\
        m &= \overline{\overline{\langle x\rangle_\Omega}}\\
        v &= -J_0 \overline{\overline{\langle u \rangle_\Omega}}\,,      
    \end{split}
\end{align}
computing the Gaussian average over the field $u$ yields the set of equations in Eq.~\eqref{eq:saddle_point_grand-potential} of the main text.\\

\tocless\subsubsection{Alternative expression of $P_\Omega$}\label{app:rewriting_P_Omega}

An alternative expression of the probability distribution $P_\Omega(x,h_1,h_2)$ can be obtained by evaluating the grand-potential average of a generic test function $f(x)$:
\begin{align}\label{eq:app_calculation_P_alternative}
    \begin{split}
        \langle f \rangle_\Omega &= \int \dd x\,P(x,h_1,h_2)f(x)\\
        &\propto \int \dd x\, |\p^2 \EOmegaeff(x,h_1)| \exp \left[\frac{1}{2}\left(\Delta C - \frac{\Delta A^2}{\Delta q} \right)x^2 + x F(h_1,h_2) + \mu \mIMF(x) \right]\\
        &\times \exp\left[- \frac{\left(\p_x \EOmegaeff\right)^2}{2\Delta q} +\frac{\Delta A x}{\Delta q}\p_x \EOmegaeff \right]f(x)\\
        &= \int \dd x\,\int \dd h_0\, \delta(h_0 - \p_x \EOmegaeff)|\p^2_x \EOmegaeff| \exp \left[\frac{1}{2}\left(\Delta C - \frac{\Delta A^2}{\Delta q} \right)x^2 + x F(h_1,h_2) + \mu \mIMF(x) \right]\\
        &\times \exp\left[- \frac{h_0^2}{2\Delta q} +\frac{\Delta A x}{\Delta q}h_0 \right]f(x)\\
        &= \int \dd h_0\, \ee^{-\frac{h_0^2}{2 \Delta q}}\sum_{y \in \Crt_x[\EOmegaeff(x,h_1 + h_0)]}\exp\left[\frac{1}{2}\left(\Delta C - \frac{\Delta A^2}{\Delta q} \right)y^2 + y F(h_1,h_2) + \mu \mIMF(y) +\frac{\Delta A y}{\Delta q}h_0 \right]f(y)\\
        &\propto \left\langle \sum_{y \in \Crt \EOmegaeff(x,h_1+h_0)}\ee^{\frac{1}{2}\left(\Delta C - \frac{\Delta A^2}{\Delta q} \right)y^2 + y F(h_1,h_2) + \mu \mIMF(y) +\frac{\Delta A y}{\Delta q}h_0} f(y)\right\rangle_{h_0}\,,
    \end{split}
\end{align}
where we made use of the properties of the Dirac delta, and we have defined in the last line a Gaussian average over the field $h_0$, with variance $\Delta q$. We can thus give an alternative expression of the single-spin grand-potential probability density as
\begin{equation}\label{eq:app_P_Omega_alternative}
    P_\Omega(x,h_1,h_2) \propto \left\langle\sum_{y \in \Crt\left[\EOmegaeff(x,h_1+h_0)\right]} \ee^{\frac{1}{2}\left(\Delta C - \frac{\Delta A^2}{\Delta q} \right)y^2 + y F(h_1,h_2) + \mu \mIMF(y) +\frac{\Delta A y}{\Delta q}h_0}\delta (x-y) \right\rangle_{h_0}\, 
\end{equation}
\end{widetext}
where $\EOmegaeff(x,h)$ is defined in~\eqref{eq:Eomegaeff}.\\

\tocless\subsubsection{The effective energy is convex if and only if supersymmetry is preserved}\label{app:convexity_implies_susy}

If the effective grand-potential energy $\EOmegaeff(x,h_1)$ is a convex function of $x$ for all the realizations of the noise $h_1$, i.e. if $a+t_R\leq 0$, then supersymmetry is preserved, namely $\Delta A=A_o=C_o=\Delta C=v=0$. This can be seen through a self-consistent condition, by evaluating the saddle point equations around the supersymmetric point. For the parameter $v$, we get in fact,
\begin{align}\label{eq:v_is_0_in_SUSY}
    \begin{split}
        v &= \frac{J_0}{\Delta q}\overline{\overline{\langle \p_x \EOmegaeff\rangle_\Omega}}\\
        &= \frac{J_0}{\Delta q}\overline{\overline{ \left\langle \sum_{y \in \Crt_x\left[\EOmegaeff(x,h_1+h_0)\right]} h_0 \ee^{\mu \mIMF(y)}\right\rangle_{h_0}}}\\
        &= \frac{J_0}{\Delta q}\overline{\overline{\left\langle h_0 \right\rangle_{h_0}}} = 0\,. 
    \end{split}
\end{align}
In the second equality, we used the alternative definition of the single-spin distribution $P_\Omega$ in the supersymmetric phase, given by Eq.~\eqref{eq:app_P_Omega_alternative}. In the third equality, we used the fact that, since $\EOmegaeff(x,h)$ is a convex function of $x$ for all $h$, there is one critical point for every realization of the Gaussian field $h_0$, and the mean field index evaluates to zero because of the convexity of the effective energy. In the last equality, we used the fact that the mean of $h_0$ is, by definition, $0$. This calculation implies that the self-consistent conditions $C_o=A_o=0$ are verified too, since these order parameters are all proportional to $\langle\p_x \EOmegaeff\rangle_\Omega$. The self-consistent condition for the parameter $\Delta C$ reads instead
\begin{align}
    \begin{split}
        &\Delta C =-\frac{1}{\Delta q}\left[1 - \Delta q^{-1}\overline{\overline{\langle \left(\p_x \EOmegaeff\right)^2\rangle_\Omega}}\right]\\
        &= -\frac{1}{\Delta q}\\
        &+ \Delta q^{-2}\overline{\overline{\left\langle\sum_{y \in \Crt_x\left[\EOmegaeff(x,h_1+h_0)\right]} h_0^2 \ee^{\mu \mIMF(y)}\right\rangle_{h_0}}}\\
        &= -\frac{1}{\Delta q} + \frac{1}{\Delta q^2}\overline{\overline{\langle  h_0^2\rangle_{h_0}}}\\
        &= 0\,,
    \end{split}
\end{align}
where we used a similar reasoning to the one adopted in Eq.~\eqref{eq:v_is_0_in_SUSY}. Finally, we verify the self-consistent condition for $\Delta A$ in the following way,
\begin{widetext}
\begin{align}
    \begin{split}
        \Delta A &= \overline{\overline{\Delta q^{-1}\langle x\p_x \EOmegaeff\rangle_\Omega}} - t_R =\overline{\overline{\Biggl\langle \sum_{y \in \Crt_x\left[\EOmegaeff(x,h_1+h_0)\right]} \Biggl[\frac{h_0y}{\Delta q} - \frac{1}{\p_x^2 \EOmegaeff(y,h_1+h_0)}\Biggr]\Biggr\rangle_{h_0}}}\\
        &= \overline{\overline{\Biggl\langle \sum_{y \in \Crt_x\left[\EOmegaeff(x,h_1+h_0)\right]} \Biggl[\frac{\dd y}{\dd h_0} - \frac{1}{\p_x^2 \EOmegaeff(y,h_1+h_0)}\Biggr]\Biggr\rangle_{h_0}}}\\
        &=-\overline{\overline{\left\langle  \sum_{y \in \Crt_x\left[\EOmegaeff(x,h_1+h_0)\right]} \frac{\dd}{\dd h_0} \left(\p_x\EOmegaeff(x,h_1+h_0)\rvert_{x=y(h_0)}\right)\right\rangle_{h_0}}} = 0\,,
    \end{split}
\end{align}
\end{widetext}
where the third equality follows from integration by parts, while in the final equality we used the fact that $y$ is a critical point of the function $\EOmegaeff$. These calculations demonstrate that when $a+t_R\leq 0$, the supersymmetry condition is self-consistently verified. Note that this is true regardless of the magnetization $m$ being zero or not. Moreover, these self-consistent conditions hold also in the case $\Delta q \to 0$, when $h_0$ concentrates around the origin. 

On the other hand, the self consistent condition for a supersymmetric phase is violated when the mean field potential $\EOmegaeff(x,h)$ is non-convex for some value of $h$, i.e. when $a+t_R>0$. Let us compute, in this case, the quantity $\Delta C + C_o$. There is a range of values of $h_0$ for which the number of critical points of $\EOmegaeff(x,h_1+h_0)$ is three: two minima and a saddle. The equality obtained in the supersymmetric case becomes therefore an upper bound, where only one minimum among the three critical points is counted. We thus obtain
\begin{align}
    \begin{split}
        \Delta C + C_o &= -\frac{1}{\Delta q}\left[1 - \frac{1}{\Delta q}\overline{\overline{\langle (\p_x \EOmegaeff)^2\rangle_\Omega}}\right]\\
        &> -\frac{1}{\Delta q} + \frac{1}{\Delta q}\overline{\overline{\langle h^2_0 \rangle_{h_0}}} =0\,,
    \end{split}
\end{align}
which demonstrates the violation of the supersymmetric self-consistent condition.

\subsection{$\Omega(\mu)=0$ in the supersymmetric phase}\label{app:susy_implies_grand-potential_zero}

In the supersymmetric phase, $\Delta A=A_o=\Delta C=C_o=v=0$. Starting from Eq.~\eqref{eq:app_grand-potential_second} evaluated within the replica-symmetric ansatz, we can integrate out the auxiliary fields $u^\alpha$, as done in Eq.~\eqref{eq:app_calculation_exponentialweight_grandpotential_rs}, and obtain the expression
\begin{widetext}
\begin{align}
    \begin{split}        
        \OmegaSK(\mu) &= -\overline{\overline{\log \int \dd x \frac{\lvert \p^2_x \EOmegaeff(x,h_1)\rvert}{\sqrt{2\pi \Delta q}}\exp\left[ x F(h_1,h_2) + \mu\mIMF(x)-\frac{\left( \p_x \EOmegaeff(x,h_1)\right)^2}{2\Delta q}\right]}}\,,
    \end{split}
\end{align}
\end{widetext}
which can be rewritten, using Eq.~\eqref{eq:app_calculation_P_alternative} in the supersymmetric phase, as
\begin{equation}
     \OmegaSK(\mu) = -\overline{\overline{\log \left\langle \sum_{y \in \Crt_x\left[ \EOmegaeff(x,h_1+h_0)\right]} \ee^{\mu\mIMF(y)}\right\rangle_{h_0}}}\,.
\end{equation}
The right hand side involves a sum over the critical points of the effective energy $\EOmegaeff$ defined in Eq.~\eqref{eq:Eomegaeff}, weighted by an exponential factor according to their index. As discussed in the preceding Section, in the supersymmetric phase the effective energy $\EOmegaeff(x,h)$ is a convex function of $x$ for any realization of the field $h$. Its unique critical point is thus a minimum, which has index zero by definition. The grand potential in the supersymmetric phase thus evaluates to
\begin{equation}
    \OmegaSK(\mu) = -\overline{\overline{\log 1}} = 0\,,
\end{equation}
which is the result announced in the main text.

\subsection{Derivation of Eq.~\eqref{eq:susceptibilities}}\label{app:interpretation_SUSY}

In this Appendix we discuss how some order parameters appearing in the Kac-Rice calculation can be interpreted, in the replica symmetric phase, as susceptibilities of the grand-potential $\Omega^\text{\tiny{SK}}(\mu)$ with respect to perturbations of the single-site energy $E_I$ described by Eq.~\eqref{eq:single_site_perturbed}. The different perturbations independently change the mean-field grand-potential action $S_\Omega$ in Eq.~\eqref{eq:app_grand-potential_second}. The perturbation proportional to $\epsilon_A$ changes $S_\Omega$ by an amount $\Delta S_{\Omega,\,A}$, given by
\begin{align}
    \begin{split}
        &\Delta S_{\Omega,\,A} \equiv \epsilon_A \sum_\alpha u^\alpha x^\alpha \\
        &- \sum_\alpha \log \frac{|\p^2_x E_I(x^\alpha,a+t_R^\alpha+\epsilon_A)|}{|\p^2_x E_I(x^\alpha,a+t_R^\alpha)|} \\
        &+ \mu\sum_\alpha \Theta(-\p_x^2 E_I(x^\alpha,a+t_R^\alpha+\epsilon_A)) \\
        &- \mu\sum_\alpha \Theta(-\p_x^2 E_I(x^\alpha,a+t_R^\alpha))\,,  
    \end{split}
\end{align}
therefore, the derivative with respect to $\epsilon_A$ of the grand-potential becomes, within the replica symmetric ansatz,
\begin{align}
    \begin{split}
        \p_{\epsilon_A} \Omega^\text{\tiny{SK}}(\mu)\rvert_{\epsilon_A=0} &= -\lim_{n\to 0}\frac{1}{n} \sum_\alpha\Biggl[ \langle u^\alpha x^\alpha \rangle_\Omega \\&- \frac{1}{2}\left\langle \frac{1}{\lvert\p_x^2 E_I(x^\alpha, a+t_R^\alpha)\rvert}\right\rangle_\Omega\Biggr]\\
        &= -\overline{\overline{\langle ux\rangle_\Omega}} + \frac{1}{2}t_R \\
        &= -A_o - \Delta A\,,
    \end{split}
\end{align}
thus proving the first relation in Eq.~\eqref{eq:susceptibilities}. The perturbation proportional to $\epsilon_C$ introduces a random fields on each soft spin site. It thus couples different replicas. The associated change in the mean field energy $\Delta S_{\Omega,\,C}$ is
\begin{align}
    \begin{split}
         \Delta S_{\Omega,\,C} = \epsilon_C\sum_{\alpha,\,\beta} u^\alpha u^\beta\,.
    \end{split}
\end{align}
The derivative with respect to $\epsilon_C$ of the grand-potential becomes
\begin{equation}
    \begin{split}
        \p_{\epsilon_C} \Omega^\text{\tiny{SK}}(\mu)\rvert_{\epsilon_C=0} &= -\lim_{n\to 0} \frac{1}{n}\sum_{\alpha,\,\beta} \langle u^\alpha u^\beta\rangle_\Omega \\
        &= -\left[\overline{\overline{\langle u^2\rangle_\Omega}} - \overline{\overline{\langle u\rangle_\Omega^2}}\right]\\
        &= -\Delta C\,.
    \end{split}
\end{equation}
This is the second relation in Eq.~\eqref{eq:susceptibilities}. Finally, the perturbation proportional to $\epsilon_v$ yields a change in the mean field grand-potential energy $\Delta S_{\Omega,\,v}$ given by
\begin{equation}
    \Delta S_{\Omega,\,v} = J_0\epsilon_v \sum_\alpha u^\alpha\,, 
\end{equation}
and the associated derivative of the grand-potential reads
\begin{align}
    \begin{split}
        \p_{\epsilon_v} \Omega^\text{\tiny{SK}}(\mu)\rvert_{\epsilon_v =0} &= -\lim_{n\to 0} \frac{1}{n} J_0\sum_\alpha \langle u^\alpha\rangle_\Omega \\
        &= - J_0\overline{\overline{\langle u \rangle_\Omega}}\\
        &= -v\,,
    \end{split}    
\end{align}
which is the last relation in Eq.~\eqref{eq:susceptibilities} in the main text.\\

\subsection{Direct sampling of critical points in finite size systems}\label{app:numerical_sampling_critical_points}
The direct sampling of critical points in finite size system is performed following the algorithm of~\cite{yamamuraGeometricLandscapeAnnealing2024}. Given a system of $N$ soft spins, we sample $N_J$ realizations of the connectivity matrix $\bJSK$.  For each realization of the connectivity matrix, we initialize the system from random initial conditions, and we find the closest critical point using Newton-conjugate gradient descent on the squared norm of the gradient of the energy, $\left[\bnabla E(\bx,a)\right]^2$, with $E(\bx,a)$ given by Eq.~\eqref{eq:E_CIM}. When sampling minima, we use Newton-conjugate gradient descent on the energy $E(\bx,a)$ of the system. This procedure is iterated for $N_\text{\tiny{samples}}$ times for {\it each} realization.  A deduplication procedure is employed to avoid counting the same critical point multiple times. Pairs of critical points whose Euclidean distance is smaller then $10^{-6}$ are identified as the same critical point. To produce the scatter-plots in Fig.~\ref{fig:critical_points_SK}, we used $N=12$, $N_J=10$ and $N_\text{\tiny{samples}}=5\times 10^5$.\\

\subsection{Argument for the stability of  solutions of $\Omega(\mu)$ with $m=0$}\label{app:stability_paramagnetic_susy}

The following argument rationalizes the finding that the most abundant critical point  have zero net magnetization in the supersymmetry-broken region, even for moderately large values of the ferromagnetic alignment $J_0$. Let us consider small perturbations to the magnetization $\delta m^\alpha$ and to the associated Lagrange multiplier $i\delta v^\alpha$ around the  $m=0$, replica-symmetric, supersymmetry-breaking saddle point of the grand-potential $\OmegaSK(\mu)$ in Eq.~\eqref{eq:app_grand-potential_second}, where $m_\alpha=v_\alpha=0$. The perturbation components $\delta m_\alpha$ are real, while the components $i\delta v_\alpha$ are pure imaginary numbers, since the steepest descent path for the auxiliary variables $\bv$ across which the saddle point is evaluated is orthogonal to the real axis. The change of the grand-potential up to quadratic order in these perturbations  is 
\begin{equation}\label{eq:app_perturbation_Omega}
    n 
    \delta \OmegaSK(\mu) = \begin{bmatrix} \delta \bmm \\ \delta \bv\end{bmatrix} \cdot \begin{bmatrix} \frac{\p^2 \OmegaSK}{\p m_\alpha m_\beta} & i\frac{\p^2 \OmegaSK}{\p m_\alpha \p v_\beta} \\ i\frac{\p^2 \OmegaSK }{\p v_\alpha \p m_\beta} & -\frac{\p^2 \OmegaSK}{\p v_\alpha \p v_\beta} \end{bmatrix}\begin{bmatrix} \delta \bmm \\ \delta \bv\end{bmatrix}\,,
\end{equation}
where we have defined a $2n\times2n$ matrix, made up by four $n\times n$ blocks. Their entries read
\begin{align}\label{eq:app_entries_stability_Omega}
    \begin{split}
        \frac{\p^2 \OmegaSK}{\p m_\alpha \p m_\beta} &= -J_0^2 \overline{\overline{\langle u^2 \rangle_\Omega}}\delta_{\alpha\beta} =-J_0^2\Delta C\delta_{\alpha\beta}\\
        \frac{\p^2 \OmegaSK}{\p m_\alpha \p v_\beta} &= \left(1 -J_0 \overline{\overline{\langle u x \rangle_\Omega}}\right)\delta_{\alpha\beta} \\
        &= \left(1 - J_0 \Delta A + J_0 t_R\right)\delta_{\alpha\beta}\\
        \frac{\p^2 \OmegaSK}{\p v_\alpha \p v_\beta} &= -\overline{\overline{\langle x^2 \rangle_\Omega}}\delta_{\alpha\beta} = -\Delta q\delta_{\alpha\beta}\,,
    \end{split}
\end{align}
where we used the fact that in the replica-symmetric  phase with zero net magnetization $v=m=q_o=C_o=A_o =0$. All $n\times n$ sub-blocks are diagonal. The  $m=0$ solution is stable if all the eigenvalues of the Hessian matrix in Eq.~\eqref{eq:app_perturbation_Omega} are positive. To check whether this condition is obeyed, let us observe that when integrating numerically the saddle point equations in Eq.~\eqref{eq:saddle_point_grand-potential} for $J_0=0$, we obtain that $\Delta C<0$ both for $\mu=0$ and for a large and negative value of $\mu$, as shown in Fig.~\ref{fig:app_DeltaC_Co}.
\begin{figure}
    \includegraphics[width=\columnwidth]{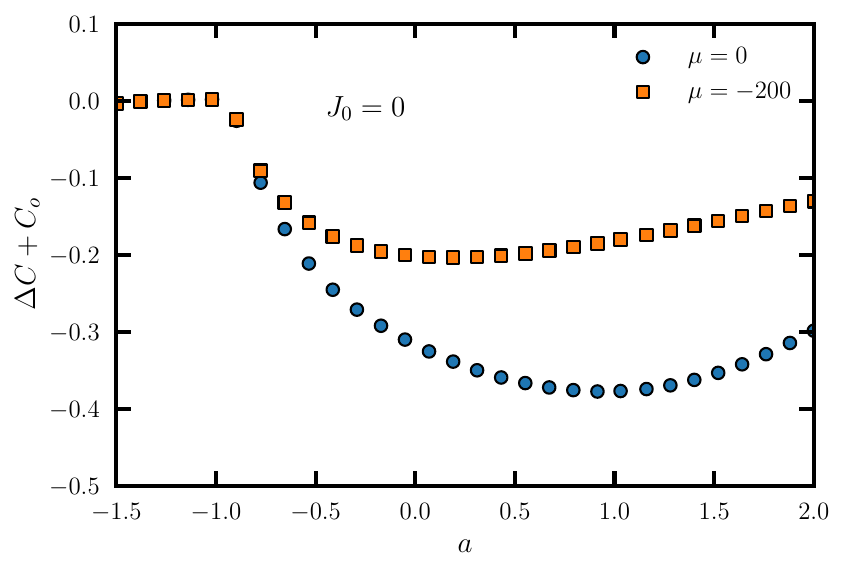}
    \caption{\textbf{Negative values of $\Delta C$ in the  supersymmetry-breaking phase with zero net magnetization.} Plot of $\Delta C$ as a function of the laser gain in the absence of ferromagnetic bias, obtained from the numerical solution of Eq.~\eqref{eq:saddle_point_grand-potential}. We use two distinct values of the chemical potential $\mu$, corresponding to the properties of the most abundant critical points and the properties of the most abundant minima. The latter are obtained by taking a large negative value of $\mu$. $\Delta C$ is less than zero in the supersymmetry-broken phase.\label{fig:app_DeltaC_Co}}
\end{figure}
We thus assume that in the  $m=0$ phase, $\Delta C<0$. Then, the top left block of the Hessian in Eq.~\eqref{eq:app_perturbation_Omega}, associated to fluctuations of the magnetization is positive definite. The matrix in Eq.~\eqref{eq:app_perturbation_Omega} has therefore one eigenvalue with a positive definite real part, and thus the  $m=0$ solution is stable as long as both eigenvalues are positive. The stability condition becomes thus
\begin{equation}
    \det \begin{bmatrix} -J_0^2\Delta C & i(1 - J_0 \Delta A +J_0 t_R)  \\ i(1 - J_0 \Delta A +J_0 t_R) & \Delta q \end{bmatrix} >0\,,
\end{equation}
which yields
\begin{equation}
    \left[1 - J_0(\Delta A + t_R)\right]^2 - J_0^2\Delta C \Delta q >0\,.
\end{equation}
This condition is satisfied for $\Delta C<0$. This stability analysis rationalizes our numerical observation that the most abundant critical points  have zero net magnetization even in presence of a net ferromagnetic coupling strength. \\

\section{Dynamical mean field theory}\label{app:dmft}
In this Appendix we derive the dynamical mean field equations for the coherent Ising machine, presented in Eq.~\eqref{eq:dmft}, using the dynamical cavity method~\cite{agoritsas2018out}. The starting point is the gradient descent dynamics for a system of $N$ spins given by Eq.~\eqref{eq:gradient_descent}, which we rewrite here as
\begin{equation}\label{eq:gd_spin_i}
    \tau \dot x_i = -\p_x E_I(x_i,a(t)) + \frac{J_0}{N}\sum_j x_j + \frac{1}{\sqrt{N}}\sum_j J_{ij}x_j\,. 
\end{equation}
We want to describe the effective dynamics of a single spin of the system. To do so, we consider a system made by $N+1$ spins, where an additional degree of freedom $x_0(t)$ has been added. We refer to this additional spin as the cavity variable. The new spin interacts with the other spins in the system through the ferromagnetic coupling $J_0/N$ and through a set of symmetric, disordered couplings $J_{0j}=J_{j0}$, which are independent, identically distributed Gaussian variables of mean $0$ and unit variance. The cavity variable perturbs the dynamics of the other spins in the system through a field $h_i(t)$, which reads 
\begin{align}\label{eq:h_i}
    \begin{split}
        h_i(t) &\equiv \frac{J_0}{N}x_0(t) +\frac{J_{i0}}{\sqrt{N}} x_0(t) \\\ &= \frac{J_{i0}}{\sqrt{N}} x_0(t) + O(N^{-1})\,,
    \end{split}
\end{align} 
in the second line, we have observed that the field $h_i(t)$ is of order $O(N^{-1/2})$. Let us denote by $\tilde x_i(t)$ the trajectories of spins $i=1,\ldots,N$ perturbed by the cavity degree of freedom. The equation of motion of the perturbed spins is
\begin{equation}\label{eq:xtilde}
    \begin{split}
     \tau \dot{\widetilde{x}}_i &= -\p_x E_I(\tilde{x}_i,a(t)) + \frac{J_0}{N}\sum_{j\neq0} \tilde{x}_j + \frac{1}{\sqrt{N}}\sum_{j\neq0} J_{ij}\tilde{x}_j \\
     &+ h_i(t)\,, 
     \end{split}
\end{equation}
with $h_i(t)$ given by Eq.~\eqref{eq:h_i}. On the other hand, the equation of motion of the cavity variable is thus 
\begin{equation}\label{eq:gd_cavity}
    \tau\dot x_0 = -E_I'(x_0,a(t)) + \frac{J_0}{N}\sum_{i\neq 0} \tilde x_i + \frac{1}{\sqrt{N}}\sum_{i\neq0} J_{0i}\tilde x_i\,. 
\end{equation}
All the terms in Eq.~\eqref{eq:gd_spin_i},~\eqref{eq:xtilde} and Eq.~\eqref{eq:gd_cavity} are of order $O(1)$. On the other hand we see from Eq.~\eqref{eq:h_i} that  $h_i(t)  \sim O(N^{-1/2})$. We can thus treat the interaction between spin $i$ and spin $0$ as a small perturbation to the trajectory of spin $i$ in the system without the cavity variable, $x_i(t)$. Using linear response theory, we thus write 
\begin{align}\label{eq:linear_response}
    \begin{split}
        \tilde x_i&(t) \approx x_i(t) + \sum_j \int_0^t \dd\tau\, \frac{\delta x_i(t)}{\delta h_j(\tau)}\Bigg\rvert_{h_j=0} h_j(\tau) \\
        &= x_i(t) + \frac{1}{\sqrt{N}}\sum_j \int_0^t \dd\tau\, \frac{\delta x_i(t)}{\delta h_j(\tau)}\Bigg\rvert_{h_j=0} J_{j0}x_0(\tau)\,,
    \end{split}
\end{align}
Where $\frac{\delta x_i(t)}{\delta h_j(t')}\Big\lvert_{h_j=0}$ is the variation of the value $x_i(t)$ under a small change in the field $h_j(t')$ acting on spin $j$, evaluated in the absence of any external field. Substituting Eq.~\eqref{eq:linear_response} into Eq.~\eqref{eq:gd_cavity} we obtain, to leading order in $N$, 
\begin{align}\label{eq:gd_cavity_intermediate}
    \begin{split}
        \tau\dot x_0 &\approx -\p_x E_I(x_0,a(t)) + \frac{J_0}{N} \sum_{i\neq 0} x_i(t)  \\
        &+ \frac{1}{\sqrt{N}}\sum_{i\neq 0} J_{0i} x_i \\
        &+ \frac{1}{N}\sum_{i,j\neq 0} \int_0^t \dd\tau \frac{\delta x_i(t)}{\delta h_j(\tau)}J_{0j}J_{i0}x_0(\tau)\,.
    \end{split}
\end{align}
This equation describes the dynamics of the cavity variable in terms of the dynamics of the unperturbed spins $x_i(t)$, which are independent from the dynamics of the cavity. The spins $x_i(t)$ can be considered as a set random variables under different realizations of the disordered couplings $J_{ij}$. In the thermodynamic limit, the cross-correlations among two spin $x_i$ and $x_j$ read
\begin{align}\label{eq:correlation_xixj}
    \begin{split}
        & \left\langle \left(x_i(t) - \left\langle x_i(t)\right\rangle_\bJ\right)\left(x_j(t') - \left\langle x_j(t')\right\rangle_\bJ\right)\right\rangle_\bJ \\
        &= \frac{1}{N}\sum_{k,l} \left\langle J_{ik}J_{jl}x_i(t)x_j(t')\right\rangle_\bJ \\        &=\delta_{ij}\left\langle x_i(t)x_j(t')\right\rangle_\bJ + O(N^{-1})\,.
    \end{split}
\end{align}
This equation demonstrates that in the thermodynamic limit the dynamics of two different spins under different realizations of the quenched disorder are independent of each other. The term $N^{-1}J_0\sum_{i\neq0} x_i(t)$ in Eq.~\eqref{eq:gd_cavity_intermediate} can then be written as
\begin{align}\label{eq:J0Nterm}
    \begin{split}
        \frac{J_0}{N}\sum_i x_i(t) &= \frac{J_0}{N} \sum_i \left\langle x_i(t)\right\rangle_\bJ \\
        &+ \frac{J_0}{N}\sum_i x_i(t) - \left\langle x_i(t)\right\rangle_\bJ \\
        &= \frac{J_0}{N}\sum_i \left\langle x_i(t)\right\rangle_\bJ  + O(N^{-1/2})\\
        &\equiv J_0 m(t) + O(N^{-1/2})\,.
    \end{split}
\end{align}
In the second equality, we applied the central limit theorem, while the third equality defines the time-dependent magnetization of the system $m(t)$. Similarly, the response term in Eq.~\eqref{eq:gd_cavity_intermediate} concentrates around its average value in the thermodynamic limit:
\begin{align}\label{eq:response}
    \begin{split}
        &\frac{1}{N} 
        \int_0^t \dd\tau\, \sum_{i,j} \frac{\delta x_i(t)}{\delta h_j(\tau)}\Biggr\rvert_{h_j=0} J_{0i}J_{j0} x_0(\tau)\\ &\approx \int_0^t\dd\tau\, \left\langle\frac{1}{N}\sum_{i}\frac{\delta x_i(t)}{\delta h_i(\tau)}\Biggr\rvert_{h_j=0}\right\rangle_\bJ x_0(\tau)\\
        &\equiv \int_0^t \dd\tau\, R(t,\tau) x_0(\tau)\,,
    \end{split}
\end{align}
where the second line defines the response function $R(t,\tau)$. Finally, the second term in Eq.~\eqref{eq:gd_cavity_intermediate} can by written, using the central limit theorem, as a Gaussian noise $\xi(t)$ with mean $\langle\xi(t)\rangle_\bJ =0$ and variance 
\begin{align}\label{eq:noise_correlation}
    \begin{split}               \left\langle\xi(t)\xi(t')\right\rangle_\bJ &= \frac{1}{N}\sum_{i,j}\left\langle J_{0i}J_{0j}x_i(t) x_j(t')\right\rangle_\bJ\\
    &= \frac{1}{N}\sum_i \left\langle x_i(t) x_i(t')\right\rangle_\bJ \equiv C(t,t')\,,
    \end{split}
\end{align}
where the last equality defines the correlation function $C(t,t')$. Finally, we observe that, since the system is fully connected, the dynamics of one spin is equivalent to the dynamics of any other spin. This is true for the dynamics of the cavity variable as well. The index $0$ in Eq.~\eqref{eq:gd_cavity_intermediate} can then be dropped, as well as the index $i$ in the second-to-last lines of Eqs.~\eqref{eq:J0Nterm},~\eqref{eq:response}, and ~\eqref{eq:noise_correlation}. We thus obtain an effective equation of motion for a spin $x(t)$ in the coherent Ising machine, namely
\begin{align}\label{eq::app_dmft}
    \begin{split}
        \tau\dot x(t) &=-\p_x E_I(x,a) \\
        &+ J_0m(t) + \int_0^t\dd\tau R(t,\tau) x(\tau) + \xi(\tau)\\
        \left\langle \xi(t)\xi(t')\right\rangle_\bJ &= C(t,t') = \left\langle x(t)x(t')\right\rangle_\bJ\\
        m(t) &= \left\langle x(t)\right\rangle_\bJ\\
        R(t,t') &= \left\langle \frac{\delta x(t)}{\delta h(t')}\Biggr\rvert_{h=0}\right\rangle_\bJ = \left\langle\frac{\delta x(t)}{\delta \xi(t')}\right\rangle_\bJ\,.
    \end{split}
\end{align}
In the last line, we used the fact that the variation of $x(t)$ with respect to a variation in an external field $h(t')$ can equivalently be obtained by making an infinitesimal change in the noise at time $t'$. Equation ~\eqref{eq::app_dmft} is the dynamical mean field theory of the coherent Ising machine.  It provides a set of self-consistent equations that can be solved to obtain the temporal correlation $C(t,t')$, magnetization $m(t)$, and response $R(t,t')$ given initial conditions on $x$. We show how to find such solutions next.\\ 

\subsection{Numerical solution of the dynamical mean field theory}\label{app:dmft_numerics}
Equation~\eqref{eq::app_dmft} can be solved numerically for any initial condition $x(0)$ on a discrete temporal grid, $t_i=i\Delta t$ with $i=0\,,\ldots, N_\text{steps}$, by means of an iterative self-consistent scheme~\cite{eissfeller1992new, eissfeller1994mean, royNumericalImplementationDynamical2019}. Starting from an initial guess of the function $m(t_i)$, $C(t_i,t_j)$ and $R(t_i,t_j)$ we can generate a set of $N_\text{traj}$ independent trajectories $\{x_\alpha(t_i)\}_{\alpha=1}^{N_\text{traj}}$ using an Euler scheme. To do this, we first generate and store $N_\text{traj} \times N_\text{steps}$ realizations of the noise, with correlations $\langle \xi_\alpha(t_i) \xi_{\beta}(t_j)\rangle = \delta_{\alpha\beta} C(t_i,t_j)$. We can then use the self-consistent equations for $m(t)$ and $C(t,t')$ given by Eq.~\eqref{eq::app_dmft} to compute the new values of the correlation function $C_\text{new}(t_i,t_j)$ and of the magnetization $m_\text{new}(t_i)$, as
\begin{align}
    \begin{split}
        C_\text{new}(t_i,t_j) &= \frac{1}{N_\text{traj}} \sum_{\alpha=1}^{N_\text{traj}} x_\alpha(t_i)x_\alpha(t_j)\\
        m_\text{new}(t_i) &= \frac{1}{N_\text{traj}} \sum_{\alpha=1}^{N_\text{traj}} x_\alpha(t_i)\,.
    \end{split}
\end{align}
To compute the response function of the system, we can use the Novikov relation~\cite{novikov1965functionals, royNumericalImplementationDynamical2019}, which relates the response function to the realizations of the noise along a trajectory as
\begin{equation}
    R(t,t') = \left\langle x(t)\int \dd\tau\, C^{-1}(t',\tau)\xi(\tau)\right\rangle_\bJ\,, 
\end{equation}
where $C^{-1}(t,t')$ is the inverse of the correlation function $C(t,t')$, defined by the identity $\int \dd \tau C(t,\tau)C^{-1}(\tau,t') =\delta(t-t')$. For a discrete time grid and a finite number of trajectories, Novikov relation  reads
\begin{align}
    \begin{split}
    R_\text{new}(t_i,t_j) &= \frac{1}{N_\text{traj} \Delta t}\sum_{\alpha=1}^{N_\text{traj}}x_\alpha(t_i) \\
    &\times \sum_{k=1}^{N_\text{steps}} C^{-1}(t_i,t_k)\xi_\alpha(t_k)\,.
    \end{split}
\end{align}
Once the new values of the order parameters $C_\text{new}$, $R_\text{new}$ and $m_\text{new}$ are known, we update the current values of the order parameter using a linear interpolation between the current values and the new values with parameter $\gamma$, namely
\begin{align}
    \begin{split}
        C(t_i,t_j) &\leftarrow \gamma C_\text{new}(t_i,t_j) + (1-\gamma)C(t_i,t_j)\\
        R(t_i,t_j) &\leftarrow \gamma R_\text{new}(t_i,t_j) + (1-\gamma)R(t_i,t_j)\\
        m(t_i) &\leftarrow \gamma m_\text{new}(t_i) + (1-\gamma)m(t_i)\,.
    \end{split}
\end{align}
This interpolation scheme is  used to avoid large jumps as well as oscillations between  updates. Once the order parameters are updated, the procedure is repeated until convergence is achieved. As a convergence criterion, we impose 
\begin{align}
    \begin{split}
        &\frac{1}{N_\text{steps}}\sum_i (m(t_i)-m_\text{new}(t_i))^2 \\
        &+ \frac{1}{N_\text{steps}^2}\sum_{i,j} \left(C(t_i,t_j) - C_\text{new}(t_i,t_j)\right)^2 \leq \epsilon\,,
    \end{split}
\end{align}
with $\epsilon>0$. The complexity of this algorithm is of  order $O(N_\text{traj}\times N_\text{steps}^2)$. The results presented in the main text are obtained using $N_\text{traj}=20\,000$, $T/\Delta t = 5\,000$, $\Delta t =0.15$, $\gamma=0.3$, $\epsilon=10^{-4}$. 

\begin{widetext}
\section{Wishart planted ensemble}\label{app:wpe_replicas}

In this Section, we compute the average of the single spin distribution at the global minima of the CIM-WPE in Eq.~\eqref{eq:Pwpe}. We begin by rewriting the single spin distribution as the derivative of a modified free energy function,
\begin{align}\label{eq:app_delta_WPE}
    \begin{split}
    \frac{1}{N}  \sum
    _{i=1}^N \langle\delta(x_i -x) \rangle_{\bx, \bz^\mu}^{\text{\tiny{WPE}}} &= \frac{1}{N} \Biggl \langle \frac{\int \prod_{i=1}^N \dd y_i\, \sum_i \delta(y_i - x) \ee^{-\beta E^{\text{\tiny{WPE}}}(\by)}}{\int \prod_i \dd y_i\, \ee^{-\beta E^{\text{\tiny{WPE}}}(\by)}} \Biggr \rangle _{\bz^\mu} \\
    &= \frac{1}{N} \Biggl \langle \partial_\ell \log \int \dd\mathbf{y}\, \ee^{-\beta E^{\text{\tiny{WPE}}} + \ell \sum_i \delta(y_i -x)} \Biggr \rangle_{\bz^\mu}\Biggl |_{\ell = 0} \\
    &\equiv \frac{1}{N} \partial_\ell \langle \log Z_\ell^{\text{\tiny{WPE}}} \rangle_{\bz^\mu} |_{\ell = 0}\,,
    \end{split}
\end{align}
where the last line defines a modified partition function $Z_\ell^\text{\tiny{WPE}}$. To compute the average over the disorder, we employ the replica trick described in Eq.~\eqref{eq:replicatrick}, which yields
\begin{align}\label{eq:app_wpe_replicated_partition}
    \begin{split}
        \frac{1}{N} \partial_\ell \langle \log Z_\ell^{\text{\tiny{WPE}}} \rangle_{\bz^\mu} |_{\ell = 0} &= \frac{1}{N} \partial_\ell \lim_{n \to 0} \frac{1}{n} \langle (Z_\ell^{\text{\tiny{WPE}}})^n \rangle_{\bz^\mu}\rvert_{\ell=0} \\
        &= \frac{1}{N} \partial_\ell \lim_{n\to 0}\frac{1}{n} \Biggl \langle \int \prod_{\alpha=1}^n \dd \mathbf{y}^\alpha\, \ee^{-\beta \sum_\alpha E^{\text{\tiny{WPE}}}(\mathbf{y}^\alpha) + \ell \sum_\alpha \sum_i \delta(y_i^\alpha - x)} \Biggr \rangle_{\bz^\mu}\Biggl|_{\ell = 0}\,.
    \end{split}
\end{align}
Upon substitution of Eq.~\eqref{eq:E_WPE} and Eq.~\eqref{eq:def_WPE} into the replicated partition function in Eq.~\eqref{eq:app_wpe_replicated_partition} we obtain
\begin{align} \label{eq:wpereplicas1}
     \left\langle\left( Z^\text{\tiny{WPE}}_\ell \right)^n\right\rangle_{\bz^\mu} = \int_{\by^\alpha} \,\left\langle \exp \left[ -\frac{\beta}{2N} \sum_{\alpha=1}^n\sum_{\mu=1}^M(\bw^\mu \cdot \mathbf{y}^\alpha)^2 + \frac{\beta}{2N} \sum_{\alpha,\, \mu}\sum_{i=1}^N (w^\mu_i y_i^\alpha)^2 - \sum_{\alpha,\,i}\left[\beta E_I(y_i^\alpha,a) - \ell \delta(y_i^\alpha - x)\right]\right] \right \rangle_{\bz^\mu}\,,
\end{align}
where we used the shorthanded notation $\int_{\by^\alpha} \equiv \int \prod_{\alpha=1}^n\dd\by^\alpha$ to denote an integral over the replicated space of soft-spin systems. The second term in the exponential comes from the fact that the connectivity matrix in the Wishart planted ensemble is defined to be traceless.  We use Hubbard-Stratonovich transformation in Eq.~\eqref{eq:app_Hubbard_Stratonovich}  to linearize the term $(\bw^\mu \cdot \mathbf{y}^\alpha)^2$ while introducing a set of $n\times M$ continuous degrees of freedom $\{\phi^\alpha_\mu\}$ for $\alpha=1,\ldots,\,n$ and $\mu=1,\ldots,\,M$. We obtain
\begin{align}\label{eq:wpeZ_zexplicit}
    \begin{split}
     \langle (Z^\text{\tiny{WPE}}_\ell)^n \rangle_{\bz^\mu}  &= \Biggl \langle \int_{\by^\alpha,\,\bphi^\alpha} \exp \Biggl[-\sum_{\mu \alpha} \frac{(\phi_\mu^\alpha)^2}{2} -i \sqrt{\frac{\beta}{N}}\sum_{\mu \alpha} \phi_\mu^\alpha (\bw^\mu \cdot \mathbf{y}^\alpha)  \\
     &+ \frac{\beta}{2N} \sum_{\alpha,\, \mu,\, i} ( w_i^\mu y_i^\alpha)^2 -\sum_{\alpha,\,i}\left[\beta E_I(y_i^\alpha,a) - \ell \delta(y_i^\alpha - x)\right]\Biggr] \Biggr \rangle_{\bz^\mu}\,,
    \end{split}
\end{align}
where we employed a shorthanded notation for the integral $\int_{\by^\alpha,\,\bphi^\alpha} \equiv \int \prod_{\alpha} \frac{\dd \by^\alpha \dd\bphi^\alpha}{\sqrt{(2\pi)^N}}$. The patterns $\bw^\mu$ are constructed from the realizations of the quenched disorder $\bz^\mu$, as described in Eq.~\eqref{eq:wmu} of the main text. Writing explicitly the average over the quenched disorder we obtain 
\begin{align}\label{eq:app_Zn_with_z}
    \begin{split}
     \langle (Z^\text{\tiny{WPE}}_\ell)^n \rangle_{\bz^\mu} &=   \int \prod_{\mu=1}^M \frac{\dd \mathbf{z}^\mu}{(\sqrt{2\pi})^N} \int_{\by^\alpha,\,\bphi^\alpha}  \exp\left[-\frac{1}{2}\sum_{\mu,\alpha} (\phi_{\mu}^\alpha)^2 - \sum_{\alpha,i}\left[ \beta E_I(y_i^\alpha, a)  - \ell \delta(y_i^\alpha - x)\right]\right] \\
     &\times\prod_{\mu=1}^M \exp \left[-\frac{1}{2}\sum_{i,j} z_i^\mu \left(\delta_{ij} - \frac{\beta}{N} \left( \sum_{k,\alpha} \Sigma^{1/2}_{ik} \Sigma_{jk}^{1/2} (y_k^\alpha)^2\right)\right) z_j^\mu -i \sqrt{\frac{\beta}{N}} \sum_\alpha \phi_\mu^\alpha \by^\alpha \cdot \bSigma^{1/2} \mathbf{z}^\mu \right]\,,\end{split}
\end{align}
where we see that the traceless property of the connectivity matrix $\bJWPE$ has the effect of changing the correlations of the quenched disorder. The inverse of the covariance matrix of the quenched variables $\{\bz^\mu\}$ is now given by the matrix
\begin{align}
    \begin{split}
    \delta_{ij} &- \frac{\beta}{N}\sum_{k,\alpha} \Sigma_{ik}^{1/2} \Sigma_{jk}^{1/2} (y_k^\alpha)^2 = \delta_{ij} - \frac{\beta}{N-1}\sum_{\alpha,k} \left(\delta_{ik} - \frac{1}{N}\right)\left(\delta_{jk} - \frac{1}{N}\right)(y_k^\alpha)^2 \\
    &= \delta_{ij} - \frac{\beta}{N-1}\sum_{\alpha=1}^n \left(\delta_{ij} (y_i^\alpha)^2 - \frac{1}{N}\left((y_j^\alpha)^2+(y_i^\alpha)^2\right)+ \frac{1}{N^2}\sum_{k} (y_k^\alpha)^2\right)\\
    &\approx \delta_{ij}\left(1 - \frac{\beta}{N}\sum_{\alpha=1}^n (y^\alpha_i)^2\right)\,,
    \end{split}
\end{align}
where in the last passage we have neglected all the terms of order $O(N^{-2})$ or higher. By defining the variable $\bGamma^\mu$ as
\begin{equation}\label{eq:Gammamu}
    \bGamma^\mu \equiv \sqrt{\frac{\beta}{N}}\sum_\alpha \phi_\mu^\alpha \bSigma^{1/2}\by^\alpha  
\end{equation} 
we evaluate the Gaussian integral over the quenched disorder $\bz^\mu$ in Eq.~\eqref{eq:wpeZ_zexplicit}, and we expand the argument of the exponential to the leading order in powers of $N^{-1}$, obtaining 
\begin{align}\label{eq:app_int_z}
    \begin{split}
        \prod_{\mu=1}^M \int \frac{\dd \bz^\mu}{(2\pi)^{N/2}}\,&\exp\left[-\frac{1}{2}\sum_i (z_i^\mu)^2\left[1 - \frac{\beta}{N}\sum_\alpha (y_i^\alpha)^2 \right] - i \bGamma^\mu \cdot \bz^\mu \right]\\
        &= \exp\left[-\frac{M}{2}\sum_i \log\left(1 - \frac{\beta}{N}\sum_\alpha (y_i^\alpha)^2\right)\right]\prod_\mu \exp\left[-\frac{1}{2}\sum_i \frac{(\Gamma_i^\mu)^2}{1 - \frac{\beta}{N}\sum_\alpha (y_i^\alpha)^2}\right]\\
        &\approx \exp\left[\frac{M\beta}{2N}\sum_{\alpha,i} (y_i^\alpha)^2 - \frac{1}{2}\bGamma^\mu\cdot \bGamma^\mu\right]\\
        &= \exp\left[\frac{\alpha \beta}{2}\sum_{\alpha,i} (y_i^\alpha)^2 - \frac{1}{2}\bGamma^\mu\cdot \bGamma^\mu\right]\,.
    \end{split}
\end{align}
From this expression, we observe that the lack of a diagonal element in $\mathbf{J}^{\text{\tiny{WPE}}}$ contributes to a shift in the effective laser gain by an amount $\alpha=M / N$, the fraction of patterns stored in the connectivity matrix. Plugging Eq.~\eqref{eq:app_int_z} into Eq.~\eqref{eq:app_Zn_with_z} yields
\begin{align}
    \begin{split}
    \left\langle\left( Z^\text{\tiny{WPE}}_\ell\right)^n\right\rangle_{\bz^\mu}&=\int_{\by^\alpha, \bphi^\alpha}  \exp \left[ {-\sum_{\mu \alpha} \frac{(\phi_{\mu}^\alpha)^2}{2} - \beta \sum_{\alpha,i} E_I(y_i^\alpha,a+\alpha)- \frac{1}{2} \sum_{i\mu} (\Gamma_i^\mu)^2 + \ell \sum_{\alpha,i} \delta(y_i^\alpha - x)} \right]\\
     &= \int_{\by^\alpha, \bphi^\alpha}\left( \prod_\mu \exp \left[ {-\sum_{ \alpha} \frac{1}{2} (\phi_\mu^\alpha)^2 - \frac{\beta}{2N} \sum_{\alpha,\beta} \left( \phi_\mu^\alpha \phi_\mu^\beta \mathbf{y}^\alpha \cdot \mathbf{y^\beta} - \frac{\phi_\mu^\alpha \phi_\mu^\beta}{N} (\mathbf{y}^\alpha \cdot \mathbf{t})(\mathbf{y}^\beta \cdot \mathbf{t})\right)} \right] \right)\\
     &\times \ee^{- \beta \sum_{i\alpha} E_I(y_i^\alpha, a+\alpha) + \ell \sum_{\alpha,i} \delta(y_i^\alpha - x) }\,,
     \end{split}
\end{align}
where in the second line we have used the definition of $\bGamma^\mu$ in Eq.~\eqref{eq:Gammamu}. Integrating out the Gaussian variables $\phi^{\alpha}_\mu$
\begin{align}
    \left\langle\left( Z^\text{\tiny{WPE}}_\ell\right)^n\right\rangle_{\bz^\mu} &= \int_{\by^\alpha} \exp \left[-\frac{M}{2} \text{Tr} \log(\mathbf{1}_n + \beta \mathbf{K}) -\beta \sum_{\alpha,i}E_I(y_i^\alpha,a+\alpha) + \ell \sum_{\alpha, i} \delta(y_i^\alpha - x) \right], 
\end{align}

where the matrix $\mathbf{1}_n$ and the matrix $\bK$ are $n\times n$ matrices. The former is the identity matrix in replica space, while the latter is defined as
\begin{align}
    K_{\alpha \beta} &= \frac{1}{N} \mathbf{y}^\alpha\cdot \mathbf{y}^\beta - \frac{1}{N^2}  (\mathbf{y}^\alpha \cdot \mathbf{t})(\mathbf{y}^\beta \cdot \mathbf{t})\,.
\end{align}

Now we use the identity
\begin{align}
    f(a) = \int \dd x\, \delta(x-a) f(a) = \int \dd x\, \delta(x-a) f(x).
\end{align}
to insert a set of order parameters into the replicated partition function $\langle\left( Z_\ell\right)^n\rangle_{\bz_\mu}$. The first one is an $n\times n$ overlap matrix $\mathbf{Q}$, while the second one is a $n$-dimensional magnetization vector $\bmm$. We thus obtain 

\begin{align}
    \begin{split}
     \left\langle\left( Z^\text{\tiny{WPE}}_\ell\right)^n\right\rangle_{\bz^\mu} &=   \int    \left( \prod_\alpha \dd\mathbf{y}^\alpha \right) \int \left( \prod_{\alpha\leq\beta} \dd Q_{\alpha \beta}  \right) \delta\left( Q_{\alpha \beta} - N^{-1} \sum_i y_i^\alpha y_i^\beta \right) \left( \prod_\alpha \dd m_\alpha \right)\delta \left(m_\alpha -N^{-1} \sum_i t_i y_i^\alpha \right) \\
    & \times \exp \left[-\frac{M}{2} \text{Tr}\log(\mathbf{1} + \beta \mathbf{K}) \right] \ee^{-\beta \sum_{i\alpha} E_I(y_i^\alpha,a+\alpha) + \ell \sum_{\alpha, i} \delta(y_i^\alpha- x)}.
    \end{split}
\end{align}

Using the Fourier representation of the Delta distribution, we now introduce a set of Lagrange multipliers,  $r_{\alpha\beta}$ and $\widehat{m}_\alpha$, for each entries of  $\mathbf{Q}$ and $\bmm$, respectively.  
\begin{align}
    \begin{split}
     \left\langle\left( Z^\text{\tiny{WPE}}_\ell\right)^n\right\rangle_{\bz^\mu} &=    \int    \prod_\alpha \dd\mathbf{y}^\alpha \prod_{\alpha\leq \beta} \frac{i\dd r_{\alpha \beta}}{2\pi } \prod_{\alpha \leq\beta} Q_{\alpha \beta} \prod_\alpha \dd m_\alpha \prod_\alpha \frac{i\dd  \widehat{m}_\alpha }{2\pi } \ee^{-\beta \sum_{\alpha, i} E_I(y_i^\alpha,a+\alpha) + \ell \sum_{\alpha, i} \delta(y_i^\alpha- x)} \\
     &\times \exp \left[ -\frac{M}{2} \text{Tr} \log (\mathbf{1} + \beta \mathbf{Q} - \beta \bmm \otimes \bmm)-\sum_{\alpha \beta}\frac{1}{2}r_{\alpha \beta}(NQ_{\alpha \beta} - \sum_{i}y_i^\alpha y_i^\beta) -\sum_\alpha \widehat{m}_\alpha (Nm_\alpha - \sum_i t_i y_i^\alpha)  \right]
     \end{split}
\end{align}

After rescaling $r_{\alpha\beta} \to \alpha \beta^2 r_{\alpha\beta}$, $\widehat{m}_\gamma \to \alpha \beta \widehat{m}_\gamma $, and performing more steps of algebra, we obtain
\begin{align}\label{eq:app_ZWPEl}
    \begin{split}
        \left\langle\left( Z^\text{\tiny{WPE}}_\ell\right)^n\right\rangle_{\bz^\mu} &=  \int \dd\mathbf{\Gamma} \exp N \Biggl[ -\frac{\alpha}{2} \text{Tr}\log(\mathbf{1} + \beta \mathbf{Q} - \beta \bmm \otimes \bmm) - \frac{\alpha \beta^2 }{2} \sum_{\alpha \beta} r_{\alpha \beta}Q_{\alpha \beta} - \alpha \beta \sum_\alpha \widehat{m}_\alpha m_\alpha  + \log Z_{\text{\tiny{MF}},\ell}^{\text{\tiny{WPE}}}\Biggr] \Biggr\rvert_{\ell = 0}\\
        &\equiv \int \dd\bGamma\, \ee^{N S^\text{\tiny{WPE}}[\bGamma]}\,,
    \end{split}
\end{align}
where we denoted by $\mathbf{\Gamma} = \{\mathbf{r}, \mathbf{Q}, \bmm, \hat{\bmm}\}$ the set of various order parameters, and by $\dd \bGamma$ the product of their volume factors in phase space. We also defined an action $S^\text{\tiny{WPE}}(\bGamma)$ and a mean field partition function $Z^\text{\tiny{WPE}}_{\text{\tiny{MF}},\,\ell}$ as
\begin{align}\label{eq:SWPE}
    \begin{split}
    S^{\text{\tiny{WPE}}}[\mathbf{\Gamma}] &= -\frac{\alpha}{2} \text{Tr}\log(\mathbf{1} + \beta \mathbf{Q} - \beta \bmm \otimes \bmm) - \frac{\alpha \beta^2 }{2} \sum_{\alpha \beta} r_{\alpha \beta}Q_{\alpha \beta} - \alpha \beta  \sum_\alpha \widehat{m}_\alpha m_\alpha  + \log Z_{\text{\tiny{MF}}, \ell}^{\text{\tiny{WPE}}}\,, \\Z_{\text{\tiny{MF}}, \ell}^{\text{\tiny{WPE}}} &= \int \prod_\alpha \dd y^\alpha\,\ee^{\frac{\alpha \beta^2 }{2} \sum_{\alpha \beta} r_{\alpha \beta}y^\alpha y^\beta + \alpha \beta\sum_{\alpha} \widehat{m}_\alpha  y^\alpha - \beta \sum_{\alpha} E_I(y^\alpha,a+\alpha) + \ell \sum_{\alpha} \delta(y^\alpha- x)  }\,. 
    \end{split}
\end{align}
Substituting Eq.~\eqref{eq:app_ZWPEl} into Eq.~\eqref{eq:app_delta_WPE} and taking the limit $N\to\infty$, we obtain
\begin{align}\label{eq:delta_as_S}
    \lim_{N\to \infty}\frac{1}{N}  \sum
    _{i=1}^N \langle\delta(x_i -x) \rangle_{\bx, \bz^\mu}^{\text{\tiny{WPE}}} = \lim_{N\to\infty}\frac{1}{N}\partial_\ell \lim_{n\to 0} \frac{1}{n} \log   \int \dd\mathbf{\Gamma}\, \ee^{NS^{\text{\tiny{WPE}}}[\mathbf{\Gamma}]}.
\end{align}
As in Appendix~\ref{app:Gz_calculation}, we ignore the subleading normalization factors in $\dd\mathbf{\Gamma}$. In the limit $N\to\infty$, the integral on the right hand side of Eq.~\eqref{eq:delta_as_S} is evaluated at the saddle point of the action $S^\text{\tiny{WPE}}(\bGamma)$. The saddle point equations are given in Eq.~\eqref{eq:saddle_point_wpe} of the main text. After the saddle point evaluation, we obtain
\begin{align} \label{eq:wpe_free_energy_saddlepoint}
    \begin{split}
    \lim_{N\to\infty}\frac{1}{N}  \sum
    _{i=1}^N \langle\delta(x_i -x) \rangle_{\bx, \bz^\mu}^{\text{\tiny{WPE}}} &= \partial_\ell \lim_{n\to 0} \frac{1}{n}  \Biggl[ -\frac{\alpha}{2} \text{Tr}\log(\mathbf{1} + \beta \mathbf{Q} - \beta \bmm \otimes \bmm) \\
    &- \frac{\alpha \beta^2 }{2} \sum_{\alpha \beta} r_{\alpha \beta}Q_{\alpha \beta} - \alpha \beta  \sum_\alpha \widehat{m}_\alpha m_\alpha  + \log Z_{\text{\tiny{MF}}, \ell}^{\text{\tiny{WPE}}}\Biggr] \Biggr|_{\ell = 0}.
    \end{split}
\end{align}
The derivative over $\ell$ eliminates all but the last term on the right hand side of the above equation, thus yielding 
\begin{align} \label{eq:wpe_P_final}
    \lim_{N\to \infty}\frac{1}{N}  \sum
    _{i=1}^N \langle\delta(x_i -x) \rangle_{\bx, \bz^\mu}^{\text{\tiny{WPE}}} &= \lim_{n \to 0} \frac{1}{n} \frac{1}{Z_{\text{\tiny{MF}}}^{\text{\tiny{WPE}}}}  \int \prod_\alpha \dd y^\alpha \left( \sum_{\alpha} \delta(y^\alpha- x) \right)\ee^{-\beta E_{\text{\tiny{MF}}}^{\text{\tiny{WPE}}}(\{y^\alpha \}) },
\end{align}

where the evaluation at $\ell = 0$ has been performed. This is Eq.~\eqref{eq:P_WPE_MF} in the main text. In the next Section, we evaluate the expression above using the replica-symmetric ansatz.\\

\subsection{Replica-symmetric ansatz}\label{app:wpe_rs}

In the replica-symmetric phase, we take $Q_{\alpha \beta} = q_d\delta_{\alpha \beta} + q_o(1-\delta_{\alpha \beta}),\ r_{\alpha \beta} = r_d\delta_{\alpha \beta} + r_o(1-\delta_{\alpha \beta})$. We take $m_\alpha = m, \widehat{m}_\alpha = \widehat{m}$. The logarithm of the modified mean field partition function $\log Z_{\text{\tiny{MF}},\,\ell}^{\text{\tiny{WPE}}}$ becomes
\begin{align}
    \begin{split}
    \log Z_{\text{\tiny{MF}},\,\ell}^{\text{\tiny{WPE}}} &= \log \int \prod_\alpha \dd y^\alpha\, \ee^{\frac{\alpha \beta^2 }{2} r_o\left(\sum_\alpha y^\alpha \right)^2 + \frac{\alpha \beta^2}{2} (r_d - r_o)\sum_\alpha (y^\alpha)^2 + \alpha \beta \widehat{m} \sum_{\alpha}  y^\alpha - \beta \sum_{\alpha} E_I(y^\alpha,a+\alpha) + \ell \sum_\alpha \delta(y^\alpha-x)   }  \\
    &=  \log \overline{\prod_\alpha \left[ \int  \dd y^\alpha\,  \ee^{-\beta \sqrt{\alpha r_o}hy^\alpha + \frac{\alpha \beta^2}{2} (r_d -r_o) (y^\alpha)^2 + \alpha \beta \widehat{m}  y^\alpha - \beta E_I(y^a,a+\alpha) + \ell \delta(y^\alpha-x)} \right]}\\
    &= \log \overline{\left[ \int  \dd y^\alpha\,  \ee^{-\beta \sqrt{\alpha r_o}h y^\alpha + \frac{\alpha \beta^2}{2} (r_d -r_o) (y^\alpha)^2 + \alpha \beta \widehat{m} y^\alpha - \beta E_I(y^a,a+\alpha) + \ell  \delta(y^\alpha-x)}\right]^n}\\
    &= n \,\overline{\log \left[ \int  \dd y^\alpha\,\ee^{-\beta \sqrt{\alpha r_o}hy^\alpha + \frac{\alpha \beta^2}{2} (r_d -r_o) (y^\alpha)^2 + \alpha \beta \widehat{m} y^\alpha - \beta E_I(y^a,a+\alpha) + \ell  \delta(y^\alpha-x)}  \right]} + O(n)\\
    &\equiv n\, \overline{\log\int \dd y\,\ee^{-\beta \ERSWPE(x,h) + \ell \delta(y-x)}} + O(n)\,.
    \end{split}
\end{align}
\end{widetext}
In the second line, we performed a Hubbard-Stratonovich transformation to linearize the squared sum over the replicated spins $y^\alpha$, thus introducing a Gaussian random field $h$ of mean zero and unit variance. The overline denotes an average over this Gaussian distribution. In the third line, we used the fact that integral over the different replicas factorizes into the product of $n$ identical integrals. In the fourth line, we expanded to the lowest order in $n$. The last line defined an effective single-site mean field energy $\ERSWPE(x,h)$, which is given in Eq.~\eqref{eq:E_WPE_RS} of the main text. Through these manipulations, Eq.~\eqref{eq:wpe_P_final} in the replica-symmetric phase becomes
\begin{align}
    \begin{split}
    P^\text{\tiny{WPE}}_\text{\tiny{MF}}(x)&= \overline{\ee^{-\beta E_{\text{\tiny{eff}}}^{\text{\tiny{WPE}}}(x,h)}[Z_{\text{\tiny{RS}}}^{\text{\tiny{WPE}}}]^{-1}} \\
    &\equiv P_{\text{\tiny{RS}}}^{\text{\tiny{WPE}}}(x)\,,
    \end{split}
\end{align}
which is Eq.~\eqref{eq:PRSWPE} in the main text. Following relatively similar manipulations to the ones performed  above and in Appendices~\ref{app:Gz_calculation} and~\ref{app:saddle_point_rs}, we obtain the saddle point equations ofEq.~\eqref{eq:saddle_point_wpe} in the main text.\\

\subsection{Stability of the replica symmetric solution}\label{app:replicon_wpe}

To investigate the stability of the replica-symmetric phase, we study the Hessian of the action $S^{\text{\tiny{WPE}}}$ with respect to perturbations in the space of overlaps $Q_{\alpha\beta}$ and Lagrange multipliers $r_{\alpha\beta}$.The Hessian matrix has the following block structure,
\begin{align}\label{eq:app_Hessian_S}
    \begin{split}
     \begin{bmatrix}
        \frac{\partial ^2 S^{\text{\tiny{WPE}}}}{\partial Q_{\alpha \beta} \partial Q_{\gamma \delta}} & \frac{\partial ^2 S^{\text{\tiny{WPE}}} }{\partial Q_{\alpha \beta} \partial r_{\gamma \delta}} \\
        \frac{\partial ^2 S^{\text{\tiny{WPE}}} }{\partial Q_{\alpha \beta} \partial r_{\gamma \delta}} & \frac{\partial ^2 S^{\text{\tiny{WPE}}} }{\partial r_{\alpha \beta} \partial r_{\gamma \delta}} 
    \end{bmatrix}\,.
    \end{split}
\end{align}
Let us observe first that the top right and bottom left blocks are diagonal, because of the term proportional to $Q_{\alpha\beta}r_{\alpha\beta}$ in the action $S^\text{\tiny{WPE}}$ in Eq.~\eqref{eq:SWPE}. The structure of the Hessian matrix is then identical to the one that arises when studying the stability of the replica symmetric phase in the Hopfield or anti-Hopfield models~\cite{amit1987statistical, nokura1998spin}. The condition for the stability of the convex phase derived in those cases applies also here, and it reads
\begin{widetext}
\begin{equation}\label{eq:app_stabiltiy_wpe}
        1 - \frac{1}{\alpha\beta^2}\left[\frac{\p^2 S^\text{\tiny{WPE}}}{\p Q_{\alpha\beta}\p Q_{\alpha\beta}} - 2\frac{\p^2 S^\text{\tiny{WPE}}}{\p Q_{\alpha\beta}\p Q_{\alpha\gamma}} + \frac{\p^2 S^\text{\tiny{WPE}}}{\p Q_{\alpha\beta}\p Q_{\gamma\delta}} \right]\left[\frac{\p^2 S^\text{\tiny{WPE}}}{\p r_{\alpha\beta}\p r_{\alpha\beta}} - 2\frac{\p^2 S^\text{\tiny{WPE}}}{\p r_{\alpha\beta}\p r_{\alpha\gamma}} + \frac{\p^2 S^\text{\tiny{WPE}}}{\p r_{\alpha\beta}\p r_{\gamma\delta}}\right] <0.
\end{equation}
\end{widetext}
We compute first the terms involving derivatives with respect to $Q_{\alpha\beta}$, which depend only on the logarithmic term in the action $S^\text{\tiny{WPE}}$ in Eq.~\eqref{eq:SWPE}. To derive this, we first use the following identity, valid for any continuous function $f$ and any symmetric matrix $\bA$,
\begin{align}
    \begin{split}
    \partial_{A_{\alpha\beta}} \Tr f(\bA) &= f'(\bA)_{\alpha\beta} + f'(\bA)_{\beta\alpha} \\
    &- \delta_{\alpha\beta}f'(\bA)_{\alpha\beta}\,.
    \end{split}
\end{align}

In our case, $f(\bA) = \log (\bA+\bB)$. The first derivative of $f$ is 
\begin{align}
    \begin{split}
        f'(\bA)_{\alpha\beta} &= (\bA+\bB)^{-1}_{\alpha\beta}+(\bA+\bB)^{-1}_{\beta\alpha}\\
        &-\delta_{\alpha\beta}(\bA+\bB)^{-1}_{\alpha\beta}.
    \end{split}
\end{align}

Now we let $\bK \equiv (\bA+\bB)^{-1}$. We seek $\partial_{A_{\gamma \delta}} K_{\alpha \beta}$. Using a matrix calculus identity, we get
\begin{align}
    \partial_{A_{\gamma \delta}} K_{\alpha \beta} &= - \sum_{\rho \sigma} K_{\alpha \rho} \frac{\partial A_{\rho \sigma}}{\partial A_{\gamma \delta}} K_{\sigma \beta }
\end{align}

Because $A$ is symmetric, 
\begin{align}
    \frac{\partial A_{\rho \sigma}}{\partial A_{\gamma \delta}} = \begin{cases}
        \mathbb{I}_{\rho \gamma} \mathbb{I}_{\sigma \delta} + \mathbb{I}_{\rho \delta}\mathbb{I}_{\sigma \gamma},\ \gamma \neq  \delta \\
        \mathbb{I}_{\rho \gamma}\mathbb{I}_{\rho \gamma},\ \gamma =\delta\,,
    \end{cases}
\end{align}
where we have used the symbol $\mathbb{I}$ to represent an indicator function (i.e. a Kronecker delta) to avoid confusion with the $\delta$ representing an index. 

We thus obtain
\begin{align}
    \begin{split}
    &\frac{\p^2\Tr f(A) }{\partial A_{\alpha \beta}\partial A_{\gamma \delta}}  = \begin{cases}
        2\partial_{A_{\gamma \delta}}K_{\alpha \beta},\ \alpha \neq \beta\\
        \partial_{A_{\gamma \delta }}K_{\alpha \alpha },\ \alpha = \beta
    \end{cases} \\
    &= \begin{cases}
        -2(K_{\alpha \gamma}K_{\delta \beta} + K_{\alpha \delta}K_{\beta \gamma}),\ \alpha \neq \beta , \gamma \neq \delta \\
        -2 K_{\alpha \gamma} K_{\gamma \beta},\ \alpha \neq \beta , \gamma =\delta \\
        -(K_{\alpha \gamma}K_{\delta \alpha} + K_{\alpha \delta}K_{\gamma \alpha}),\ \alpha =\beta,  \gamma \neq \delta \\
        - K_{\alpha \gamma}K_{\alpha \gamma},\ \alpha =\beta , \gamma =\delta.
    \end{cases}
    \end{split}
\end{align}

In our case, $\bA+\bB = \mathbf{1}_n+\beta \mathbf{Q} - \beta \bmm \otimes \bmm$. Therefore in the replica symmetric phase, $\bK$ becomes
\begin{align}
    \mathbf{K} &= \left((1 + \Delta \widetilde q)\mathbf{1}_n + \beta(q_o - m^2) \bt_n\otimes\bt_n \right)^{-1}\,,
\end{align}
where $\bt_n \equiv [1,\ldots,1]^\mathrm{T}$ is a $n$-dimensional vector with all entries equal to $1$. Using the Sherman-Morrison formula~\cite{bartlett1951inverse}, we obtain 
\begin{align}
    \begin{split}
    \mathbf{K} &= \frac{1}{1 + \Delta \widetilde q} \mathbf{1}_n - \frac{\beta(q_o - m^2)}{(1 + \Delta \widetilde q)^2}\bt_n\otimes\bt_n\\
    &= -\Delta \widetilde r \mathbf{1}_n - \beta r_o  \bt_n\otimes\bt_n\,,
    \end{split}
\end{align}
where in the second line we used the replica-symmetric saddle point equations given by Eq.~\eqref{eq:sp_wpe_rs}.
We can now compute all the matrix elements of the top left block of the Hessian of the CIM-WPE action in Eq.~\eqref{eq:app_Hessian_S}, obtaining,
\begin{align}\label{eq:app_ppS_QQ}
    \begin{split}
        \frac{\partial^2 S^\text{\tiny{WPE}}}{\partial Q_{\alpha \alpha}\p Q_{\alpha \alpha}} &= \frac{\alpha \beta^2}{2} K_{\alpha \alpha}^2 \\
        \frac{\partial^2 S^\text{\tiny{WPE}}}{\partial Q_{\alpha \alpha}\p Q_{\alpha \gamma}} &= \alpha \beta^2 K_{\alpha \alpha}K_{\gamma \alpha} \\
        \frac{\partial^2 S^\text{\tiny{WPE}}}{\partial Q_{\alpha \alpha}Q_{\gamma\delta}} &= \frac{\alpha\beta^2}{2} \left(K_{\alpha \gamma}K_{\delta \alpha} + K_{\alpha \delta}K_{\gamma \alpha} \right) \\
        \frac{\partial^2 S^\text{\tiny{WPE}}}{\partial Q_{\alpha \alpha}\p Q_{\gamma\gamma}} &= \frac{\alpha \beta^2}{2} K_{\alpha \gamma}^2\\
        \frac{\partial^2 S^\text{\tiny{WPE}}}{\partial Q_{\alpha \beta}\p Q_{\alpha \beta}} &= \alpha \beta^2 (K_{\alpha \alpha}^2 + K_{\alpha \beta}^2) \\
        \frac{\partial^2 S^\text{\tiny{WPE}}}{\partial Q_{\alpha \beta}\p Q_{\alpha \gamma}} &= \alpha \beta^2 (K_{\alpha \alpha}K_{\beta \gamma} + K_{\alpha \gamma}K_{\alpha \beta}) \\
        \frac{\partial^2 S^\text{\tiny{WPE}}}{\partial Q_{\alpha \beta}\p Q_{\gamma\delta}} &= \alpha \beta^2(K_{\alpha \gamma}K_{\delta \beta} + K_{\alpha \delta}K_{\beta \gamma})
    \end{split}
\end{align}

The elements of $\bK$ are, in the replica symmetric phase,
\begin{align}
    \begin{split}
    K_{\alpha\alpha} &= \frac{1}{1+\Delta \widetilde q} - \frac{\beta(q_o-m^2)}{(1+\Delta \widetilde q)^2} \\
    &= -\Delta \widetilde{r} - \beta r_o\\
    K_{\alpha\beta} &= - \frac{\beta(q_o-m^2)}{(1+\Delta \widetilde q)^2}\\
    &= -\beta r_o\,,
    \end{split}
\end{align}
with $\alpha\neq\beta$. 

Now we turn to the computation of derivatives with respect to the Lagrange multipliers $r_{\alpha\beta}$, which involve the term $-\log Z_{\text{\tiny{MF}}}^{\text{\tiny{WPE}}}$ in Eq.~\eqref{eq:SWPE}. Using the fact that $r_{\alpha\beta}$ is a symmetric matrix, we obtain 
\begin{align}
    \begin{split}
    -\partial_{r_{\alpha \beta}}\partial_{r_{\gamma \delta}} \log Z &= -\alpha^2 \beta^4\langle x^\alpha x^\beta x^\gamma x^\delta \rangle_\text{\tiny{MF}}^\text{\tiny{WPE}} \\
    &+ \alpha^2 \beta^4\langle x^\alpha x^\beta \rangle_\text{\tiny{MF}}^\text{\tiny{WPE}} \langle x^\gamma x^\delta \rangle_\text{\tiny{MF}}^\text{\tiny{WPE}}\,.
    \end{split}
\end{align}

In the replica-symmetric phase,  the different types of matrix elements become
\begin{align}\label{eq:app_ppSrr}
    \begin{split}
    \frac{\partial^2 S^\text{\tiny{WPE}}}{\partial r_{\alpha \alpha}\p r_{\alpha \alpha}} &= \alpha^2 \beta^4 \left[ \overline{\langle x^4\rangle_\text{\tiny{RS}}^\text{\tiny{WPE}}} - \overline{\langle x^2\rangle_\text{\tiny{RS}}^\text{\tiny{WPE}}}^2 \right]\\
    \frac{\partial^2 S^\text{\tiny{WPE}}}{\partial r_{\alpha \alpha}\p r_{\alpha \gamma}} &= \alpha^2 \beta^4 \left[\overline{\langle x^3 \rangle_\text{\tiny{RS}}^\text{\tiny{WPE}} \langle x\rangle_\text{\tiny{RS}}^\text{\tiny{WPE}}} - \overline{\langle x^2 \rangle_\text{\tiny{RS}}^\text{\tiny{WPE}}}\,\overline{ \left(\langle x \rangle_\text{\tiny{RS}}^\text{\tiny{WPE}}\right)^2}\right] \\
    \frac{\partial^2 S^\text{\tiny{WPE}}}{\partial r_{\alpha \alpha}\p r_{\gamma \delta}} &= \alpha^2\beta^4 \Bigl[ \overline{\langle x^2 \rangle_\text{\tiny{RS}}^\text{\tiny{WPE}} \left(\langle x \rangle_\text{\tiny{RS}}^\text{\tiny{WPE}}\right)^2} \\&- \overline{\langle x^2\rangle_\text{\tiny{RS}}^\text{\tiny{WPE}} }\,\overline{\left(\langle x \rangle_\text{\tiny{RS}}^\text{\tiny{WPE}}\right)^2}\Bigr] = 0\\
    \frac{\partial^2 S^\text{\tiny{WPE}}}{\partial r_{\alpha \alpha}\p r_{\gamma\gamma}} &= \alpha^2 \beta^4 \left[ \overline{\left(\langle x^2 \rangle_\text{\tiny{RS}}^\text{\tiny{WPE}}\right)^2} - \overline{\langle x^2 \rangle_\text{\tiny{RS}}^\text{\tiny{WPE}}}^2\right] = 0\\
    \frac{\partial^2 S^\text{\tiny{WPE}}}{\partial r_{\alpha \beta}\p r_{\alpha \beta}} &= \alpha^2 \beta^4 \left[\overline{\left(\langle x^2 \rangle_\text{\tiny{RS}}^\text{\tiny{WPE}}\right)^2}- \overline{\left(\langle x \rangle_\text{\tiny{RS}}^\text{\tiny{WPE}}\right)^2}^2 \right] \\
    \frac{\partial^2 S^\text{\tiny{WPE}}}{\partial r_{\alpha \beta}\p r_{\alpha \gamma}} &= \alpha^2 \beta^4 \left[\overline{\langle x^2 \rangle_\text{\tiny{RS}}^\text{\tiny{WPE}} \left(\langle x \rangle_\text{\tiny{RS}}^\text{\tiny{WPE}}\right)^2} - \overline{\left(\langle x \rangle_\text{\tiny{RS}}^\text{\tiny{WPE}}\right)^2}^2 \right] \\
    \frac{\partial^2 S^\text{\tiny{WPE}}}{\partial r_{\alpha \beta}\p r_{\gamma \delta}} &= \alpha^2\beta^4 \left[ \overline{\left(\langle x\rangle_\text{\tiny{RS}}^\text{\tiny{WPE}}\right)^4} - \overline{\left(\langle x \rangle_\text{\tiny{RS}}^\text{\tiny{WPE}}\right)^2}^2\right] = 0\,,
    \end{split}
\end{align}
where $\langle\ldots\rangle_\text{\tiny{RS}}^\text{\tiny{WPE}}$ is an average over the Boltzmann distribution of the effective mean field energy $\EWPERS(x,h)$, for a given realization of the quenched field $h$, namely,
\begin{equation}
    \langle\ldots\rangle_\text{\tiny{RS}}^\text{\tiny{WPE}} \equiv \frac{\int \dd x\,\ldots \ee^{-\beta \ERSWPE(x,h)}}{\int \dd x\, \ee^{-\beta \ERSWPE(x,h)}}\,,
\end{equation}
and we recall that the overline denotes an average over the realizations of the quenched field $h$, which follows a Gaussian distribution of zero mean and unit variance. Plugging the relevant entries of Eq.~\eqref{eq:app_ppS_QQ} and Eq.~\eqref{eq:app_ppSrr} into the stability condition in Eq.~\eqref{eq:app_stabiltiy_wpe}, and evalutaing the resulting expression in the low temperature limit $\beta \to\infty$, we obtain the stability condition
\begin{align}
    \begin{split}
        \widetilde{\Lambda}_\text{\tiny{R}}^{\text{\tiny{WPE}}} &\equiv -1 + \alpha\Delta \widetilde r^2\beta^2 \overline{\left[\left\langle x^2 - \left(\langle x \rangle^\text{\tiny WPE}_\text{\tiny{RS}}\right)^2\right\rangle_\text{\tiny{RS}}^\text{\tiny{WPE}}\right]^2} \\
        &= -1 + \alpha\Delta \widetilde{r}^2\overline{\left[\p^2_x \ERSWPE(x,h)\rvert_{x=x^*(h)}\right]^{-2}} <0\,,
    \end{split}
\end{align}
where $x^*(h)$ is the global minimum of the effective energy $\ERSWPE(x,h)$ for a fixed realization of $h$. This is the stability condition given by Eq.~\eqref{eq:wpe_replicon} of the main text.
\bibliography{biblio}
\end{document}